\documentclass[aps,prl,twocolumn,amsmath,amssymb,floatfix]{revtex4-2}

\usepackage{graphicx}
\usepackage{dcolumn}
\usepackage{bm}
\usepackage{comment}
\usepackage[T1]{fontenc}
\usepackage{dblfloatfix}
\usepackage{booktabs}
\usepackage{float}
\usepackage{amsmath}
\usepackage{nicefrac}
\usepackage{xcolor}
\usepackage[T1]{fontenc}
\usepackage[activate={true,nocompatibility}]{microtype}
\usepackage[colorlinks]{hyperref}
\usepackage{footnote}
\usepackage{ulem}
\hypersetup{
    colorlinks=true,
    linkcolor=blue,
    filecolor=black,
    citecolor=blue,
    urlcolor=blue
}
\usepackage{soul}
\newcommand{\isot}[2]{\ensuremath{^{#1}\textrm{#2}}}
\newcommand{\LaBr}{\mbox{\ensuremath{\textrm{LaBr}_3}}}

\newcommand{\gam}{\ensuremath{\gamma}}
\newcommand{\bet}{\ensuremath{\beta}}
\newcommand{\jpi}[2]{\ensuremath{\textrm{#1}^{#2}}}

\begin{document}

\preprint{APS/123-QED}

\title{Universal Effective Charges in the $sd$ and $fp$ Shells} 

\author{T.~H.~Ogunbeku$^{a}$}
\author{J.~M.~Allmond$^{b}$}
\author{T.~J.~Gray$^{b,c}$}
\author{W.-J.~Ong$^{a}$}
\author{B. A.~Brown$^{d}$}
\author{A. Gargano$^{e}$}
\author{R. Grzywacz$^{c}$}
\author{J. D. Holt$^{f,g}$}
\author{A. O. Macchiavelli$^{b}$}
\author{T. Miyagi$^{h,i,j,k}$} 
\author{S.~Neupane$^{a}$}
\author{B. C. Rasco$^{b}$}
\author{H. Schatz$^{d,l}$}
\author{B. M. Sherrill$^{d,l}$} 
\author{O. B. Tarasov$^{d}$}
\author{H. Arora$^{m}$} 
\author{A. D. Ayangeakaa$^{n,o}$} 
\author{H. C. Berg$^{d,l}$} 
\author{J. M. Berkman$^{d,p}$}
\author{D. L. Bleuel$^{a}$}
\author{K. Bosmpotinis$^{d,l}$}
\author{M. P. Carpenter$^{q}$} 
\author{G. Cerizza$^{d}$} 
\author{A. Chester$^{d}$} 
\author{J. M. Christie$^{c}$}
\author{I. Cox$^{c}$} 
\author{H. L. Crawford$^{r}$} 
\author{B. P. Crider$^{s}$} 
\author{J. Davis$^{t}$}
\author{A. A. Doetsch$^{d,l}$} 
\author{J. G. Duarte$^{a}$}
\author{A. Estrade$^{m}$} 
\author{A. Fija\l\text{kowska}$^{u}$}
\author{C. Frantzis$^{d,l}$}
\author{T. Gaballah$^{d,s}$} 
\author{E. C. Good$^{t}$} 
\author{K. Haak$^{d,l}$} 
\author{S. Hanai$^{v}$} 
\author{J. T. Harke$^{a}$}
\author{A. C. Hartley$^{d,l}$} 
\author{K. Hermansen$^{d,l}$} 
\author{D. E. M. Hoff$^{a}$}
\author{D. Hoskins III$^{c}$}
\author{J. Huffman$^{d,l}$}
\author{P. Van Isacker$^{w}$} 
\author{R. Jain$^{a}$}
\author{M. Karny$^{u}$} 
\author{T. T. King$^{b}$}
\author{N. Kitamura$^{c}$}
\author{K. Kolos$^{a}$}
\author{A. Laminack$^{b}$}
\author{S. N. Liddick$^{d,p}$}
\author{B. Longfellow$^{a}$}
\author{R. S. Lubna$^{d}$}
\author{S. Lyons$^{t}$} 
\author{M. Madurga$^{c}$}
\author{M.~J.~Mogannam$^{d,p}$} 
\author{G. Owens-Fryar$^{d,l}$}
\author{J.~R.~Palomino$^{s}$}
\author{M. M. Rajabali$^{x}$} 
\author{A. L. Richard$^{y}$} 
\author{I.J.~Richardson$^{d}$}
\author{E. K. Ronning$^{d,p}$} 
\author{G. E. Rose$^{z}$} 
\author{T. J. Ruland$^{b}$}
\author{K. P. Rykaczewski$^{b}$}
\author{N. D. Scielzo$^{a}$}
\author{D. P. Scriven$^{d}$}
\author{D. Seweryniak$^{aa}$}
\author{K. Siegl$^{c}$}
\author{M. Singh$^{c}$}
\author{A. Spyrou$^{d,l}$} 
\author{M. Stepaniuk$^{u}$}
\author{A. E. Stuchbery$^{ab}$} 
\author{A. Sweet$^{a}$}
\author{V. Tripathi$^{ac}$}
\author{A. Tsantiri$^{d,l}$} 
\author{S. Uthayakumaar$^{d}$}
\author{W. B. Walters$^{ad}$} 
\author{S. Watters$^{d,l}$}
\author{Z. Xu$^{c}$}
\author{R. Yokoyama$^{v}$}

\affiliation{$^{a}$Lawrence Livermore National Laboratory, Livermore, California 94550, USA}
\affiliation{$^{b}$Physics Division, Oak Ridge National Laboratory, Oak Ridge, Tennessee 37831, USA}
\affiliation{$^{c}$Department of Physics and Astronomy, University of Tennessee, Knoxville, Tennessee 37966, USA}
\affiliation{$^{d}$Facility for Rare Isotope Beams, Michigan State University, East Lansing, Michigan 48824, USA}
\affiliation{$^{e}$Istituto Nazionale di Fisica Nucleare, Complesso Universitario di Monte S. Angelo, Napoli I-80126, Italy}
\affiliation{$^{f}$TRIUMF, 4004 Wesbrook Mall, Vancouver, British Columbia V6T 2A3, Canada}
\affiliation{$^{g}$Department of Physics, McGill University, 3600 Rue University, Montréal, Quebec H3A 2T8, Canada}
\affiliation{$^{h}$Department of Nuclear Physics, Technische Universität Darmstadt, 64289 Darmstadt, Germany}
\affiliation{$^{i}$ExtreMe Matter Institute EMMI, GSI Helmholtzzentrum für Schwerionenforschung GmbH, 64291 Darmstadt, Germany}
\affiliation{$^{j}$Max-Planck-Institut für Kernphysik, Saupfercheckweg 1, 69117 Heidelberg, Germany}
\affiliation{$^{k}$Center for Computational Sciences, University of Tsukuba, 1-1-1 Tennodai, Tsukuba, Ibraki 305-8577, Japan}
\affiliation{$^{l}$Department of Physics and Astronomy, Michigan State University, East Lansing, Michigan 48824, USA}
\affiliation{$^{m}$Department of Physics, Central Michigan University, Mount Pleasant, Michigan 48859, USA}
\affiliation{$^{n}$Department of Physics and Astronomy, University of North Carolina at Chapel Hill, Chapel Hill, North Carolina 27599, USA}
\affiliation{$^{o}$Triangle Universities Nuclear Laboratory, Duke University, Durham, North Carolina 27708, USA}
\affiliation{$^{p}$Department of Chemistry, Michigan State University, East Lansing, Michigan 48824, USA}
\affiliation{$^{q}$Physics Division, Argonne National Laboratory, Argonne, Illinois 60439, USA}
\affiliation{$^{r}$Nuclear Science Division, Lawrence Berkeley National Laboratory, Berkeley, California 94720, USA}
\affiliation{$^{s}$Department of Physics and Astronomy, Mississippi State University, Mississippi State, Mississippi 39762, USA}
\affiliation{$^{t}$Pacific Northwest National Laboratory, Richland, Washington 99354, USA}
\affiliation{$^{u}$Faculty of Physics, University of Warsaw, PL 02-093 Warsaw, Poland}
\affiliation{$^{v}$Center for Nuclear Study, University of Tokyo, Wako, Saitama 351-0198, Japan}
\affiliation{$^{w}$Grand Accélérateur National d’Ions Lourds, CEA/DSM-CNRS/IN2P3, Boulevard Henri Becquerel, F-14076 Caen, France}
\affiliation{$^{x}$Physics Department, Tennessee Technological University, Cookeville, Tennessee 38505, USA}
\affiliation{$^{y}$Department of Physics and Astronomy, Ohio University, Athens, Ohio 45701, USA}
\affiliation{$^{z}$University of California, Berkeley, Berkeley, California 94704, USA}
\affiliation{$^{aa}$Physics Division, Argonne National Laboratory, Argonne, Illinois 60439, USA}
\affiliation{$^{ab}$Department of Nuclear Physics and Accelerator Applications, Research School of Physics, The Australian National University, ACT 2601, Canberra, Australia}
\affiliation{$^{ac}$Department of Physics, Florida State University, Tallahassee, FL, 32306, USA}
\affiliation{$^{ad}$Department of Chemistry and Biochemistry, University of Maryland, College Park, Maryland 20742, USA}

\date{\today}

\begin{abstract}
The 247-keV state in $^{54}$Sc, populated in the $\beta$ decay of $^{54}$Ca, is reported here as a nanosecond isomer with a half-life of 26.0(22) ns. 
The state is interpreted as the $1^+$ member of the $\pi f_{7/2}\otimes\nu f_{5/2}$ spin-coupled multiplet, which decays to the $3^+,\pi f_{7/2} \otimes \nu p_{1/2}$ ground state. 
The new half-life corresponds to a pure $E2$ transition with a strength of 1.93(16) W.u., providing the most precise, unambiguous $B(E2)$ value in the neutron-rich $fp$ region to date for a nucleus with valence protons above $Z=20$. 
Notably, it is roughly four times larger than the $B(E2; 1/2^{-} \rightarrow 5/2^{-})$ value in $^{55}$Ca. 
The results, as compared to semi-empirical and $ab~initio$ shell-model calculations, indicate (1) a weak $N=34$ sub-shell gap relative to $N = 32$, (2) a large $E2$ enhancement in Sc as compared to Ca due to $1p-1h$ proton excitations across $Z=28$, and (3) empirical effective proton and neutron charges, $e_\pi$ = 1.30(8)$e$ and $e_\nu$ = 0.452(7)$e$, respectively, that are in contrast to reports of $e_\pi \approx 1.1-1.15e$ and $e_\nu \approx 0.6-0.8e$ for $fp$-shell nuclei near $N = Z$. 
We demonstrate that these reports are erroneous and that, in fact, a universal set of effective charges can be used across the $sd$ and $fp$ shells.\\
                              
\end{abstract}

\maketitle

The nuclear shell model provides a fundamental view of nuclear structure \cite{Mayer1949,Haxel1949}. 
It presumes independent particle motion in a spherical mean field with strong spin-orbit coupling, leading to significant energy gaps at ``magic'' proton and neutron numbers, $Z, N = $ 2, 8, 20, 28, 50, 82 and 126. 
Modeling nuclear structure then involves understanding the correlations or effective interactions between valence nucleons, which mix single-particle configurations, within a finite, truncated model space. 
However, exotic nuclei with increasing neutron excess relative to stable nuclei become sensitive to different aspects of the nuclear forces  \cite{Warner2004,Dobaczewski2007,satula2009,Otsuka2020}. 
This leads to an evolution of the single-particle orbitals, causing standard magic numbers to disappear and new ones to emerge \cite{Otsuka2005,Otsuka2010}. 
All effects from outside of the truncated model space are then treated with effective operators, which have been reported to have isospin and orbital dependencies beyond the $sd$ shell \cite{Valiente-Dobon2009,Jungclaus2024}. 

Considerable interest has been directed towards the emergence of new sub-shell gaps at $N = 32$ and 34 in neutron-rich nuclei within the $fp$ shell. These gaps are currently understood to be driven by the elevation of the $\nu 1f_{5/2}$ orbital as protons are removed from the $\pi 1f_{7/2}$ orbital. 
The existence of the $N = 32$ sub-shell gap has been extensively confirmed near $Z = 20$ by the measurement of high $E(2^+_1)$ values in \isot{50}{Ar} \cite{Steppenbeck2015}, \isot{52}{Ca} \cite{Huck1985,Gade2006}, \isot{54}{Ti} \cite{Janssens2002,Liddick2004}, and \isot{56}{Cr} \cite{Prisciandaro2001,Mantica2003,Burger2005}. 
These findings have been further supported by mass measurements of \isot{51-54}{Ca} \cite{Gallant2012,Wienholtz2013}, \isot{52,53}{K} \cite{Rosenbusch2015}, and \isot{52-54}{Sc} \cite{Xu2015,Xu2019}. 
However, mass measurements of Ti and V isotopes \cite{Leistenschneider2018,Porter2022} have introduced some ambiguity.

The $N = 34$ sub-shell closure is less clear. Initial evidence came from \isot{54}{Ca}, where a high $E(2^+_1)$ of 2043(19)~keV was measured~\cite{Steppenback2013}. 
This was further supported by direct mass measurements of \isot{55-57}{Ca} \cite{Michimasa2018}. 
In \isot{52}{Ar}, the $2_{1}^{+}$ state was measured at 1656(18) keV \cite{Liu2019}. 
Spectroscopic strengths from knockout reactions supported sub-shell closures at $N = 32$ and $N = 34$~\cite{Enciu2022,Li2024}. 
However, the persistence of the $N = 34$ sub-shell gap above $Z = 20$ remains uncertain~\cite{Leistenschneider2021,Steppenbeck2017,Liddick2004,Dinca2005,Iimura2023,Porter2022,Zhu2006}.

Another point of interest for the region is the unexpectedly large charge radii of \isot{49-52}{Ca} \cite{GarciaRuiz2016}, where neutron filling shifts from the $\nu 1f_{7/2}$ to the $\nu 2p_{3/2}$ orbital. 
The sudden rise beyond $N=28$ is striking given the equivalent charge radii of \isot{40}{Ca} and \isot{48}{Ca}. 
In fact, the isotope shifts between \isot{40}{Ca} and \isot{48}{Ca} seem to be described solely by the $E2$ strength \cite{brown2022}. 
The change in charge radii beyond $N=28$ has been attributed to core polarization and a $p$-orbit halo, both of which can induce neutron skins \cite{Bonnard2016}. 
The effective neutron charge, determined from a $B(E2;2^+\rightarrow0^+)$ measurement in \isot{50}{Ca}, also showed a sudden change with the occupation of the $\nu 2p_{3/2}$ orbital \cite{Valiente-Dobon2009}.
A comparable result was reported from inelastic proton scattering on \isot{50}{Ca} \cite{riley2014}.

In this letter, we investigate \isot{54}{Sc} by decay spectroscopy at the Facility for Rare Isotope Beams (FRIB) using the FRIB Decay Station initiator (FDSi) \cite{FDSi1,FDSi2}.
$^{54}$\textrm{Sc} is positioned one proton above the $Z=20$ shell closure and one neutron between the  $N=32,34$ sub-shell closures.
This gives rise to relatively simple single proton-neutron ($PN$) coupled states at low excitation energies that are sensitive to the magnitudes of the $N = 32,34$ sub-shell gaps. 
A new nanosecond isomer is reported that corresponds to the most precise, unambiguous $E2$ transition strength in the neutron-rich $fp$ region to date in a nucleus with valence protons above $Z = 20$. 
The new B($E2$) value and resulting effective charges provide a unique opportunity to probe core-polarization effects in neutron-rich $fp$-shell nuclei near $N=32,34$.

The present results are from two experiments conducted at FRIB using the two-focal-plane configuration of the FDSi. 
In both experiments, a secondary cocktail beam of fully stripped ions produced in the fragmentation reaction between an \isot{82}{Se} primary beam and a \isot{9}{Be} target were delivered to the FDSi.
At the first focal plane, \gam~rays and neutrons were detected using the DEcay Germanium Array initiator (DEGAi) and the Versatile Array of Neutron Detectors at Low Energy (VANDLE) \cite{VANDLE1,VANDLE2}, respectively.
The Modular Total Absorption Spectrometer (MTAS) \cite{Karny2016,Karny2020} was situated at the second focal plane.
At both focal planes, cocktail beams centered around \isot{52,54}{K} were implanted within a position-sensitive yttrium orthosilicate (YSO) detector~\cite{IMPLANT,singh2025}. See Refs.~\cite{Crawford2022, Gray2023, Cox2024} for additional experimental details.

\begin{figure}[t]
  \centering
    \includegraphics[width=\columnwidth]{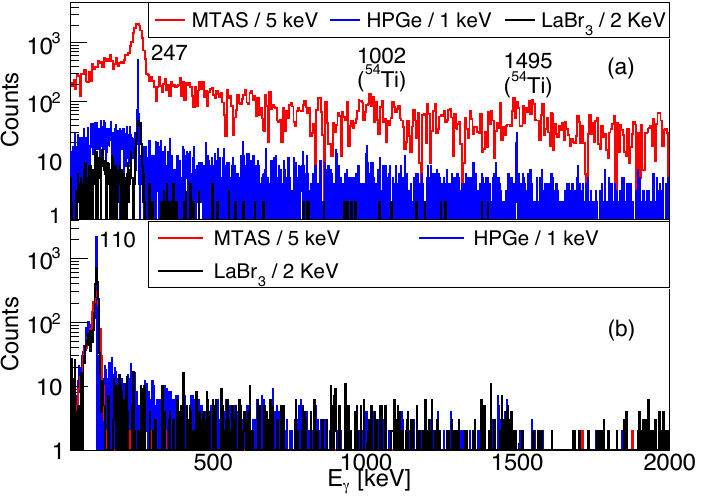}
        \caption{(a) \bet-delayed \gam-ray spectra measured within 300 ms of \isot{54}{Ca} ion implantation.
        (b) \gam-ray spectra collected within 20 $\mu$s of \isot{54m}{Sc} implantation.}
    \label{fig:gamma_panel}
\end{figure}

The \gam ~rays observed following the \bet ~decay of \isot{54}{Ca} and implantation of \isot{54m}{Sc} are presented in Figs.~\ref{fig:gamma_panel}(a) and (b), respectively.
The 247-keV \bet-delayed \gam-ray transition, identified in this work with an absolute intensity of $62(8)\%$, was previously reported in Refs.~\cite{Mantica2008, Crawford2010}. The transition was attributed to a 247-keV state in \isot{54}{Sc} with a spin-parity of $J^\pi=$ \jpi{1}{+} based on the selectivity of the \jpi{0}{+} parent decay.
The statistics of the present study are higher by an order of magnitude than Ref.~\cite{Crawford2010}, but no additional discrete \gam~rays in \isot{54}{Sc} were observed.

\begin{figure}[t]
  \centering 
     \includegraphics[width=\columnwidth]{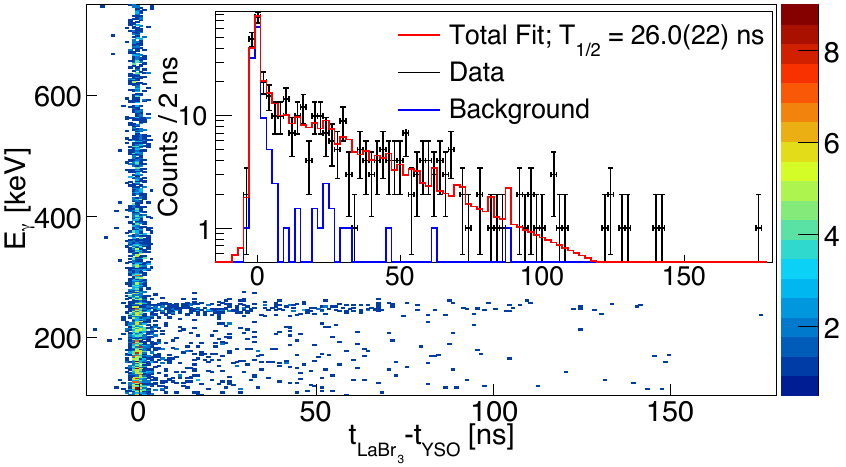}
        \caption{\bet-\gam ~time-difference distribution as a function of \gam-ray energy in the \LaBr ~detectors following \isot{54}{Ca} implantation. 
        \textit{\textbf{Inset}}: Time-difference distribution corresponding to a 247-keV energy gate.}    
    \label{fig:beta-gamma-panel_}
\end{figure}

The fast-timing capabilities of the YSO and \LaBr ~scintillators at the first focal plane of the FDSi have enabled the identification of the 247-keV, \jpi{1}{+} state in \isot{54}{Sc} as a nanosecond isomer, as shown in Fig.~\ref{fig:beta-gamma-panel_}.
A maximum-likelihood fit to the \bet-\gam ~time difference distribution yielded a 26.0(22)$_{\mathrm{stat.}}$(3)$_{\mathrm{sys.}}$ ns half-life.
The fit incorporated an exponentially modified Gaussian function with a background distribution derived from regions adjacent to the 247-keV peak.

The presence of a low-lying microsecond isomer in \isot{54}{Sc} was initially reported in Ref.~\cite{Grzywacz1998} with $J^\pi=(4^+, 5^+)$, and confirmed~\cite{Liddick2004, Crawford2010} based on the detection of a 110-keV \gam-ray peak following \isot{54m}{Sc} implantation.
In this study, the microsecond \isot{54m}{Sc} isomer is confirmed as shown in Fig.~\ref{fig:gamma_panel}(b) and no additional \gam~rays are observed in association with its decay.
A fit to the implant-\gam ~time-difference distributions yielded half-lives of $2.74(4)~\mu$s, $2.80(6)~\mu$s, and $2.72(9)~\mu$s for the HPGe, \LaBr, and MTAS detectors, respectively. 
The weighted average, $2.75(3)\mu$s, agrees with previous values of $7(5)\mu$s~\cite{Grzywacz1998} and $2.77(2)\mu$s~\cite{Crawford2010}.

The low-lying states of \isot{54}{Sc} can be schematically interpreted with respect to a \isot{52}{Ca} ($Z=20, N=32$) core, outside of which the odd \jpi{7/2}{-} proton from \isot{53}{Sc} and odd \jpi{1/2}{-} neutron from \isot{53}{Ca} spin couple to a \jpi{3}{+}, \jpi{4}{+} doublet. 
The population of the first \jpi{2}{+} and \jpi{4}{+} states in \isot{54}{Ti} following the \bet~decay of \isot{54}{Sc} supports a \jpi{3}{+} ground-state assignment for \isot{54}{Sc}, consistent with discussions in Refs. \cite{Liddick2004, Mantica2008, Crawford2010}. 
Furthermore, this population is identical between the \isot{54}{Sc}$\rightarrow$\isot{54}{Ti} and \isot{54}{Ca}$\rightarrow$\isot{54}{Sc}$\rightarrow$\isot{54}{Ti} decay chains, suggesting the absence of a \bet-decaying isomer.
The next lowest multiplet of states would then arise from a single-particle neutron excitation across $N=34$, corresponding to the excited \jpi{5/2}{-} state in \isot{53}{Ca} coupled to the \jpi{7/2}{-} ground state in \isot{53}{Sc}. The $\pi f_{7/2}\otimes\nu f_{5/2}$ coupling forms a sextet of states with $J^\pi=1^+-~6^+$.

The 247-keV \gam~ray de-exciting the nanosecond isomer corresponds to a \jpi{1}{+}$\rightarrow$ \jpi{3}{+} transition between the two multiplets with an experimental $B(E2)$ of $1.93(16)_{\mathrm{stat.}}(2)_{\mathrm{sys.}}$ W.u.\ determined in this work. 
The 110-keV \gam~ray de-exciting the microsecond isomer may represent either (a) a pure \jpi{2}{+}$\rightarrow$ \jpi{4}{+} or \jpi{6}{+}$\rightarrow$ \jpi{4}{+} $E2$ transition between the two multiplets with $B(E2) = 0.86(1)$~W.u., or (b) a mixed \jpi{4}{+}$\rightarrow$ \jpi{3}{+} $E2/M1$ transition within the $\pi f_{7/2}\otimes\nu p_{1/2}$ multiplet with $B(M1) \leq 5.92(7) \times 10^{-6}$~W.u. and $B(E2) \leq 0.86(1)$~W.u. 
We adopt the latter, which is consistent with discussions in Ref.~\cite{Crawford2010}, as the \jpi{2}{+} and \jpi{6}{+} states are expected to lie well above the antiparallel \jpi{1}{+} state, and the in-multiplet $M1$ transition is expected to be strongly hindered; the strength to first order is given by
$B(M1; 4^+\rightarrow 3^+) \approx \frac{21}{64\pi}(g^\pi_{f_{7/2}} - g^\nu_{p_{1/2}})^2 \nonumber = 8.4\times10^{-3} ~\text{W.u.}$

\begin{figure}[t]
    \centering
    \includegraphics[width=\columnwidth]{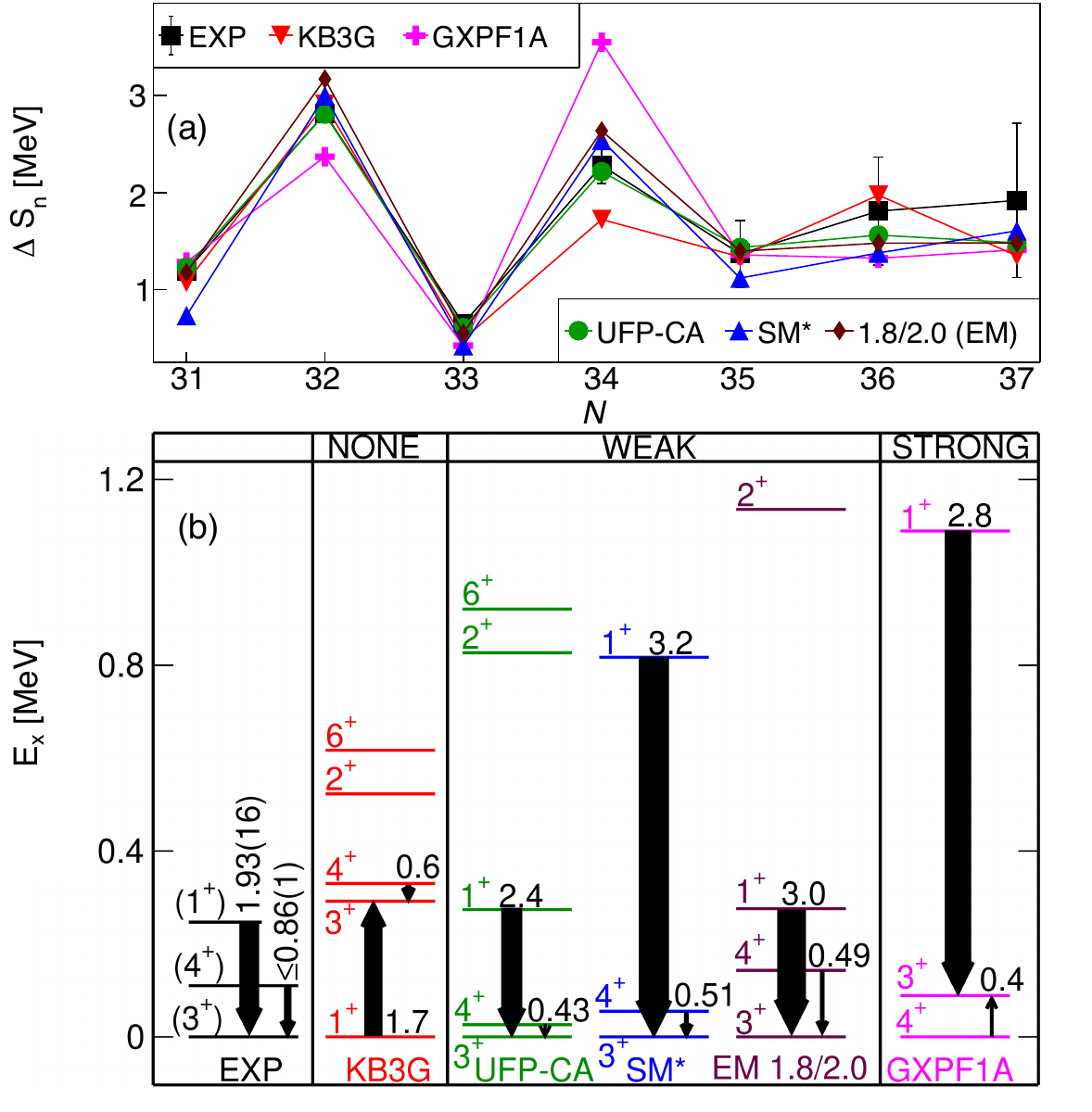}  
    \caption{Comparison of experimental and theoretical (a) neutron shell gaps along the Ca isotopic chain and (b) energy levels of \isot{54}{Sc}.
    Theoretical interactions are grouped based on their predictions of the strength of the sub-shell gap at $N = 34$.
    The $B(E2)$ strengths are given in W.u. and calculated with $e_\pi = 1.5 e$, $e_\nu = 0.5 e$.}
    \label{fig:level_scheme}
\end{figure}

Differences in neutron separation energies along the Ca isotopic chain, which emphasize the magnitude of the shell gaps, and level schemes for \isot{54}{Sc} are shown in Figs.~\ref{fig:level_scheme}(a) and (b), respectively.
Predictions by the KB3G \cite{Poves2001} and GXPF1A \cite{Honma2005} interactions, used in previous \isot{54}{Sc} studies, are compared with recent UFP-CA (empirically adjusted) \cite{Magilligan2021}, SM$^{*}$ (perturbative \textit{ab initio}) \cite{Coraggio2020,Coraggio2021} and VS-IMSRG 1.8/2.0 EM (\textit{ab initio}) \cite{hergert2016,Stroberg2019,imsrg} interactions.
The UFP-CA Hamiltonian for the $T = 1$ neutron-neutron interaction was obtained from constraints to data for the Ca isotopes \cite{Magilligan2021}.
For application to the Sc isotopes, the $T = 0$ part of the GXPF1A Hamiltonian \cite{Honma2005} was added.
The $T = 1$ part of the GXPF1A Hamiltonian was obtained from energy data for nuclei towards the middle of the $fp$ shell.
Thus, the $T = 1$ parts of the UFP-CA and GXPF1A interactions are nucleus-dependent and the UFP-CA Hamiltonian should only be used for nuclei near $Z = 20$ with $N > 28$.
All calculations were performed with KSHELL \cite{Shimizu2019} in the full \textit{fp} model space. For the SM$^{*}$ interaction, the neutron $g_{9/2}$ orbit was also included. 
The KB3G and GXPF1A interactions predict $N=34$ gaps for Ca that are non-existent and too strong, respectively, as shown in Fig. \ref{fig:level_scheme} (a). 
The sensitivity of the low-energy structure of \isot{54}{Sc} to the magnitude of the $N=34$ gap is reflected in the two corresponding level schemes, namely a $1^+,\pi f_{7/2}\otimes\nu f_{5/2}$ state that is too low or too high, as illustrated in Fig. \ref{fig:level_scheme} (b). 
The recent interactions (UFP-CA, SM$^{*}$ and VS-IMSRG 1.8/2.0 EM) predict a weak $N=34$ gap relative to $N=32$ and better describe the experimental data and schematic interpretations. 
We note that the $B(E2;1^+\rightarrow3^+)$ strengths predicted by the empirically adjusted interactions (KB3G, UFP-CA, and GXPF1A) also increase with the magnitude of the $N=34$ gap and that the value for UFP-CA, which best describes the level energies, is larger than experiment. 

\begin{figure}[t]
  \centering
    \includegraphics[width=\columnwidth]{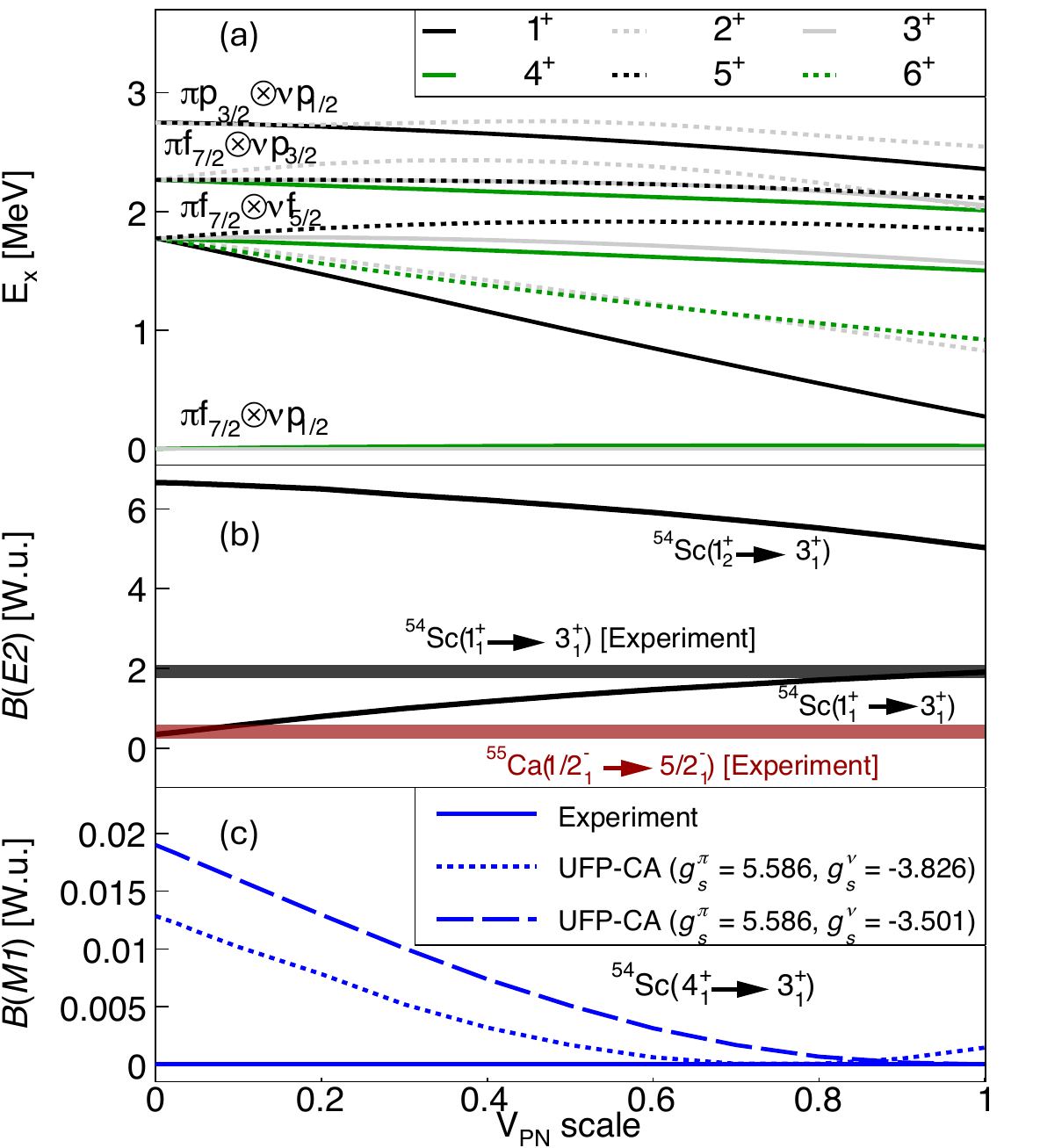}
        \caption{UFP-CA calculations of \isot{54}{Sc} as a function of $PN$ strength for (a) the first four $PN$ spin-coupled multiplets, and (b) the $B(E2)$ values with newly fitted effective charges e$_\pi$ = 1.30$e$, e$_\nu$ = 0.452$e$. 
        (c) $B(M1;4^+\rightarrow3^+)$ values with free and quenched spin $g$ factors.} 
    \label{fig:pnevolution}
\end{figure}

The UFP-CA results are further investigated in Fig.~\ref{fig:pnevolution} as a function of the proton-neutron ($PN$) interaction strength. All $PN$ two-body matrix elements were multiplied by a scale parameter, ~$V_{PN}$ scale. In the weak coupling limit ($V_{PN}$ scale = 0), the energies (and transitions) correspond to the excited neutron states in \isot{53}{Ca} and the excited proton states in \isot{53}{Sc}. As the $PN$ strength increases, each multiplet splits and begins to mix with nearby configurations. 
At $V_{PN}$ scale = 0, $B(E2;1_1^+\rightarrow3_{g.s.}^+)$ corresponds to the \isot{53}{Ca} $B(E2; f_{5/2}\rightarrow p_{1/2})$. 
The experimentally known $B(E2; 1/2^-\rightarrow 5/2^-)$ of \isot{55}{Ca} \cite{Koiwai2022} is shown as a proxy for this limit, represented by the red band in Fig.~\ref{fig:pnevolution} (b).
The $PN$ interaction, at its nominal value ($V_{PN}$ scale = 1), increases the $E2$ strength by a factor of four. 
The large increase arises primarily from mixing with the second $1^+$ configuration from the $\pi p_{3/2}\otimes\nu p_{1/2}$ multiplet, where proton $p_{3/2} \rightarrow f_{7/2}$ transitions across $Z=28$ have large intrinsic $E2$ strength. Note the large $B(E2;1_2^+\rightarrow3_1^+)$ strength in the weak-coupling limit, and that the sum with $B(E2;1_1^+\rightarrow3_1^+)$ remains relatively constant with the $PN$ strength. The magnitudes of the $Z=28$ and $N=34$ gaps control the proximity of the two $1^+$ configurations and therefore the mixing strength. 

The $B(M1;4^+\rightarrow3^+)$ strength vanishes completely with either small adjustments to the $PN$ interaction strength or 0.915 attenuation of the free neutron $g_s$ value; the total attenuation could be larger if the free proton $g_s$ were to be attenuated. However, the ground-state magnetic dipole moments of \isot{49,51}{Ca}, which do not depend on the proton $g_s$, fit best to a neutron $g_s$ attenuation of 0.957.

A survey of $E2$ data for neutron-rich \textrm{Ca}, \textrm{Sc}, and \textrm{Ti} isotopes is provided in Table \ref{tab:experimental_be2_fp}, including UFP-CA predictions with the ``standard'' isoscalar core-polarization effective charges of $e_\pi = 1.5 e$, $e_\nu = 0.5 e$ and newly fitted effective charges of $e_\pi = 1.30(8) e$, $e_\nu = 0.452(7) e$. Significantly larger effective charges would have been found if $1p-1h$ excitations across $Z=28$ or $N=28$ were not included, cf.~Fig.~\ref{fig:pnevolution}(b) and Refs.~\cite{brown1974,Lisetskiy2004}. 
The Blomqvist and Molinari oscillator parameter, $\hbar\omega=45A^{-1/3} - 25A^{-2/3}$ ~\cite{BLOMQVIST1968},
was used for all the calculations, where $B(E2) \propto \langle r^2 \rangle e_{\pi,\nu}^2 \propto \hbar^2\omega^2e_{\pi,\nu}^2$. 
Had $\hbar\omega'=41A^{-1/3}$ been used, the extracted effective charges would scale up by $\hbar\omega'/\hbar\omega=1.068$. 
The new effective charges improve the description of the $E2$ data in Table \ref{tab:experimental_be2_fp} from a reduced chi-squared, $\tilde{\chi}^2$, of 18.1 to 1.1. 
The new effective neutron charge, e$_\nu$, is qualitatively consistent with $\sim0.5e$ reported in Ref.~\cite{Valiente-Dobon2009} from the $B(E2;0^+\rightarrow2^+)$ of \isot{50}{Ca} but now includes fitting to the precise ground-state quadrupole moments of \isot{49,51}{Ca} \cite{GarcaRuiz2015}. 
The new $e_\pi/e_\nu$ ratio of 2.88(18), which is independent of $\hbar\omega$, is also consistent with 3.5(9) from inelastic proton scattering results on \isot{50}{Ca} \cite{riley2014}, but is now more precise by a factor of 5. 

\begin{table}[t]
     \centering
     \caption{Experimental $E2$ strengths of neutron-rich $fp$-shell isotopes near $N=32,34$ \cite{Dinca2005, Valiente-Dobon2009, GarcaRuiz2015, Koiwai2022} compared with theoretical predictions using the UFP-CA interaction with the standard (e$_\pi$ = 1.5$e$, e$_\nu$ = 0.5$e$) and fitted (e$_\pi$ = 1.30$e$, e$_\nu$ = 0.452$e$) effective charges.}
     \resizebox{\columnwidth}{!}{
     \begin{tabular}{cccccc}
     \toprule
     \midrule
     Isotope & J$_{i}^{\pi}$ $\rightarrow$ J$_{f}^{\pi}$  & \multicolumn{3}{c}{B$(E2)$ [W.u.] or Q [eb]} \\
     \cmidrule(r){3-5}
      & & EXP. & UFP-CA & UFP-CA \\
      & &  & (1.5, 0.5) & (1.3, 0.452)\\
      \midrule
      \isot{49}{Ca} & $\nicefrac{3}{2}_{\mathrm{g.s}}^{-}$ & $-0.036(3)$ & $-0.043$ & $-0.039$\\
      \rule{0pt}{2.5ex} 
      \isot{50}{Ca} & ${2}_{1}^{+}$ $\rightarrow$ ${0}_{1}^{+}$ & 0.68(2) & 0.82 & 0.68\\
      \rule{0pt}{2.5ex} 
      \isot{51}{Ca} & $\nicefrac{3}{2}_{\mathrm{g.s}}^{-}$ & +0.036(12) & +0.042 & +0.038\\
      \rule{0pt}{2.5ex} 
      \isot{55}{Ca} & $\nicefrac{1}{2}_{1}^{-}$ $\rightarrow$ $\nicefrac{5}{2}_{1}^{-}$ & 0.42(18) & 0.26 & 0.21\\
      \midrule
      \isot{51}{Sc} & $\nicefrac{11}{2}_{1}^{-}$ $\rightarrow$ $\nicefrac{7}{2}_{1}^{-}$ & 1.9(5) & 1.65 & 1.34\\ 
      \rule{0pt}{2.5ex}   
      \isot{54}{Sc} & ${1}_{1}^{+}$ $\rightarrow$ ${3}_{1}^{+}$ & 1.93(16) & 2.44 & 1.91\\ 
      \rule{0pt}{2.5ex}   
      \isot{54}{Ti} & ${2}_{1}^{+}$ $\rightarrow$ ${0}_{1}^{+}$ & 6.0(12) & 8.56 & 6.57\\ 
      \midrule
      $\tilde{\chi}^2$ & & & 18.1 & 1.1\\ 
      \midrule
     \bottomrule
     \label{tab:experimental_be2_fp}
     \end{tabular}
     }
 \end{table}

The new effective charges differ significantly from those previously established in the $fp$ shell. For example, effective charges of $e_\pi = 1.15e$, $e_\nu = 0.8e$ (ratio of 1.44) were determined for \isot{51}{Fe}/\isot{51}{Mn} at $N\approx Z$ in the $fp$ shell~\cite{duRietz2004}, and later updated to $e_\pi = 1.12e$, $e_\nu = 0.67e$ (ratio of 1.67) to best describe the stable $N=28$ and $Z=28$ chains~\cite{allmond2014}. The new values are also different from $e_\pi = 1.1e$, $e_\nu = 0.6e$ (ratio of 1.83) used recently to describe \isot{49,50}{Ti}~\cite{gray2024}. It has been argued that changes in the effective charges across the $fp$ shell are due to an isospin or orbital dependence~\cite{Valiente-Dobon2009}. However, it is worth noting that the previous and new effective charges all have similar sums, namely $e_\pi + e_\nu \approx 1.7e-1.95e$. 

\begin{figure}[t]
  \centering
    \includegraphics[width=0.7\columnwidth]{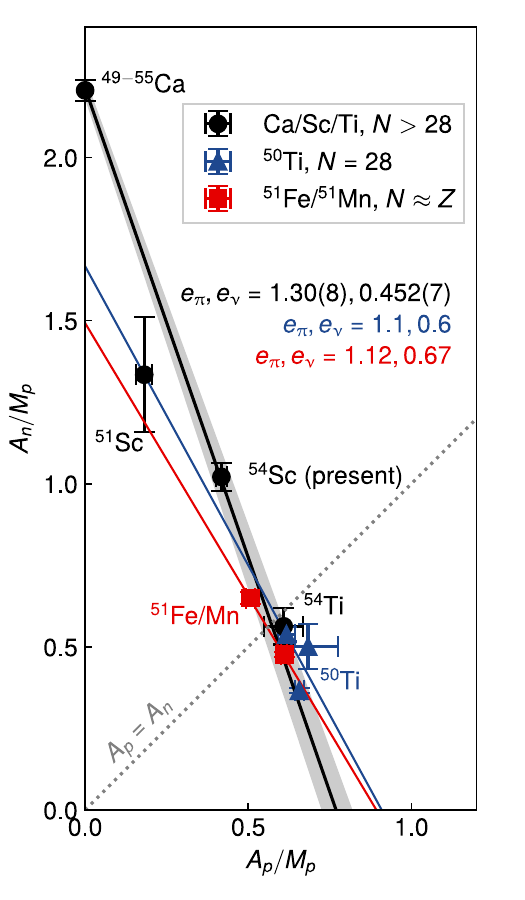}  
        \caption{Neutron versus proton transition amplitudes relative to the experimental $E2$ matrix elements for $fp$-shell nuclei listed in Table \ref{tab:experimental_be2_fp} [black line, $e_\pi = 1.30(8)e$, $e_\nu = 0.452(7)e$]. Additional data points represent transitions in \isot{50}{Ti} ($N=28$) and  \isot{51}{Fe}/\isot{51}{Mn} ($N \approx Z$) \cite{duRietz2004,allmond2014, gray2024} with lines representing previously adopted effective charges. The proton amplitudes are zero for the Ca isotopes so the weighted average was adopted. 
        See Appendix~\ref{app} for more details.
        }
    \label{fig:E2_amplitude_dependence}
\end{figure}

The $E2$ data and effective charges can be linearly systematized by plotting $A_n/M_p$ versus $A_p/M_p$, as shown in Fig.~\ref{fig:E2_amplitude_dependence}. All data points should fall on a straight line with slope $e_\pi/e_\nu$ and intercept $1/e_\nu$ \cite{brown1974}.
$A_p \textrm{ and } A_n$ represent the calculated transition amplitudes, which relate to the $E2$ matrix element by $\langle J_f^\pi||E2||J_i^\pi\rangle=A_pe_\pi + A_ne_\nu$. $M_p$ is the experimental $E2$ matrix element, where $B(E2) = M_p^2/(2J_i +1)$. $J_f^\pi \textrm{ and } J_i^\pi$ denote the spin-parity of the final and initial states, respectively. 
Fig.~\ref{fig:E2_amplitude_dependence} shows that a large majority of $fp$-shell nuclei, including those at $N=28$, have $A_p \approx A_n$ meaning they are mostly sensitive to the sum of effective charges as opposed to the ratio. All of the previous cases are consistent with the new effective charges of $e_\pi = 1.30(8)e$, $e_\nu = 0.452(7)e$.  
These values are equivalent to the universal effective charges in the $sd$ shell, $e_\pi = 1.36(5)e$ and $e_\nu = 0.45(5)e$~\cite{Richter2008}. Further, they are equivalent to microscopic derivations by Dufour and Zuker, $e_\pi = 1.31e$ and $e_\nu = 0.46e$ \cite{dufour1996}, which have been applied to $sd$~\cite{dao2022}, $sd$-$pf$~\cite{caurier2014}, $pf$~\cite{poves2020}, $fpg_{9}d_{5}$~\cite{crawford2013}, and $pf$-$sdg$~\cite{nowacki2016} valence spaces.
Therefore, universal effective charges of $e_\pi = 1.33e$, $e_\nu = 0.45e$, based on empirical fits, can be used across the $sd$ and $fp$ shells.

In summary, the structure of neutron-rich $^{54}$\textrm{Sc} is reported from two experiments at FRIB using the FDSi.
The 247-keV \jpi{1}{+} state was identified as a nanosecond isomer while the previously known microsecond isomer was interpreted as a $4^+\rightarrow3^+$ transition between the $\pi f_{7/2} \otimes \nu p_{1/2}$ spin-coupled multiplet members.
The new \jpi{1}{+} lifetime and resulting effective charge analysis demonstrates that a universal set of effective charges can be used across the $sd$ and $fp$ shells, which should enable more consistent and accurate electric-quadrupole transition and moment calculations for a large number of atomic nuclei. 
No evidence for changes in the effective charges due to an isospin or orbital dependence is found.

This material is based upon work supported in part by the U.S. Department of Energy, Office of Science, Office of Nuclear Physics under Contract Nos.~DE-AC02-06CH11357 (ANL), DE-AC02-98CH10946 (BNL), DE-AC02-05CH11231 (LBNL), DE-AC52-07NA27344 (LLNL), DE-SC0020451, DE-SC0023633 (Michigan State), DE-SC0014448 (Mississippi State), DE-AC05-00OR22725 (ORNL), and DE-FG02-96ER40983 (UTK). 
The publisher acknowledges the US government license to provide public access under the DOE Public Access Plan (http://energy.gov/downloads/doe-public-access-plan). 
This work was also supported by the U.S. National Science Foundation under Grant Nos.~PHY-23-10078, PHY-2110365, PHY-2012522 (FSU),  and PHY-1848177 (CAREER) (Mississippi State). 
Additional support was provided by the U.S. Department of Energy, National Nuclear Security Administration under Award No.~DE-NA0003180 (Michigan State) and the Stewardship Science Academic Alliances program through DOE Award Nos.~DE-NA0003899 (UTK), DE-SC0016988 (TTU) and No.~DOE-DE-NA0003906 (Michigan State), and National Science Foundation Major Research Instrumentation Program Award No.~1919735 (UTK, TTU).
This research was also supported by the Laboratory Directed Research and Development Program at Pacific Northwest National Laboratory operated by Battelle for the U.S. Department of Energy.
This work was also supported in part by the Deutsche Forschungsgemeinschaft (DFG, German Research Foundation) --- Project-ID 279384907 --- SFB 1245.
This work was in part supported by the European Research Council (ERC) under the European Union's Horizon 2020 research and innovation programme (Grant Agreement No. 101020842) and JST ERATO Grant No. JPMJER2304, Japan.
For the VS-IMSRG calculations, the \href{https://doi.org/10.1140/epja/s10050-023-01039-y}{\texttt{NuHamil}}, \href{https://github.com/ragnarstroberg/imsrg}{\texttt{IMSRG++}} and \href{https://doi.org/10.1016/j.cpc.2019.06.011}{\texttt{KSHELL}} codes were used to generate chiral EFT matrix elements, to perform the valence-space decoupling, and to solve the valence-space problems, respectively.
The VS-IMSRG calculations were in part performed with an allocation of computing resources at the J\"ulich Supercomputing Center and on Cygnus and Pegasus at the CCS, University of Tsukuba. 
This research used resources of the Facility for Rare Isotope Beams, which is a DOE Office of Science User Facility.

\newpage
\clearpage
\onecolumngrid
\appendix
\section{Comparison of experimental $E2$ strengths of $fp$-shell nuclei near $N = 32, 34$ \cite{Dinca2005, Valiente-Dobon2009, GarcaRuiz2015, Koiwai2022} to theoretical predictions using the UFP-CA interaction with adopted effective charges, $e_\pi$ = 1.30$e$, $e_\nu$ = 0.452$e$. The same quantities are compared for \isot{50}{Ti} ($N = 28$) and \isot{51}{Fe}/\isot{51}{Mn} ($N \approx Z$) \cite{gray2024, duRietz2004,allmond2014}, using the GXPF1A interaction. The $E2$ matrix element, $\langle J_f^\pi||E2||J_i^\pi\rangle=M_p$, is related to the transition probabilities through the expression, $B(E2) = M_p^2/(2J_i +1)$, where $M_p = A_pe_\pi + A_ne_\nu$.}
\label{app}
\begin{table*}[h]
     \centering
     \resizebox{\textwidth}{!}{
     \begin{tabular}{cccccccc}
     \toprule
     \midrule
     Isotope & J$_{i}^{\pi}$ $\rightarrow$ J$_{f}^{\pi}$  & \multicolumn{2}{c}{EXPERIMENT} & \multicolumn{4}{c}{THEORY (UFP-CA)} \\
     \cmidrule(r){3-4}
     \cmidrule(r){5-8}
      & & $B(E2)$ [W.u.] or $Q_s$ [eb] & $M(E2)$ [efm$^2$] & $A_p$ [efm$^2$] & $A_n$ [efm$^2$] & $M(E2)$ [efm$^2$] & $B(E2)$ [W.u.] or $Q_s$ [eb] \\
      \midrule
      \isot{49}{Ca} & $\nicefrac{3}{2}_{\mathrm{g.s}}^{-}$ & $-0.036(3)$ & $-5.08(42)$ & $0$ & $-12.31$ & $-5.56$ & $-0.039$ \\
      \rule{0pt}{2.5ex} 
      \isot{50}{Ca} & ${2}_{1}^{+}$ $\rightarrow$ ${0}_{1}^{+}$ & $0.68(2)$ & $6.10(9)$ & $0$ & 13.45 & $6.08$ & $0.68$\\
      \rule{0pt}{2.5ex} 
      \isot{51}{Ca} & $\nicefrac{3}{2}_{\mathrm{g.s}}^{-}$ & $+0.036(12)$ & $5.08(169)$ & $0$ & $11.95$ & $5.40$ & $+0.038$\\
      \rule{0pt}{2.5ex} 
      \isot{55}{Ca} & $\nicefrac{1}{2}_{1}^{-}$ $\rightarrow$ $\nicefrac{5}{2}_{1}^{-}$ & $0.42(18)$ & $3.23(69)$ & $0$ & $5.10$ & $2.30$ & $0.21$\\
      \isot{51}{Sc} & $\nicefrac{11}{2}_{1}^{-}$ $\rightarrow$ $\nicefrac{7}{2}_{1}^{-}$ & 1.9(5) & $16.0(21)$ &$2.90$ &$21.35$ & 13.42 & 1.34\\ 
      \rule{0pt}{2.5ex}   
      \isot{54}{Sc} & ${1}_{1}^{+}$ $\rightarrow$ ${3}_{1}^{+}$ & $1.93(16)$ & $8.38(35)$ & $3.46$ & $8.47$ & $8.33$ & $1.91$\\ 
      \rule{0pt}{2.5ex}   
      \isot{54}{Ti} & ${2}_{1}^{+}$ $\rightarrow$ ${0}_{1}^{+}$ & $6.0(12)$ & $19.1(19)$ & $11.62$ & $10.75$ & $19.96$ & $6.57$\\ 
      \midrule
       ${\chi}^2_{norm}$ & & &  & & & & $1.1$\\ 
      \midrule
       &  & & &\multicolumn{4}{c}{THEORY (GXPF1A)} \\
      \cmidrule(r){5-8}

      \isot{50}{Ti} & ${2}_{1}^{+}$ $\rightarrow$ ${0}_{1}^{+}$ & $6.43(52)$ & $18.76(76)$ & $11.58$ & $10.10$ & $19.61$ & $7.03$\\
      \rule{0pt}{2.5ex}
      \isot{50}{Ti} & ${4}_{1}^{+}$ $\rightarrow$ ${2}_{1}^{+}$ & $5.5(15)$ & $23.27(317)$ & $15.90$ & $11.66$ & $25.94$ & $6.83$\\ 
      \rule{0pt}{2.5ex}
      \isot{50}{Ti} & ${6}_{1}^{+}$ $\rightarrow$ ${4}_{1}^{+}$ & $3.14(13)$ & $21.13(44)$ & $13.87$ & $7.73$ & $21.53$ & $3.26$\\ 
      \rule{0pt}{2.5ex}
      \isot{51}{Mn} & $\nicefrac{27}{2}_{1}^{-}$ $\rightarrow$ $\nicefrac{23}{2}_{1}^{-}$ & $4.16(12)$ & $36.18(52)$ & $22.09$ & $17.22$ & $36.50$ & $4.23$\\ 
      \rule{0pt}{2.5ex}
      \isot{51}{Fe} & $\nicefrac{27}{2}_{1}^{-}$ $\rightarrow$ $\nicefrac{23}{2}_{1}^{-}$ & $3.68(21)$ & $34.02(97)$ & $17.22$ & $22.09$ & $32.37$ & $3.33$\\ 
      \midrule
      ${\chi}^2_{norm}$ & & &  & & & & $1.5$\\ 
      \midrule
      \bottomrule
      \label{tab:supplemental_material_1}
      \end{tabular}
     }
 \end{table*}


\begin{thebibliography}{77}%
\makeatletter
\providecommand \@ifxundefined [1]{%
 \@ifx{#1\undefined}
}%
\providecommand \@ifnum [1]{%
 \ifnum #1\expandafter \@firstoftwo
 \else \expandafter \@secondoftwo
 \fi
}%
\providecommand \@ifx [1]{%
 \ifx #1\expandafter \@firstoftwo
 \else \expandafter \@secondoftwo
 \fi
}%
\providecommand \natexlab [1]{#1}%
\providecommand \enquote  [1]{``#1''}%
\providecommand \bibnamefont  [1]{#1}%
\providecommand \bibfnamefont [1]{#1}%
\providecommand \citenamefont [1]{#1}%
\providecommand \href@noop [0]{\@secondoftwo}%
\providecommand \href [0]{\begingroup \@sanitize@url \@href}%
\providecommand \@href[1]{\@@startlink{#1}\@@href}%
\providecommand \@@href[1]{\endgroup#1\@@endlink}%
\providecommand \@sanitize@url [0]{\catcode `\\12\catcode `\$12\catcode `\&12\catcode `\#12\catcode `\^12\catcode `\_12\catcode `\%12\relax}%
\providecommand \@@startlink[1]{}%
\providecommand \@@endlink[0]{}%
\providecommand \url  [0]{\begingroup\@sanitize@url \@url }%
\providecommand \@url [1]{\endgroup\@href {#1}{\urlprefix }}%
\providecommand \urlprefix  [0]{URL }%
\providecommand \Eprint [0]{\href }%
\providecommand \doibase [0]{https://doi.org/}%
\providecommand \selectlanguage [0]{\@gobble}%
\providecommand \bibinfo  [0]{\@secondoftwo}%
\providecommand \bibfield  [0]{\@secondoftwo}%
\providecommand \translation [1]{[#1]}%
\providecommand \BibitemOpen [0]{}%
\providecommand \bibitemStop [0]{}%
\providecommand \bibitemNoStop [0]{.\EOS\space}%
\providecommand \EOS [0]{\spacefactor3000\relax}%
\providecommand \BibitemShut  [1]{\csname bibitem#1\endcsname}%
\let\auto@bib@innerbib\@empty
\bibitem [{\citenamefont {Mayer}(1949)}]{Mayer1949}%
  \BibitemOpen
  \bibfield  {author} {\bibinfo {author} {\bibfnamefont {M.~G.}\ \bibnamefont {Mayer}},\ }\href {https://doi.org/10.1103/PhysRev.75.1969} {\bibfield  {journal} {\bibinfo  {journal} {Phys. Rev.}\ }\textbf {\bibinfo {volume} {75}},\ \bibinfo {pages} {1969} (\bibinfo {year} {1949})}\BibitemShut {NoStop}%
\bibitem [{\citenamefont {Haxel}\ \emph {et~al.}(1949)\citenamefont {Haxel}, \citenamefont {Jensen},\ and\ \citenamefont {Suess}}]{Haxel1949}%
  \BibitemOpen
  \bibfield  {author} {\bibinfo {author} {\bibfnamefont {O.}~\bibnamefont {Haxel}}, \bibinfo {author} {\bibfnamefont {J.~H.~D.}\ \bibnamefont {Jensen}},\ and\ \bibinfo {author} {\bibfnamefont {H.~E.}\ \bibnamefont {Suess}},\ }\href {https://doi.org/10.1103/PhysRev.75.1766.2} {\bibfield  {journal} {\bibinfo  {journal} {Phys. Rev.}\ }\textbf {\bibinfo {volume} {75}},\ \bibinfo {pages} {1766} (\bibinfo {year} {1949})}\BibitemShut {NoStop}%
\bibitem [{\citenamefont {Warner}(2004)}]{Warner2004}%
  \BibitemOpen
  \bibfield  {author} {\bibinfo {author} {\bibfnamefont {D.}~\bibnamefont {Warner}},\ }\href {https://doi.org/10.1038/430517a} {\bibfield  {journal} {\bibinfo  {journal} {Nature}\ }\textbf {\bibinfo {volume} {430}},\ \bibinfo {pages} {517} (\bibinfo {year} {2004})}\BibitemShut {NoStop}%
\bibitem [{\citenamefont {Dobaczewski}\ \emph {et~al.}(2007)\citenamefont {Dobaczewski}, \citenamefont {Michel}, \citenamefont {Nazarewicz}, \citenamefont {Płoszajczak},\ and\ \citenamefont {Rotureau}}]{Dobaczewski2007}%
  \BibitemOpen
  \bibfield  {author} {\bibinfo {author} {\bibfnamefont {J.}~\bibnamefont {Dobaczewski}}, \bibinfo {author} {\bibfnamefont {N.}~\bibnamefont {Michel}}, \bibinfo {author} {\bibfnamefont {W.}~\bibnamefont {Nazarewicz}}, \bibinfo {author} {\bibfnamefont {M.}~\bibnamefont {Płoszajczak}},\ and\ \bibinfo {author} {\bibfnamefont {J.}~\bibnamefont {Rotureau}},\ }\href {https://doi.org/https://doi.org/10.1016/j.ppnp.2007.01.022} {\bibfield  {journal} {\bibinfo  {journal} {Progress in Particle and Nuclear Physics}\ }\textbf {\bibinfo {volume} {59}},\ \bibinfo {pages} {432} (\bibinfo {year} {2007})}\BibitemShut {NoStop}%
\bibitem [{\citenamefont {Satu\l{}a}\ \emph {et~al.}(2009)\citenamefont {Satu\l{}a}, \citenamefont {Zalewski}, \citenamefont {Dobaczewski}, \citenamefont {Olbratowski}, \citenamefont {Rafalski}, \citenamefont {Werner},\ and\ \citenamefont {Wyss}}]{satula2009}%
  \BibitemOpen
  \bibfield  {author} {\bibinfo {author} {\bibfnamefont {W.}~\bibnamefont {Satu\l{}a}}, \bibinfo {author} {\bibfnamefont {M.}~\bibnamefont {Zalewski}}, \bibinfo {author} {\bibfnamefont {J.}~\bibnamefont {Dobaczewski}}, \bibinfo {author} {\bibfnamefont {P.}~\bibnamefont {Olbratowski}}, \bibinfo {author} {\bibfnamefont {M.}~\bibnamefont {Rafalski}}, \bibinfo {author} {\bibfnamefont {T.~R.}\ \bibnamefont {Werner}},\ and\ \bibinfo {author} {\bibfnamefont {R.~A.}\ \bibnamefont {Wyss}},\ }\href {https://doi.org/10.1142/S0218301309012902} {\bibfield  {journal} {\bibinfo  {journal} {International Journal of Modern Physics E}\ }\textbf {\bibinfo {volume} {18}},\ \bibinfo {pages} {808} (\bibinfo {year} {2009})}\BibitemShut {NoStop}%
\bibitem [{\citenamefont {Otsuka}\ \emph {et~al.}(2020)\citenamefont {Otsuka}, \citenamefont {Gade}, \citenamefont {Sorlin}, \citenamefont {Suzuki},\ and\ \citenamefont {Utsuno}}]{Otsuka2020}%
  \BibitemOpen
  \bibfield  {author} {\bibinfo {author} {\bibfnamefont {T.}~\bibnamefont {Otsuka}}, \bibinfo {author} {\bibfnamefont {A.}~\bibnamefont {Gade}}, \bibinfo {author} {\bibfnamefont {O.}~\bibnamefont {Sorlin}}, \bibinfo {author} {\bibfnamefont {T.}~\bibnamefont {Suzuki}},\ and\ \bibinfo {author} {\bibfnamefont {Y.}~\bibnamefont {Utsuno}},\ }\href {https://doi.org/10.1103/RevModPhys.92.015002} {\bibfield  {journal} {\bibinfo  {journal} {Rev. Mod. Phys.}\ }\textbf {\bibinfo {volume} {92}},\ \bibinfo {pages} {015002} (\bibinfo {year} {2020})}\BibitemShut {NoStop}%
\bibitem [{\citenamefont {Otsuka}\ \emph {et~al.}(2005)\citenamefont {Otsuka}, \citenamefont {Suzuki}, \citenamefont {Fujimoto}, \citenamefont {Grawe},\ and\ \citenamefont {Akaishi}}]{Otsuka2005}%
  \BibitemOpen
  \bibfield  {author} {\bibinfo {author} {\bibfnamefont {T.}~\bibnamefont {Otsuka}}, \bibinfo {author} {\bibfnamefont {T.}~\bibnamefont {Suzuki}}, \bibinfo {author} {\bibfnamefont {R.}~\bibnamefont {Fujimoto}}, \bibinfo {author} {\bibfnamefont {H.}~\bibnamefont {Grawe}},\ and\ \bibinfo {author} {\bibfnamefont {Y.}~\bibnamefont {Akaishi}},\ }\href {https://doi.org/10.1103/PhysRevLett.95.232502} {\bibfield  {journal} {\bibinfo  {journal} {Phys. Rev. Lett.}\ }\textbf {\bibinfo {volume} {95}},\ \bibinfo {pages} {232502} (\bibinfo {year} {2005})}\BibitemShut {NoStop}%
\bibitem [{\citenamefont {Otsuka}\ \emph {et~al.}(2010)\citenamefont {Otsuka}, \citenamefont {Suzuki}, \citenamefont {Honma}, \citenamefont {Utsuno}, \citenamefont {Tsunoda}, \citenamefont {Tsukiyama},\ and\ \citenamefont {Hjorth-Jensen}}]{Otsuka2010}%
  \BibitemOpen
  \bibfield  {author} {\bibinfo {author} {\bibfnamefont {T.}~\bibnamefont {Otsuka}}, \bibinfo {author} {\bibfnamefont {T.}~\bibnamefont {Suzuki}}, \bibinfo {author} {\bibfnamefont {M.}~\bibnamefont {Honma}}, \bibinfo {author} {\bibfnamefont {Y.}~\bibnamefont {Utsuno}}, \bibinfo {author} {\bibfnamefont {N.}~\bibnamefont {Tsunoda}}, \bibinfo {author} {\bibfnamefont {K.}~\bibnamefont {Tsukiyama}},\ and\ \bibinfo {author} {\bibfnamefont {M.}~\bibnamefont {Hjorth-Jensen}},\ }\href {https://doi.org/10.1103/PhysRevLett.104.012501} {\bibfield  {journal} {\bibinfo  {journal} {Phys. Rev. Lett.}\ }\textbf {\bibinfo {volume} {104}},\ \bibinfo {pages} {012501} (\bibinfo {year} {2010})}\BibitemShut {NoStop}%
\bibitem [{\citenamefont {Valiente-Dob\'on}\ \emph {et~al.}(2009)\citenamefont {Valiente-Dob\'on}, \citenamefont {Mengoni}, \citenamefont {Gadea}, \citenamefont {Farnea}, \citenamefont {Lenzi}, \citenamefont {Lunardi}, \citenamefont {Dewald}, \citenamefont {Pissulla}, \citenamefont {Szilner}, \citenamefont {Broda}, \citenamefont {Recchia}, \citenamefont {Algora}, \citenamefont {Angus}, \citenamefont {Bazzacco}, \citenamefont {Benzoni}, \citenamefont {Bizzeti}, \citenamefont {Bizzeti-Sona}, \citenamefont {Boutachkov}, \citenamefont {Corradi}, \citenamefont {Crespi}, \citenamefont {de~Angelis}, \citenamefont {Fioretto}, \citenamefont {G\"orgen}, \citenamefont {Gorska}, \citenamefont {Gottardo}, \citenamefont {Grodner}, \citenamefont {Guiot}, \citenamefont {Howard}, \citenamefont {Kr\'olas}, \citenamefont {Leoni}, \citenamefont {Mason}, \citenamefont {Menegazzo}, \citenamefont {Montanari}, \citenamefont {Montagnoli}, \citenamefont {Napoli}, \citenamefont {Obertelli}, \citenamefont {Paw\l{}at}, \citenamefont
  {Pollarolo}, \citenamefont {Rubio}, \citenamefont {\ifmmode~\mbox{\c{S}}\else \c{S}\fi{}ahin}, \citenamefont {Scarlassara}, \citenamefont {Silvestri}, \citenamefont {Stefanini}, \citenamefont {Smith}, \citenamefont {Steppenbeck}, \citenamefont {Ur}, \citenamefont {Wady}, \citenamefont {Wrzesi\ifmmode~\acute{n}\else \'{n}\fi{}ski}, \citenamefont {Maglione},\ and\ \citenamefont {Hamamoto}}]{Valiente-Dobon2009}%
  \BibitemOpen
  \bibfield  {author} {\bibinfo {author} {\bibfnamefont {J.~J.}\ \bibnamefont {Valiente-Dob\'on}}, \bibinfo {author} {\bibfnamefont {D.}~\bibnamefont {Mengoni}}, \bibinfo {author} {\bibfnamefont {A.}~\bibnamefont {Gadea}}, \bibinfo {author} {\bibfnamefont {E.}~\bibnamefont {Farnea}}, \bibinfo {author} {\bibfnamefont {S.~M.}\ \bibnamefont {Lenzi}}, \bibinfo {author} {\bibfnamefont {S.}~\bibnamefont {Lunardi}}, \bibinfo {author} {\bibfnamefont {A.}~\bibnamefont {Dewald}}, \bibinfo {author} {\bibfnamefont {T.}~\bibnamefont {Pissulla}}, \bibinfo {author} {\bibfnamefont {S.}~\bibnamefont {Szilner}}, \bibinfo {author} {\bibfnamefont {R.}~\bibnamefont {Broda}}, \bibinfo {author} {\bibfnamefont {F.}~\bibnamefont {Recchia}}, \bibinfo {author} {\bibfnamefont {A.}~\bibnamefont {Algora}}, \bibinfo {author} {\bibfnamefont {L.}~\bibnamefont {Angus}}, \bibinfo {author} {\bibfnamefont {D.}~\bibnamefont {Bazzacco}}, \bibinfo {author} {\bibfnamefont {G.}~\bibnamefont {Benzoni}}, \bibinfo {author} {\bibfnamefont {P.~G.}\
  \bibnamefont {Bizzeti}}, \bibinfo {author} {\bibfnamefont {A.~M.}\ \bibnamefont {Bizzeti-Sona}}, \bibinfo {author} {\bibfnamefont {P.}~\bibnamefont {Boutachkov}}, \bibinfo {author} {\bibfnamefont {L.}~\bibnamefont {Corradi}}, \bibinfo {author} {\bibfnamefont {F.}~\bibnamefont {Crespi}}, \bibinfo {author} {\bibfnamefont {G.}~\bibnamefont {de~Angelis}}, \bibinfo {author} {\bibfnamefont {E.}~\bibnamefont {Fioretto}}, \bibinfo {author} {\bibfnamefont {A.}~\bibnamefont {G\"orgen}}, \bibinfo {author} {\bibfnamefont {M.}~\bibnamefont {Gorska}}, \bibinfo {author} {\bibfnamefont {A.}~\bibnamefont {Gottardo}}, \bibinfo {author} {\bibfnamefont {E.}~\bibnamefont {Grodner}}, \bibinfo {author} {\bibfnamefont {B.}~\bibnamefont {Guiot}}, \bibinfo {author} {\bibfnamefont {A.}~\bibnamefont {Howard}}, \bibinfo {author} {\bibfnamefont {W.}~\bibnamefont {Kr\'olas}}, \bibinfo {author} {\bibfnamefont {S.}~\bibnamefont {Leoni}}, \bibinfo {author} {\bibfnamefont {P.}~\bibnamefont {Mason}}, \bibinfo {author} {\bibfnamefont
  {R.}~\bibnamefont {Menegazzo}}, \bibinfo {author} {\bibfnamefont {D.}~\bibnamefont {Montanari}}, \bibinfo {author} {\bibfnamefont {G.}~\bibnamefont {Montagnoli}}, \bibinfo {author} {\bibfnamefont {D.~R.}\ \bibnamefont {Napoli}}, \bibinfo {author} {\bibfnamefont {A.}~\bibnamefont {Obertelli}}, \bibinfo {author} {\bibfnamefont {T.}~\bibnamefont {Paw\l{}at}}, \bibinfo {author} {\bibfnamefont {G.}~\bibnamefont {Pollarolo}}, \bibinfo {author} {\bibfnamefont {B.}~\bibnamefont {Rubio}}, \bibinfo {author} {\bibfnamefont {E.}~\bibnamefont {\ifmmode~\mbox{\c{S}}\else \c{S}\fi{}ahin}}, \bibinfo {author} {\bibfnamefont {F.}~\bibnamefont {Scarlassara}}, \bibinfo {author} {\bibfnamefont {R.}~\bibnamefont {Silvestri}}, \bibinfo {author} {\bibfnamefont {A.~M.}\ \bibnamefont {Stefanini}}, \bibinfo {author} {\bibfnamefont {J.~F.}\ \bibnamefont {Smith}}, \bibinfo {author} {\bibfnamefont {D.}~\bibnamefont {Steppenbeck}}, \bibinfo {author} {\bibfnamefont {C.~A.}\ \bibnamefont {Ur}}, \bibinfo {author} {\bibfnamefont {P.~T.}\
  \bibnamefont {Wady}}, \bibinfo {author} {\bibfnamefont {J.}~\bibnamefont {Wrzesi\ifmmode~\acute{n}\else \'{n}\fi{}ski}}, \bibinfo {author} {\bibfnamefont {E.}~\bibnamefont {Maglione}},\ and\ \bibinfo {author} {\bibfnamefont {I.}~\bibnamefont {Hamamoto}},\ }\href {https://doi.org/10.1103/PhysRevLett.102.242502} {\bibfield  {journal} {\bibinfo  {journal} {Phys. Rev. Lett.}\ }\textbf {\bibinfo {volume} {102}},\ \bibinfo {pages} {242502} (\bibinfo {year} {2009})}\BibitemShut {NoStop}%
\bibitem [{\citenamefont {Jungclaus}\ \emph {et~al.}(2024)\citenamefont {Jungclaus}, \citenamefont {G\'orska}, \citenamefont {Miko\l{}ajczuk}, \citenamefont {Acosta}, \citenamefont {Taprogge}, \citenamefont {Nishimura}, \citenamefont {Doornenbal}, \citenamefont {Lorusso}, \citenamefont {Simpson}, \citenamefont {S\"oderstr\"om}, \citenamefont {Sumikama}, \citenamefont {Xu}, \citenamefont {Kumar}, \citenamefont {Mart\'{\i}nez-Pinedo}, \citenamefont {Nowacki}, \citenamefont {Van~Isacker}, \citenamefont {Baba}, \citenamefont {Browne}, \citenamefont {Fukuda}, \citenamefont {Gernh\"auser}, \citenamefont {Gey}, \citenamefont {Inabe}, \citenamefont {Isobe}, \citenamefont {Jung}, \citenamefont {Kameda}, \citenamefont {Kim}, \citenamefont {Kim}, \citenamefont {Kojouharov}, \citenamefont {Kubo}, \citenamefont {Kurz}, \citenamefont {Kwon}, \citenamefont {Li}, \citenamefont {Sakurai}, \citenamefont {Schaffner}, \citenamefont {Shimizu}, \citenamefont {Steiger}, \citenamefont {Suzuki}, \citenamefont {Takeda}, \citenamefont
  {Vajta}, \citenamefont {Watanabe}, \citenamefont {Wu}, \citenamefont {Yagi}, \citenamefont {Yoshinaga}, \citenamefont {Benzoni}, \citenamefont {B\"onig}, \citenamefont {Chae}, \citenamefont {Daugas}, \citenamefont {Drouet}, \citenamefont {Gadea}, \citenamefont {Ilieva}, \citenamefont {Kondev}, \citenamefont {Kr\"oll}, \citenamefont {Lane}, \citenamefont {Montaner-Piz\'a}, \citenamefont {Moschner}, \citenamefont {Naqvi}, \citenamefont {Niikura}, \citenamefont {Nishibata}, \citenamefont {Odahara}, \citenamefont {Orlandi}, \citenamefont {Patel}, \citenamefont {Podoly\'ak},\ and\ \citenamefont {Wendt}}]{Jungclaus2024}%
  \BibitemOpen
  \bibfield  {author} {\bibinfo {author} {\bibfnamefont {A.}~\bibnamefont {Jungclaus}}, \bibinfo {author} {\bibfnamefont {M.}~\bibnamefont {G\'orska}}, \bibinfo {author} {\bibfnamefont {M.}~\bibnamefont {Miko\l{}ajczuk}}, \bibinfo {author} {\bibfnamefont {J.}~\bibnamefont {Acosta}}, \bibinfo {author} {\bibfnamefont {J.}~\bibnamefont {Taprogge}}, \bibinfo {author} {\bibfnamefont {S.}~\bibnamefont {Nishimura}}, \bibinfo {author} {\bibfnamefont {P.}~\bibnamefont {Doornenbal}}, \bibinfo {author} {\bibfnamefont {G.}~\bibnamefont {Lorusso}}, \bibinfo {author} {\bibfnamefont {G.~S.}\ \bibnamefont {Simpson}}, \bibinfo {author} {\bibfnamefont {P.-A.}\ \bibnamefont {S\"oderstr\"om}}, \bibinfo {author} {\bibfnamefont {T.}~\bibnamefont {Sumikama}}, \bibinfo {author} {\bibfnamefont {Z.}~\bibnamefont {Xu}}, \bibinfo {author} {\bibfnamefont {P.}~\bibnamefont {Kumar}}, \bibinfo {author} {\bibfnamefont {G.}~\bibnamefont {Mart\'{\i}nez-Pinedo}}, \bibinfo {author} {\bibfnamefont {F.}~\bibnamefont {Nowacki}}, \bibinfo {author}
  {\bibfnamefont {P.}~\bibnamefont {Van~Isacker}}, \bibinfo {author} {\bibfnamefont {H.}~\bibnamefont {Baba}}, \bibinfo {author} {\bibfnamefont {F.}~\bibnamefont {Browne}}, \bibinfo {author} {\bibfnamefont {N.}~\bibnamefont {Fukuda}}, \bibinfo {author} {\bibfnamefont {R.}~\bibnamefont {Gernh\"auser}}, \bibinfo {author} {\bibfnamefont {G.}~\bibnamefont {Gey}}, \bibinfo {author} {\bibfnamefont {N.}~\bibnamefont {Inabe}}, \bibinfo {author} {\bibfnamefont {T.}~\bibnamefont {Isobe}}, \bibinfo {author} {\bibfnamefont {H.~S.}\ \bibnamefont {Jung}}, \bibinfo {author} {\bibfnamefont {D.}~\bibnamefont {Kameda}}, \bibinfo {author} {\bibfnamefont {G.~D.}\ \bibnamefont {Kim}}, \bibinfo {author} {\bibfnamefont {Y.-K.}\ \bibnamefont {Kim}}, \bibinfo {author} {\bibfnamefont {I.}~\bibnamefont {Kojouharov}}, \bibinfo {author} {\bibfnamefont {T.}~\bibnamefont {Kubo}}, \bibinfo {author} {\bibfnamefont {N.}~\bibnamefont {Kurz}}, \bibinfo {author} {\bibfnamefont {Y.~K.}\ \bibnamefont {Kwon}}, \bibinfo {author} {\bibfnamefont
  {Z.}~\bibnamefont {Li}}, \bibinfo {author} {\bibfnamefont {H.}~\bibnamefont {Sakurai}}, \bibinfo {author} {\bibfnamefont {H.}~\bibnamefont {Schaffner}}, \bibinfo {author} {\bibfnamefont {Y.}~\bibnamefont {Shimizu}}, \bibinfo {author} {\bibfnamefont {K.}~\bibnamefont {Steiger}}, \bibinfo {author} {\bibfnamefont {H.}~\bibnamefont {Suzuki}}, \bibinfo {author} {\bibfnamefont {H.}~\bibnamefont {Takeda}}, \bibinfo {author} {\bibfnamefont {Z.}~\bibnamefont {Vajta}}, \bibinfo {author} {\bibfnamefont {H.}~\bibnamefont {Watanabe}}, \bibinfo {author} {\bibfnamefont {J.}~\bibnamefont {Wu}}, \bibinfo {author} {\bibfnamefont {A.}~\bibnamefont {Yagi}}, \bibinfo {author} {\bibfnamefont {K.}~\bibnamefont {Yoshinaga}}, \bibinfo {author} {\bibfnamefont {G.}~\bibnamefont {Benzoni}}, \bibinfo {author} {\bibfnamefont {S.}~\bibnamefont {B\"onig}}, \bibinfo {author} {\bibfnamefont {K.~Y.}\ \bibnamefont {Chae}}, \bibinfo {author} {\bibfnamefont {J.-M.}\ \bibnamefont {Daugas}}, \bibinfo {author} {\bibfnamefont {F.}~\bibnamefont
  {Drouet}}, \bibinfo {author} {\bibfnamefont {A.}~\bibnamefont {Gadea}}, \bibinfo {author} {\bibfnamefont {S.}~\bibnamefont {Ilieva}}, \bibinfo {author} {\bibfnamefont {F.~G.}\ \bibnamefont {Kondev}}, \bibinfo {author} {\bibfnamefont {T.}~\bibnamefont {Kr\"oll}}, \bibinfo {author} {\bibfnamefont {G.~J.}\ \bibnamefont {Lane}}, \bibinfo {author} {\bibfnamefont {A.}~\bibnamefont {Montaner-Piz\'a}}, \bibinfo {author} {\bibfnamefont {K.}~\bibnamefont {Moschner}}, \bibinfo {author} {\bibfnamefont {F.}~\bibnamefont {Naqvi}}, \bibinfo {author} {\bibfnamefont {M.}~\bibnamefont {Niikura}}, \bibinfo {author} {\bibfnamefont {H.}~\bibnamefont {Nishibata}}, \bibinfo {author} {\bibfnamefont {A.}~\bibnamefont {Odahara}}, \bibinfo {author} {\bibfnamefont {R.}~\bibnamefont {Orlandi}}, \bibinfo {author} {\bibfnamefont {Z.}~\bibnamefont {Patel}}, \bibinfo {author} {\bibfnamefont {Z.}~\bibnamefont {Podoly\'ak}},\ and\ \bibinfo {author} {\bibfnamefont {A.}~\bibnamefont {Wendt}},\ }\href
  {https://doi.org/10.1103/PhysRevLett.132.222501} {\bibfield  {journal} {\bibinfo  {journal} {Phys. Rev. Lett.}\ }\textbf {\bibinfo {volume} {132}},\ \bibinfo {pages} {222501} (\bibinfo {year} {2024})}\BibitemShut {NoStop}%
\bibitem [{\citenamefont {Steppenbeck}\ \emph {et~al.}(2015)\citenamefont {Steppenbeck}, \citenamefont {Takeuchi}, \citenamefont {Aoi}, \citenamefont {Doornenbal}, \citenamefont {Matsushita}, \citenamefont {Wang}, \citenamefont {Utsuno}, \citenamefont {Baba}, \citenamefont {Go}, \citenamefont {Lee}, \citenamefont {Matsui}, \citenamefont {Michimasa}, \citenamefont {Motobayashi}, \citenamefont {Nishimura}, \citenamefont {Otsuka}, \citenamefont {Sakurai}, \citenamefont {Shiga}, \citenamefont {Shimizu}, \citenamefont {S\"oderstr\"om}, \citenamefont {Sumikama}, \citenamefont {Taniuchi}, \citenamefont {Valiente-Dob\'on},\ and\ \citenamefont {Yoneda}}]{Steppenbeck2015}%
  \BibitemOpen
  \bibfield  {author} {\bibinfo {author} {\bibfnamefont {D.}~\bibnamefont {Steppenbeck}}, \bibinfo {author} {\bibfnamefont {S.}~\bibnamefont {Takeuchi}}, \bibinfo {author} {\bibfnamefont {N.}~\bibnamefont {Aoi}}, \bibinfo {author} {\bibfnamefont {P.}~\bibnamefont {Doornenbal}}, \bibinfo {author} {\bibfnamefont {M.}~\bibnamefont {Matsushita}}, \bibinfo {author} {\bibfnamefont {H.}~\bibnamefont {Wang}}, \bibinfo {author} {\bibfnamefont {Y.}~\bibnamefont {Utsuno}}, \bibinfo {author} {\bibfnamefont {H.}~\bibnamefont {Baba}}, \bibinfo {author} {\bibfnamefont {S.}~\bibnamefont {Go}}, \bibinfo {author} {\bibfnamefont {J.}~\bibnamefont {Lee}}, \bibinfo {author} {\bibfnamefont {K.}~\bibnamefont {Matsui}}, \bibinfo {author} {\bibfnamefont {S.}~\bibnamefont {Michimasa}}, \bibinfo {author} {\bibfnamefont {T.}~\bibnamefont {Motobayashi}}, \bibinfo {author} {\bibfnamefont {D.}~\bibnamefont {Nishimura}}, \bibinfo {author} {\bibfnamefont {T.}~\bibnamefont {Otsuka}}, \bibinfo {author} {\bibfnamefont {H.}~\bibnamefont
  {Sakurai}}, \bibinfo {author} {\bibfnamefont {Y.}~\bibnamefont {Shiga}}, \bibinfo {author} {\bibfnamefont {N.}~\bibnamefont {Shimizu}}, \bibinfo {author} {\bibfnamefont {P.-A.}\ \bibnamefont {S\"oderstr\"om}}, \bibinfo {author} {\bibfnamefont {T.}~\bibnamefont {Sumikama}}, \bibinfo {author} {\bibfnamefont {R.}~\bibnamefont {Taniuchi}}, \bibinfo {author} {\bibfnamefont {J.~J.}\ \bibnamefont {Valiente-Dob\'on}},\ and\ \bibinfo {author} {\bibfnamefont {K.}~\bibnamefont {Yoneda}},\ }\href {https://doi.org/10.1103/PhysRevLett.114.252501} {\bibfield  {journal} {\bibinfo  {journal} {Phys. Rev. Lett.}\ }\textbf {\bibinfo {volume} {114}},\ \bibinfo {pages} {252501} (\bibinfo {year} {2015})}\BibitemShut {NoStop}%
\bibitem [{\citenamefont {Huck}\ \emph {et~al.}(1985)\citenamefont {Huck}, \citenamefont {Klotz}, \citenamefont {Knipper}, \citenamefont {Mieh\'e}, \citenamefont {Richard-Serre}, \citenamefont {Walter}, \citenamefont {Poves}, \citenamefont {Ravn},\ and\ \citenamefont {Marguier}}]{Huck1985}%
  \BibitemOpen
  \bibfield  {author} {\bibinfo {author} {\bibfnamefont {A.}~\bibnamefont {Huck}}, \bibinfo {author} {\bibfnamefont {G.}~\bibnamefont {Klotz}}, \bibinfo {author} {\bibfnamefont {A.}~\bibnamefont {Knipper}}, \bibinfo {author} {\bibfnamefont {C.}~\bibnamefont {Mieh\'e}}, \bibinfo {author} {\bibfnamefont {C.}~\bibnamefont {Richard-Serre}}, \bibinfo {author} {\bibfnamefont {G.}~\bibnamefont {Walter}}, \bibinfo {author} {\bibfnamefont {A.}~\bibnamefont {Poves}}, \bibinfo {author} {\bibfnamefont {H.~L.}\ \bibnamefont {Ravn}},\ and\ \bibinfo {author} {\bibfnamefont {G.}~\bibnamefont {Marguier}},\ }\href {https://doi.org/10.1103/PhysRevC.31.2226} {\bibfield  {journal} {\bibinfo  {journal} {Phys. Rev. C}\ }\textbf {\bibinfo {volume} {31}},\ \bibinfo {pages} {2226} (\bibinfo {year} {1985})}\BibitemShut {NoStop}%
\bibitem [{\citenamefont {Gade}\ \emph {et~al.}(2006)\citenamefont {Gade}, \citenamefont {Janssens}, \citenamefont {Bazin}, \citenamefont {Broda}, \citenamefont {Brown}, \citenamefont {Campbell}, \citenamefont {Carpenter}, \citenamefont {Cook}, \citenamefont {Deacon}, \citenamefont {Dinca}, \citenamefont {Fornal}, \citenamefont {Freeman}, \citenamefont {Glasmacher}, \citenamefont {Hansen}, \citenamefont {Kay}, \citenamefont {Mantica}, \citenamefont {Mueller}, \citenamefont {Terry}, \citenamefont {Tostevin},\ and\ \citenamefont {Zhu}}]{Gade2006}%
  \BibitemOpen
  \bibfield  {author} {\bibinfo {author} {\bibfnamefont {A.}~\bibnamefont {Gade}}, \bibinfo {author} {\bibfnamefont {R.~V.~F.}\ \bibnamefont {Janssens}}, \bibinfo {author} {\bibfnamefont {D.}~\bibnamefont {Bazin}}, \bibinfo {author} {\bibfnamefont {R.}~\bibnamefont {Broda}}, \bibinfo {author} {\bibfnamefont {B.~A.}\ \bibnamefont {Brown}}, \bibinfo {author} {\bibfnamefont {C.~M.}\ \bibnamefont {Campbell}}, \bibinfo {author} {\bibfnamefont {M.~P.}\ \bibnamefont {Carpenter}}, \bibinfo {author} {\bibfnamefont {J.~M.}\ \bibnamefont {Cook}}, \bibinfo {author} {\bibfnamefont {A.~N.}\ \bibnamefont {Deacon}}, \bibinfo {author} {\bibfnamefont {D.-C.}\ \bibnamefont {Dinca}}, \bibinfo {author} {\bibfnamefont {B.}~\bibnamefont {Fornal}}, \bibinfo {author} {\bibfnamefont {S.~J.}\ \bibnamefont {Freeman}}, \bibinfo {author} {\bibfnamefont {T.}~\bibnamefont {Glasmacher}}, \bibinfo {author} {\bibfnamefont {P.~G.}\ \bibnamefont {Hansen}}, \bibinfo {author} {\bibfnamefont {B.~P.}\ \bibnamefont {Kay}}, \bibinfo {author}
  {\bibfnamefont {P.~F.}\ \bibnamefont {Mantica}}, \bibinfo {author} {\bibfnamefont {W.~F.}\ \bibnamefont {Mueller}}, \bibinfo {author} {\bibfnamefont {J.~R.}\ \bibnamefont {Terry}}, \bibinfo {author} {\bibfnamefont {J.~A.}\ \bibnamefont {Tostevin}},\ and\ \bibinfo {author} {\bibfnamefont {S.}~\bibnamefont {Zhu}},\ }\href {https://doi.org/10.1103/PhysRevC.74.021302} {\bibfield  {journal} {\bibinfo  {journal} {Phys. Rev. C}\ }\textbf {\bibinfo {volume} {74}},\ \bibinfo {pages} {021302} (\bibinfo {year} {2006})}\BibitemShut {NoStop}%
\bibitem [{\citenamefont {Janssens}\ \emph {et~al.}(2002)\citenamefont {Janssens}, \citenamefont {Fornal}, \citenamefont {Mantica}, \citenamefont {Brown}, \citenamefont {Broda}, \citenamefont {Bhattacharyya}, \citenamefont {Carpenter}, \citenamefont {Cinausero}, \citenamefont {Daly}, \citenamefont {Davies}, \citenamefont {Glasmacher}, \citenamefont {Grabowski}, \citenamefont {Groh}, \citenamefont {Honma}, \citenamefont {Kondev}, \citenamefont {Królas}, \citenamefont {Lauritsen}, \citenamefont {Liddick}, \citenamefont {Lunardi}, \citenamefont {Marginean}, \citenamefont {Mizusaki}, \citenamefont {Morrissey}, \citenamefont {Morton}, \citenamefont {Mueller}, \citenamefont {Otsuka}, \citenamefont {Pawlat}, \citenamefont {Seweryniak}, \citenamefont {Schatz}, \citenamefont {Stolz}, \citenamefont {Tabor}, \citenamefont {Ur}, \citenamefont {Viesti}, \citenamefont {Wiedenhöver},\ and\ \citenamefont {Wrzesiński}}]{Janssens2002}%
  \BibitemOpen
  \bibfield  {author} {\bibinfo {author} {\bibfnamefont {R.~V.~F.}\ \bibnamefont {Janssens}}, \bibinfo {author} {\bibfnamefont {B.}~\bibnamefont {Fornal}}, \bibinfo {author} {\bibfnamefont {P.~F.}\ \bibnamefont {Mantica}}, \bibinfo {author} {\bibfnamefont {B.~A.}\ \bibnamefont {Brown}}, \bibinfo {author} {\bibfnamefont {R.}~\bibnamefont {Broda}}, \bibinfo {author} {\bibfnamefont {P.}~\bibnamefont {Bhattacharyya}}, \bibinfo {author} {\bibfnamefont {M.~P.}\ \bibnamefont {Carpenter}}, \bibinfo {author} {\bibfnamefont {M.}~\bibnamefont {Cinausero}}, \bibinfo {author} {\bibfnamefont {P.~J.}\ \bibnamefont {Daly}}, \bibinfo {author} {\bibfnamefont {A.~D.}\ \bibnamefont {Davies}}, \bibinfo {author} {\bibfnamefont {T.}~\bibnamefont {Glasmacher}}, \bibinfo {author} {\bibfnamefont {Z.~W.}\ \bibnamefont {Grabowski}}, \bibinfo {author} {\bibfnamefont {D.~E.}\ \bibnamefont {Groh}}, \bibinfo {author} {\bibfnamefont {M.}~\bibnamefont {Honma}}, \bibinfo {author} {\bibfnamefont {F.~G.}\ \bibnamefont {Kondev}}, \bibinfo
  {author} {\bibfnamefont {W.}~\bibnamefont {Królas}}, \bibinfo {author} {\bibfnamefont {T.}~\bibnamefont {Lauritsen}}, \bibinfo {author} {\bibfnamefont {S.~N.}\ \bibnamefont {Liddick}}, \bibinfo {author} {\bibfnamefont {S.}~\bibnamefont {Lunardi}}, \bibinfo {author} {\bibfnamefont {N.}~\bibnamefont {Marginean}}, \bibinfo {author} {\bibfnamefont {T.}~\bibnamefont {Mizusaki}}, \bibinfo {author} {\bibfnamefont {D.~J.}\ \bibnamefont {Morrissey}}, \bibinfo {author} {\bibfnamefont {A.~C.}\ \bibnamefont {Morton}}, \bibinfo {author} {\bibfnamefont {W.~F.}\ \bibnamefont {Mueller}}, \bibinfo {author} {\bibfnamefont {T.}~\bibnamefont {Otsuka}}, \bibinfo {author} {\bibfnamefont {T.}~\bibnamefont {Pawlat}}, \bibinfo {author} {\bibfnamefont {D.}~\bibnamefont {Seweryniak}}, \bibinfo {author} {\bibfnamefont {H.}~\bibnamefont {Schatz}}, \bibinfo {author} {\bibfnamefont {A.}~\bibnamefont {Stolz}}, \bibinfo {author} {\bibfnamefont {S.~L.}\ \bibnamefont {Tabor}}, \bibinfo {author} {\bibfnamefont {C.~A.}\ \bibnamefont {Ur}},
  \bibinfo {author} {\bibfnamefont {G.}~\bibnamefont {Viesti}}, \bibinfo {author} {\bibfnamefont {I.}~\bibnamefont {Wiedenhöver}},\ and\ \bibinfo {author} {\bibfnamefont {J.}~\bibnamefont {Wrzesiński}},\ }\href {https://doi.org/https://doi.org/10.1016/S0370-2693(02)02682-5} {\bibfield  {journal} {\bibinfo  {journal} {Physics Letters B}\ }\textbf {\bibinfo {volume} {546}},\ \bibinfo {pages} {55} (\bibinfo {year} {2002})}\BibitemShut {NoStop}%
\bibitem [{\citenamefont {Liddick}\ \emph {et~al.}(2004)\citenamefont {Liddick}, \citenamefont {Mantica}, \citenamefont {Broda}, \citenamefont {Brown}, \citenamefont {Carpenter}, \citenamefont {Davies}, \citenamefont {Fornal}, \citenamefont {Glasmacher}, \citenamefont {Groh}, \citenamefont {Honma}, \citenamefont {Horoi}, \citenamefont {Janssens}, \citenamefont {Mizusaki}, \citenamefont {Morrissey}, \citenamefont {Morton}, \citenamefont {Mueller}, \citenamefont {Otsuka}, \citenamefont {Pavan}, \citenamefont {Schatz}, \citenamefont {Stolz}, \citenamefont {Tabor}, \citenamefont {Tomlin},\ and\ \citenamefont {Wiedeking}}]{Liddick2004}%
  \BibitemOpen
  \bibfield  {author} {\bibinfo {author} {\bibfnamefont {S.~N.}\ \bibnamefont {Liddick}}, \bibinfo {author} {\bibfnamefont {P.~F.}\ \bibnamefont {Mantica}}, \bibinfo {author} {\bibfnamefont {R.}~\bibnamefont {Broda}}, \bibinfo {author} {\bibfnamefont {B.~A.}\ \bibnamefont {Brown}}, \bibinfo {author} {\bibfnamefont {M.~P.}\ \bibnamefont {Carpenter}}, \bibinfo {author} {\bibfnamefont {A.~D.}\ \bibnamefont {Davies}}, \bibinfo {author} {\bibfnamefont {B.}~\bibnamefont {Fornal}}, \bibinfo {author} {\bibfnamefont {T.}~\bibnamefont {Glasmacher}}, \bibinfo {author} {\bibfnamefont {D.~E.}\ \bibnamefont {Groh}}, \bibinfo {author} {\bibfnamefont {M.}~\bibnamefont {Honma}}, \bibinfo {author} {\bibfnamefont {M.}~\bibnamefont {Horoi}}, \bibinfo {author} {\bibfnamefont {R.~V.~F.}\ \bibnamefont {Janssens}}, \bibinfo {author} {\bibfnamefont {T.}~\bibnamefont {Mizusaki}}, \bibinfo {author} {\bibfnamefont {D.~J.}\ \bibnamefont {Morrissey}}, \bibinfo {author} {\bibfnamefont {A.~C.}\ \bibnamefont {Morton}}, \bibinfo {author}
  {\bibfnamefont {W.~F.}\ \bibnamefont {Mueller}}, \bibinfo {author} {\bibfnamefont {T.}~\bibnamefont {Otsuka}}, \bibinfo {author} {\bibfnamefont {J.}~\bibnamefont {Pavan}}, \bibinfo {author} {\bibfnamefont {H.}~\bibnamefont {Schatz}}, \bibinfo {author} {\bibfnamefont {A.}~\bibnamefont {Stolz}}, \bibinfo {author} {\bibfnamefont {S.~L.}\ \bibnamefont {Tabor}}, \bibinfo {author} {\bibfnamefont {B.~E.}\ \bibnamefont {Tomlin}},\ and\ \bibinfo {author} {\bibfnamefont {M.}~\bibnamefont {Wiedeking}},\ }\href {https://doi.org/10.1103/PhysRevC.70.064303} {\bibfield  {journal} {\bibinfo  {journal} {Phys. Rev. C}\ }\textbf {\bibinfo {volume} {70}},\ \bibinfo {pages} {064303} (\bibinfo {year} {2004})}\BibitemShut {NoStop}%
\bibitem [{\citenamefont {Prisciandaro}\ \emph {et~al.}(2001)\citenamefont {Prisciandaro}, \citenamefont {Mantica}, \citenamefont {Brown}, \citenamefont {Anthony}, \citenamefont {Cooper}, \citenamefont {Garcia}, \citenamefont {Groh}, \citenamefont {Komives}, \citenamefont {Kumarasiri}, \citenamefont {Lofy}, \citenamefont {Oros-Peusquens}, \citenamefont {Tabor},\ and\ \citenamefont {Wiedeking}}]{Prisciandaro2001}%
  \BibitemOpen
  \bibfield  {author} {\bibinfo {author} {\bibfnamefont {J.~I.}\ \bibnamefont {Prisciandaro}}, \bibinfo {author} {\bibfnamefont {P.~F.}\ \bibnamefont {Mantica}}, \bibinfo {author} {\bibfnamefont {B.~A.}\ \bibnamefont {Brown}}, \bibinfo {author} {\bibfnamefont {D.~W.}\ \bibnamefont {Anthony}}, \bibinfo {author} {\bibfnamefont {M.~W.}\ \bibnamefont {Cooper}}, \bibinfo {author} {\bibfnamefont {A.}~\bibnamefont {Garcia}}, \bibinfo {author} {\bibfnamefont {D.~E.}\ \bibnamefont {Groh}}, \bibinfo {author} {\bibfnamefont {A.}~\bibnamefont {Komives}}, \bibinfo {author} {\bibfnamefont {W.}~\bibnamefont {Kumarasiri}}, \bibinfo {author} {\bibfnamefont {P.~A.}\ \bibnamefont {Lofy}}, \bibinfo {author} {\bibfnamefont {A.~M.}\ \bibnamefont {Oros-Peusquens}}, \bibinfo {author} {\bibfnamefont {S.~L.}\ \bibnamefont {Tabor}},\ and\ \bibinfo {author} {\bibfnamefont {M.}~\bibnamefont {Wiedeking}},\ }\href {https://doi.org/https://doi.org/10.1016/S0370-2693(01)00565-2} {\bibfield  {journal} {\bibinfo  {journal} {Physics Letters B}\
  }\textbf {\bibinfo {volume} {510}},\ \bibinfo {pages} {17} (\bibinfo {year} {2001})}\BibitemShut {NoStop}%
\bibitem [{\citenamefont {Mantica}\ \emph {et~al.}(2003)\citenamefont {Mantica}, \citenamefont {Morton}, \citenamefont {Brown}, \citenamefont {Davies}, \citenamefont {Glasmacher}, \citenamefont {Groh}, \citenamefont {Liddick}, \citenamefont {Morrissey}, \citenamefont {Mueller}, \citenamefont {Schatz}, \citenamefont {Stolz}, \citenamefont {Tabor}, \citenamefont {Honma}, \citenamefont {Horoi},\ and\ \citenamefont {Otsuka}}]{Mantica2003}%
  \BibitemOpen
  \bibfield  {author} {\bibinfo {author} {\bibfnamefont {P.~F.}\ \bibnamefont {Mantica}}, \bibinfo {author} {\bibfnamefont {A.~C.}\ \bibnamefont {Morton}}, \bibinfo {author} {\bibfnamefont {B.~A.}\ \bibnamefont {Brown}}, \bibinfo {author} {\bibfnamefont {A.~D.}\ \bibnamefont {Davies}}, \bibinfo {author} {\bibfnamefont {T.}~\bibnamefont {Glasmacher}}, \bibinfo {author} {\bibfnamefont {D.~E.}\ \bibnamefont {Groh}}, \bibinfo {author} {\bibfnamefont {S.~N.}\ \bibnamefont {Liddick}}, \bibinfo {author} {\bibfnamefont {D.~J.}\ \bibnamefont {Morrissey}}, \bibinfo {author} {\bibfnamefont {W.~F.}\ \bibnamefont {Mueller}}, \bibinfo {author} {\bibfnamefont {H.}~\bibnamefont {Schatz}}, \bibinfo {author} {\bibfnamefont {A.}~\bibnamefont {Stolz}}, \bibinfo {author} {\bibfnamefont {S.~L.}\ \bibnamefont {Tabor}}, \bibinfo {author} {\bibfnamefont {M.}~\bibnamefont {Honma}}, \bibinfo {author} {\bibfnamefont {M.}~\bibnamefont {Horoi}},\ and\ \bibinfo {author} {\bibfnamefont {T.}~\bibnamefont {Otsuka}},\ }\href
  {https://doi.org/10.1103/PhysRevC.67.014311} {\bibfield  {journal} {\bibinfo  {journal} {Phys. Rev. C}\ }\textbf {\bibinfo {volume} {67}},\ \bibinfo {pages} {014311} (\bibinfo {year} {2003})}\BibitemShut {NoStop}%
\bibitem [{\citenamefont {Bürger}\ \emph {et~al.}(2005)\citenamefont {Bürger}, \citenamefont {Saito}, \citenamefont {Grawe}, \citenamefont {Hübel}, \citenamefont {Reiter}, \citenamefont {Gerl}, \citenamefont {Górska}, \citenamefont {Wollersheim}, \citenamefont {Al-Khatib}, \citenamefont {Banu}, \citenamefont {Beck}, \citenamefont {Becker}, \citenamefont {Bednarczyk}, \citenamefont {Benzoni}, \citenamefont {Bracco}, \citenamefont {Brambilla}, \citenamefont {Bringel}, \citenamefont {Camera}, \citenamefont {Clément}, \citenamefont {Doornenbal}, \citenamefont {Geissel}, \citenamefont {Görgen}, \citenamefont {Grębosz}, \citenamefont {Hammond}, \citenamefont {Hellström}, \citenamefont {Honma}, \citenamefont {Kavatsyuk}, \citenamefont {Kavatsyuk}, \citenamefont {Kmiecik}, \citenamefont {Kojouharov}, \citenamefont {Korten}, \citenamefont {Kurz}, \citenamefont {Lozeva}, \citenamefont {Maj}, \citenamefont {Mandal}, \citenamefont {Million}, \citenamefont {Muralithar}, \citenamefont {Neußer}, \citenamefont
  {Nowacki}, \citenamefont {Otsuka}, \citenamefont {Podolyák}, \citenamefont {Saito}, \citenamefont {Singh}, \citenamefont {Weick}, \citenamefont {Wheldon}, \citenamefont {Wieland},\ and\ \citenamefont {Winkler}}]{Burger2005}%
  \BibitemOpen
  \bibfield  {author} {\bibinfo {author} {\bibfnamefont {A.}~\bibnamefont {Bürger}}, \bibinfo {author} {\bibfnamefont {T.~R.}\ \bibnamefont {Saito}}, \bibinfo {author} {\bibfnamefont {H.}~\bibnamefont {Grawe}}, \bibinfo {author} {\bibfnamefont {H.}~\bibnamefont {Hübel}}, \bibinfo {author} {\bibfnamefont {P.}~\bibnamefont {Reiter}}, \bibinfo {author} {\bibfnamefont {J.}~\bibnamefont {Gerl}}, \bibinfo {author} {\bibfnamefont {M.}~\bibnamefont {Górska}}, \bibinfo {author} {\bibfnamefont {H.~J.}\ \bibnamefont {Wollersheim}}, \bibinfo {author} {\bibfnamefont {A.}~\bibnamefont {Al-Khatib}}, \bibinfo {author} {\bibfnamefont {A.}~\bibnamefont {Banu}}, \bibinfo {author} {\bibfnamefont {T.}~\bibnamefont {Beck}}, \bibinfo {author} {\bibfnamefont {F.}~\bibnamefont {Becker}}, \bibinfo {author} {\bibfnamefont {P.}~\bibnamefont {Bednarczyk}}, \bibinfo {author} {\bibfnamefont {G.}~\bibnamefont {Benzoni}}, \bibinfo {author} {\bibfnamefont {A.}~\bibnamefont {Bracco}}, \bibinfo {author} {\bibfnamefont {S.}~\bibnamefont
  {Brambilla}}, \bibinfo {author} {\bibfnamefont {P.}~\bibnamefont {Bringel}}, \bibinfo {author} {\bibfnamefont {F.}~\bibnamefont {Camera}}, \bibinfo {author} {\bibfnamefont {E.}~\bibnamefont {Clément}}, \bibinfo {author} {\bibfnamefont {P.}~\bibnamefont {Doornenbal}}, \bibinfo {author} {\bibfnamefont {H.}~\bibnamefont {Geissel}}, \bibinfo {author} {\bibfnamefont {A.}~\bibnamefont {Görgen}}, \bibinfo {author} {\bibfnamefont {J.}~\bibnamefont {Grębosz}}, \bibinfo {author} {\bibfnamefont {G.}~\bibnamefont {Hammond}}, \bibinfo {author} {\bibfnamefont {M.}~\bibnamefont {Hellström}}, \bibinfo {author} {\bibfnamefont {M.}~\bibnamefont {Honma}}, \bibinfo {author} {\bibfnamefont {M.}~\bibnamefont {Kavatsyuk}}, \bibinfo {author} {\bibfnamefont {O.}~\bibnamefont {Kavatsyuk}}, \bibinfo {author} {\bibfnamefont {M.}~\bibnamefont {Kmiecik}}, \bibinfo {author} {\bibfnamefont {I.}~\bibnamefont {Kojouharov}}, \bibinfo {author} {\bibfnamefont {W.}~\bibnamefont {Korten}}, \bibinfo {author} {\bibfnamefont {N.}~\bibnamefont
  {Kurz}}, \bibinfo {author} {\bibfnamefont {R.}~\bibnamefont {Lozeva}}, \bibinfo {author} {\bibfnamefont {A.}~\bibnamefont {Maj}}, \bibinfo {author} {\bibfnamefont {S.}~\bibnamefont {Mandal}}, \bibinfo {author} {\bibfnamefont {B.}~\bibnamefont {Million}}, \bibinfo {author} {\bibfnamefont {S.}~\bibnamefont {Muralithar}}, \bibinfo {author} {\bibfnamefont {A.}~\bibnamefont {Neußer}}, \bibinfo {author} {\bibfnamefont {F.}~\bibnamefont {Nowacki}}, \bibinfo {author} {\bibfnamefont {T.}~\bibnamefont {Otsuka}}, \bibinfo {author} {\bibfnamefont {Z.}~\bibnamefont {Podolyák}}, \bibinfo {author} {\bibfnamefont {N.}~\bibnamefont {Saito}}, \bibinfo {author} {\bibfnamefont {A.~K.}\ \bibnamefont {Singh}}, \bibinfo {author} {\bibfnamefont {H.}~\bibnamefont {Weick}}, \bibinfo {author} {\bibfnamefont {C.}~\bibnamefont {Wheldon}}, \bibinfo {author} {\bibfnamefont {O.}~\bibnamefont {Wieland}},\ and\ \bibinfo {author} {\bibfnamefont {M.}~\bibnamefont {Winkler}},\ }\href
  {https://doi.org/https://doi.org/10.1016/j.physletb.2005.07.004} {\bibfield  {journal} {\bibinfo  {journal} {Physics Letters B}\ }\textbf {\bibinfo {volume} {622}},\ \bibinfo {pages} {29} (\bibinfo {year} {2005})}\BibitemShut {NoStop}%
\bibitem [{\citenamefont {Gallant}\ \emph {et~al.}(2012)\citenamefont {Gallant}, \citenamefont {Bale}, \citenamefont {Brunner}, \citenamefont {Chowdhury}, \citenamefont {Ettenauer}, \citenamefont {Lennarz}, \citenamefont {Robertson}, \citenamefont {Simon}, \citenamefont {Chaudhuri}, \citenamefont {Holt}, \citenamefont {Kwiatkowski}, \citenamefont {Man\'e}, \citenamefont {Men\'endez}, \citenamefont {Schultz}, \citenamefont {Simon}, \citenamefont {Andreoiu}, \citenamefont {Delheij}, \citenamefont {Pearson}, \citenamefont {Savajols}, \citenamefont {Schwenk},\ and\ \citenamefont {Dilling}}]{Gallant2012}%
  \BibitemOpen
  \bibfield  {author} {\bibinfo {author} {\bibfnamefont {A.~T.}\ \bibnamefont {Gallant}}, \bibinfo {author} {\bibfnamefont {J.~C.}\ \bibnamefont {Bale}}, \bibinfo {author} {\bibfnamefont {T.}~\bibnamefont {Brunner}}, \bibinfo {author} {\bibfnamefont {U.}~\bibnamefont {Chowdhury}}, \bibinfo {author} {\bibfnamefont {S.}~\bibnamefont {Ettenauer}}, \bibinfo {author} {\bibfnamefont {A.}~\bibnamefont {Lennarz}}, \bibinfo {author} {\bibfnamefont {D.}~\bibnamefont {Robertson}}, \bibinfo {author} {\bibfnamefont {V.~V.}\ \bibnamefont {Simon}}, \bibinfo {author} {\bibfnamefont {A.}~\bibnamefont {Chaudhuri}}, \bibinfo {author} {\bibfnamefont {J.~D.}\ \bibnamefont {Holt}}, \bibinfo {author} {\bibfnamefont {A.~A.}\ \bibnamefont {Kwiatkowski}}, \bibinfo {author} {\bibfnamefont {E.}~\bibnamefont {Man\'e}}, \bibinfo {author} {\bibfnamefont {J.}~\bibnamefont {Men\'endez}}, \bibinfo {author} {\bibfnamefont {B.~E.}\ \bibnamefont {Schultz}}, \bibinfo {author} {\bibfnamefont {M.~C.}\ \bibnamefont {Simon}}, \bibinfo {author}
  {\bibfnamefont {C.}~\bibnamefont {Andreoiu}}, \bibinfo {author} {\bibfnamefont {P.}~\bibnamefont {Delheij}}, \bibinfo {author} {\bibfnamefont {M.~R.}\ \bibnamefont {Pearson}}, \bibinfo {author} {\bibfnamefont {H.}~\bibnamefont {Savajols}}, \bibinfo {author} {\bibfnamefont {A.}~\bibnamefont {Schwenk}},\ and\ \bibinfo {author} {\bibfnamefont {J.}~\bibnamefont {Dilling}},\ }\href {https://doi.org/10.1103/PhysRevLett.109.032506} {\bibfield  {journal} {\bibinfo  {journal} {Phys. Rev. Lett.}\ }\textbf {\bibinfo {volume} {109}},\ \bibinfo {pages} {032506} (\bibinfo {year} {2012})}\BibitemShut {NoStop}%
\bibitem [{\citenamefont {Wienholtz}\ \emph {et~al.}(2013)\citenamefont {Wienholtz}, \citenamefont {Beck}, \citenamefont {Blaum}, \citenamefont {Borgmann}, \citenamefont {Breitenfeldt}, \citenamefont {Cakirli}, \citenamefont {George}, \citenamefont {Herfurth}, \citenamefont {Holt}, \citenamefont {Kowalska}, \citenamefont {Kreim}, \citenamefont {Lunney}, \citenamefont {Manea}, \citenamefont {Menéndez}, \citenamefont {Neidherr}, \citenamefont {Rosenbusch}, \citenamefont {Schweikhard}, \citenamefont {Schwenk}, \citenamefont {Simonis}, \citenamefont {Stanja}, \citenamefont {Wolf},\ and\ \citenamefont {Zuber}}]{Wienholtz2013}%
  \BibitemOpen
  \bibfield  {author} {\bibinfo {author} {\bibfnamefont {F.}~\bibnamefont {Wienholtz}}, \bibinfo {author} {\bibfnamefont {D.}~\bibnamefont {Beck}}, \bibinfo {author} {\bibfnamefont {K.}~\bibnamefont {Blaum}}, \bibinfo {author} {\bibfnamefont {C.}~\bibnamefont {Borgmann}}, \bibinfo {author} {\bibfnamefont {M.}~\bibnamefont {Breitenfeldt}}, \bibinfo {author} {\bibfnamefont {R.~B.}\ \bibnamefont {Cakirli}}, \bibinfo {author} {\bibfnamefont {S.}~\bibnamefont {George}}, \bibinfo {author} {\bibfnamefont {F.}~\bibnamefont {Herfurth}}, \bibinfo {author} {\bibfnamefont {J.~D.}\ \bibnamefont {Holt}}, \bibinfo {author} {\bibfnamefont {M.}~\bibnamefont {Kowalska}}, \bibinfo {author} {\bibfnamefont {S.}~\bibnamefont {Kreim}}, \bibinfo {author} {\bibfnamefont {D.}~\bibnamefont {Lunney}}, \bibinfo {author} {\bibfnamefont {V.}~\bibnamefont {Manea}}, \bibinfo {author} {\bibfnamefont {J.}~\bibnamefont {Menéndez}}, \bibinfo {author} {\bibfnamefont {D.}~\bibnamefont {Neidherr}}, \bibinfo {author} {\bibfnamefont
  {M.}~\bibnamefont {Rosenbusch}}, \bibinfo {author} {\bibfnamefont {L.}~\bibnamefont {Schweikhard}}, \bibinfo {author} {\bibfnamefont {A.}~\bibnamefont {Schwenk}}, \bibinfo {author} {\bibfnamefont {J.}~\bibnamefont {Simonis}}, \bibinfo {author} {\bibfnamefont {J.}~\bibnamefont {Stanja}}, \bibinfo {author} {\bibfnamefont {R.~N.}\ \bibnamefont {Wolf}},\ and\ \bibinfo {author} {\bibfnamefont {K.}~\bibnamefont {Zuber}},\ }\href {https://doi.org/10.1038/nature12226} {\bibfield  {journal} {\bibinfo  {journal} {Nature}\ }\textbf {\bibinfo {volume} {498}},\ \bibinfo {pages} {346} (\bibinfo {year} {2013})}\BibitemShut {NoStop}%
\bibitem [{\citenamefont {Rosenbusch}\ \emph {et~al.}(2015)\citenamefont {Rosenbusch}, \citenamefont {Ascher}, \citenamefont {Atanasov}, \citenamefont {Barbieri}, \citenamefont {Beck}, \citenamefont {Blaum}, \citenamefont {Borgmann}, \citenamefont {Breitenfeldt}, \citenamefont {Cakirli}, \citenamefont {Cipollone}, \citenamefont {George}, \citenamefont {Herfurth}, \citenamefont {Kowalska}, \citenamefont {Kreim}, \citenamefont {Lunney}, \citenamefont {Manea}, \citenamefont {Navr\'atil}, \citenamefont {Neidherr}, \citenamefont {Schweikhard}, \citenamefont {Som\`a}, \citenamefont {Stanja}, \citenamefont {Wienholtz}, \citenamefont {Wolf},\ and\ \citenamefont {Zuber}}]{Rosenbusch2015}%
  \BibitemOpen
  \bibfield  {author} {\bibinfo {author} {\bibfnamefont {M.}~\bibnamefont {Rosenbusch}}, \bibinfo {author} {\bibfnamefont {P.}~\bibnamefont {Ascher}}, \bibinfo {author} {\bibfnamefont {D.}~\bibnamefont {Atanasov}}, \bibinfo {author} {\bibfnamefont {C.}~\bibnamefont {Barbieri}}, \bibinfo {author} {\bibfnamefont {D.}~\bibnamefont {Beck}}, \bibinfo {author} {\bibfnamefont {K.}~\bibnamefont {Blaum}}, \bibinfo {author} {\bibfnamefont {C.}~\bibnamefont {Borgmann}}, \bibinfo {author} {\bibfnamefont {M.}~\bibnamefont {Breitenfeldt}}, \bibinfo {author} {\bibfnamefont {R.~B.}\ \bibnamefont {Cakirli}}, \bibinfo {author} {\bibfnamefont {A.}~\bibnamefont {Cipollone}}, \bibinfo {author} {\bibfnamefont {S.}~\bibnamefont {George}}, \bibinfo {author} {\bibfnamefont {F.}~\bibnamefont {Herfurth}}, \bibinfo {author} {\bibfnamefont {M.}~\bibnamefont {Kowalska}}, \bibinfo {author} {\bibfnamefont {S.}~\bibnamefont {Kreim}}, \bibinfo {author} {\bibfnamefont {D.}~\bibnamefont {Lunney}}, \bibinfo {author} {\bibfnamefont
  {V.}~\bibnamefont {Manea}}, \bibinfo {author} {\bibfnamefont {P.}~\bibnamefont {Navr\'atil}}, \bibinfo {author} {\bibfnamefont {D.}~\bibnamefont {Neidherr}}, \bibinfo {author} {\bibfnamefont {L.}~\bibnamefont {Schweikhard}}, \bibinfo {author} {\bibfnamefont {V.}~\bibnamefont {Som\`a}}, \bibinfo {author} {\bibfnamefont {J.}~\bibnamefont {Stanja}}, \bibinfo {author} {\bibfnamefont {F.}~\bibnamefont {Wienholtz}}, \bibinfo {author} {\bibfnamefont {R.~N.}\ \bibnamefont {Wolf}},\ and\ \bibinfo {author} {\bibfnamefont {K.}~\bibnamefont {Zuber}},\ }\href {https://doi.org/10.1103/PhysRevLett.114.202501} {\bibfield  {journal} {\bibinfo  {journal} {Phys. Rev. Lett.}\ }\textbf {\bibinfo {volume} {114}},\ \bibinfo {pages} {202501} (\bibinfo {year} {2015})}\BibitemShut {NoStop}%
\bibitem [{\citenamefont {Xu}\ \emph {et~al.}(2015)\citenamefont {Xu}, \citenamefont {Wang}, \citenamefont {Zhang}, \citenamefont {Xu}, \citenamefont {Shuai}, \citenamefont {Tu}, \citenamefont {Litvinov}, \citenamefont {Zhou}, \citenamefont {Sun}, \citenamefont {Yuan}, \citenamefont {Xia}, \citenamefont {Yang}, \citenamefont {Blaum}, \citenamefont {Chen}, \citenamefont {Chen}, \citenamefont {Fu}, \citenamefont {Ge}, \citenamefont {Hu}, \citenamefont {Huang}, \citenamefont {Liu}, \citenamefont {Lam}, \citenamefont {Ma}, \citenamefont {Mao}, \citenamefont {Uesaka}, \citenamefont {Xiao}, \citenamefont {Xing}, \citenamefont {Yamaguchi}, \citenamefont {Yamaguchi}, \citenamefont {Zeng}, \citenamefont {Yan}, \citenamefont {Zhao}, \citenamefont {Zhao}, \citenamefont {Zhang},\ and\ \citenamefont {Zhan}}]{Xu2015}%
  \BibitemOpen
  \bibfield  {author} {\bibinfo {author} {\bibfnamefont {X.}~\bibnamefont {Xu}}, \bibinfo {author} {\bibfnamefont {M.}~\bibnamefont {Wang}}, \bibinfo {author} {\bibfnamefont {Y.-H.}\ \bibnamefont {Zhang}}, \bibinfo {author} {\bibfnamefont {H.-S.}\ \bibnamefont {Xu}}, \bibinfo {author} {\bibfnamefont {P.}~\bibnamefont {Shuai}}, \bibinfo {author} {\bibfnamefont {X.-L.}\ \bibnamefont {Tu}}, \bibinfo {author} {\bibfnamefont {Y.~A.}\ \bibnamefont {Litvinov}}, \bibinfo {author} {\bibfnamefont {X.-H.}\ \bibnamefont {Zhou}}, \bibinfo {author} {\bibfnamefont {B.-H.}\ \bibnamefont {Sun}}, \bibinfo {author} {\bibfnamefont {Y.-J.}\ \bibnamefont {Yuan}}, \bibinfo {author} {\bibfnamefont {J.-W.}\ \bibnamefont {Xia}}, \bibinfo {author} {\bibfnamefont {J.-C.}\ \bibnamefont {Yang}}, \bibinfo {author} {\bibfnamefont {K.}~\bibnamefont {Blaum}}, \bibinfo {author} {\bibfnamefont {R.-J.}\ \bibnamefont {Chen}}, \bibinfo {author} {\bibfnamefont {X.-C.}\ \bibnamefont {Chen}}, \bibinfo {author} {\bibfnamefont {C.-Y.}\ \bibnamefont
  {Fu}}, \bibinfo {author} {\bibfnamefont {Z.}~\bibnamefont {Ge}}, \bibinfo {author} {\bibfnamefont {Z.-G.}\ \bibnamefont {Hu}}, \bibinfo {author} {\bibfnamefont {W.-J.}\ \bibnamefont {Huang}}, \bibinfo {author} {\bibfnamefont {D.-W.}\ \bibnamefont {Liu}}, \bibinfo {author} {\bibfnamefont {Y.-H.}\ \bibnamefont {Lam}}, \bibinfo {author} {\bibfnamefont {X.-W.}\ \bibnamefont {Ma}}, \bibinfo {author} {\bibfnamefont {R.-S.}\ \bibnamefont {Mao}}, \bibinfo {author} {\bibfnamefont {T.}~\bibnamefont {Uesaka}}, \bibinfo {author} {\bibfnamefont {G.-Q.}\ \bibnamefont {Xiao}}, \bibinfo {author} {\bibfnamefont {Y.-M.}\ \bibnamefont {Xing}}, \bibinfo {author} {\bibfnamefont {T.}~\bibnamefont {Yamaguchi}}, \bibinfo {author} {\bibfnamefont {Y.}~\bibnamefont {Yamaguchi}}, \bibinfo {author} {\bibfnamefont {Q.}~\bibnamefont {Zeng}}, \bibinfo {author} {\bibfnamefont {X.-L.}\ \bibnamefont {Yan}}, \bibinfo {author} {\bibfnamefont {H.-W.}\ \bibnamefont {Zhao}}, \bibinfo {author} {\bibfnamefont {T.-C.}\ \bibnamefont {Zhao}}, \bibinfo
  {author} {\bibfnamefont {W.}~\bibnamefont {Zhang}},\ and\ \bibinfo {author} {\bibfnamefont {W.-L.}\ \bibnamefont {Zhan}},\ }\href {https://doi.org/10.1088/1674-1137/39/10/104001} {\bibfield  {journal} {\bibinfo  {journal} {Chinese Physics C}\ }\textbf {\bibinfo {volume} {39}},\ \bibinfo {pages} {104001} (\bibinfo {year} {2015})}\BibitemShut {NoStop}%
\bibitem [{\citenamefont {Xu}\ \emph {et~al.}(2019)\citenamefont {Xu}, \citenamefont {Wang}, \citenamefont {Blaum}, \citenamefont {Holt}, \citenamefont {Litvinov}, \citenamefont {Schwenk}, \citenamefont {Simonis}, \citenamefont {Stroberg}, \citenamefont {Zhang}, \citenamefont {Xu}, \citenamefont {Shuai}, \citenamefont {Tu}, \citenamefont {Zhou}, \citenamefont {Xu}, \citenamefont {Audi}, \citenamefont {Chen}, \citenamefont {Chen}, \citenamefont {Fu}, \citenamefont {Ge}, \citenamefont {Huang}, \citenamefont {Litvinov}, \citenamefont {Liu}, \citenamefont {Lam}, \citenamefont {Ma}, \citenamefont {Mao}, \citenamefont {Ozawa}, \citenamefont {Sun}, \citenamefont {Sun}, \citenamefont {Uesaka}, \citenamefont {Xiao}, \citenamefont {Xing}, \citenamefont {Yamaguchi}, \citenamefont {Yamaguchi}, \citenamefont {Yan}, \citenamefont {Zeng}, \citenamefont {Zhao}, \citenamefont {Zhao}, \citenamefont {Zhang},\ and\ \citenamefont {Zhan}}]{Xu2019}%
  \BibitemOpen
  \bibfield  {author} {\bibinfo {author} {\bibfnamefont {X.}~\bibnamefont {Xu}}, \bibinfo {author} {\bibfnamefont {M.}~\bibnamefont {Wang}}, \bibinfo {author} {\bibfnamefont {K.}~\bibnamefont {Blaum}}, \bibinfo {author} {\bibfnamefont {J.~D.}\ \bibnamefont {Holt}}, \bibinfo {author} {\bibfnamefont {Y.~A.}\ \bibnamefont {Litvinov}}, \bibinfo {author} {\bibfnamefont {A.}~\bibnamefont {Schwenk}}, \bibinfo {author} {\bibfnamefont {J.}~\bibnamefont {Simonis}}, \bibinfo {author} {\bibfnamefont {S.~R.}\ \bibnamefont {Stroberg}}, \bibinfo {author} {\bibfnamefont {Y.~H.}\ \bibnamefont {Zhang}}, \bibinfo {author} {\bibfnamefont {H.~S.}\ \bibnamefont {Xu}}, \bibinfo {author} {\bibfnamefont {P.}~\bibnamefont {Shuai}}, \bibinfo {author} {\bibfnamefont {X.~L.}\ \bibnamefont {Tu}}, \bibinfo {author} {\bibfnamefont {X.~H.}\ \bibnamefont {Zhou}}, \bibinfo {author} {\bibfnamefont {F.~R.}\ \bibnamefont {Xu}}, \bibinfo {author} {\bibfnamefont {G.}~\bibnamefont {Audi}}, \bibinfo {author} {\bibfnamefont {R.~J.}\ \bibnamefont
  {Chen}}, \bibinfo {author} {\bibfnamefont {X.~C.}\ \bibnamefont {Chen}}, \bibinfo {author} {\bibfnamefont {C.~Y.}\ \bibnamefont {Fu}}, \bibinfo {author} {\bibfnamefont {Z.}~\bibnamefont {Ge}}, \bibinfo {author} {\bibfnamefont {W.~J.}\ \bibnamefont {Huang}}, \bibinfo {author} {\bibfnamefont {S.}~\bibnamefont {Litvinov}}, \bibinfo {author} {\bibfnamefont {D.~W.}\ \bibnamefont {Liu}}, \bibinfo {author} {\bibfnamefont {Y.~H.}\ \bibnamefont {Lam}}, \bibinfo {author} {\bibfnamefont {X.~W.}\ \bibnamefont {Ma}}, \bibinfo {author} {\bibfnamefont {R.~S.}\ \bibnamefont {Mao}}, \bibinfo {author} {\bibfnamefont {A.}~\bibnamefont {Ozawa}}, \bibinfo {author} {\bibfnamefont {B.~H.}\ \bibnamefont {Sun}}, \bibinfo {author} {\bibfnamefont {Y.}~\bibnamefont {Sun}}, \bibinfo {author} {\bibfnamefont {T.}~\bibnamefont {Uesaka}}, \bibinfo {author} {\bibfnamefont {G.~Q.}\ \bibnamefont {Xiao}}, \bibinfo {author} {\bibfnamefont {Y.~M.}\ \bibnamefont {Xing}}, \bibinfo {author} {\bibfnamefont {T.}~\bibnamefont {Yamaguchi}}, \bibinfo
  {author} {\bibfnamefont {Y.}~\bibnamefont {Yamaguchi}}, \bibinfo {author} {\bibfnamefont {X.~L.}\ \bibnamefont {Yan}}, \bibinfo {author} {\bibfnamefont {Q.}~\bibnamefont {Zeng}}, \bibinfo {author} {\bibfnamefont {H.~W.}\ \bibnamefont {Zhao}}, \bibinfo {author} {\bibfnamefont {T.~C.}\ \bibnamefont {Zhao}}, \bibinfo {author} {\bibfnamefont {W.}~\bibnamefont {Zhang}},\ and\ \bibinfo {author} {\bibfnamefont {W.~L.}\ \bibnamefont {Zhan}},\ }\href {https://doi.org/10.1103/PhysRevC.99.064303} {\bibfield  {journal} {\bibinfo  {journal} {Phys. Rev. C}\ }\textbf {\bibinfo {volume} {99}},\ \bibinfo {pages} {064303} (\bibinfo {year} {2019})}\BibitemShut {NoStop}%
\bibitem [{\citenamefont {Leistenschneider}\ \emph {et~al.}(2018)\citenamefont {Leistenschneider}, \citenamefont {Reiter}, \citenamefont {Ayet San~Andr\'es}, \citenamefont {Kootte}, \citenamefont {Holt}, \citenamefont {Navr\'atil}, \citenamefont {Babcock}, \citenamefont {Barbieri}, \citenamefont {Barquest}, \citenamefont {Bergmann}, \citenamefont {Bollig}, \citenamefont {Brunner}, \citenamefont {Dunling}, \citenamefont {Finlay}, \citenamefont {Geissel}, \citenamefont {Graham}, \citenamefont {Greiner}, \citenamefont {Hergert}, \citenamefont {Hornung}, \citenamefont {Jesch}, \citenamefont {Klawitter}, \citenamefont {Lan}, \citenamefont {Lascar}, \citenamefont {Leach}, \citenamefont {Lippert}, \citenamefont {McKay}, \citenamefont {Paul}, \citenamefont {Schwenk}, \citenamefont {Short}, \citenamefont {Simonis}, \citenamefont {Som\`a}, \citenamefont {Steinbr\"ugge}, \citenamefont {Stroberg}, \citenamefont {Thompson}, \citenamefont {Wieser}, \citenamefont {Will}, \citenamefont {Yavor}, \citenamefont {Andreoiu},
  \citenamefont {Dickel}, \citenamefont {Dillmann}, \citenamefont {Gwinner}, \citenamefont {Pla\ss{}}, \citenamefont {Scheidenberger}, \citenamefont {Kwiatkowski},\ and\ \citenamefont {Dilling}}]{Leistenschneider2018}%
  \BibitemOpen
  \bibfield  {author} {\bibinfo {author} {\bibfnamefont {E.}~\bibnamefont {Leistenschneider}}, \bibinfo {author} {\bibfnamefont {M.~P.}\ \bibnamefont {Reiter}}, \bibinfo {author} {\bibfnamefont {S.}~\bibnamefont {Ayet San~Andr\'es}}, \bibinfo {author} {\bibfnamefont {B.}~\bibnamefont {Kootte}}, \bibinfo {author} {\bibfnamefont {J.~D.}\ \bibnamefont {Holt}}, \bibinfo {author} {\bibfnamefont {P.}~\bibnamefont {Navr\'atil}}, \bibinfo {author} {\bibfnamefont {C.}~\bibnamefont {Babcock}}, \bibinfo {author} {\bibfnamefont {C.}~\bibnamefont {Barbieri}}, \bibinfo {author} {\bibfnamefont {B.~R.}\ \bibnamefont {Barquest}}, \bibinfo {author} {\bibfnamefont {J.}~\bibnamefont {Bergmann}}, \bibinfo {author} {\bibfnamefont {J.}~\bibnamefont {Bollig}}, \bibinfo {author} {\bibfnamefont {T.}~\bibnamefont {Brunner}}, \bibinfo {author} {\bibfnamefont {E.}~\bibnamefont {Dunling}}, \bibinfo {author} {\bibfnamefont {A.}~\bibnamefont {Finlay}}, \bibinfo {author} {\bibfnamefont {H.}~\bibnamefont {Geissel}}, \bibinfo {author}
  {\bibfnamefont {L.}~\bibnamefont {Graham}}, \bibinfo {author} {\bibfnamefont {F.}~\bibnamefont {Greiner}}, \bibinfo {author} {\bibfnamefont {H.}~\bibnamefont {Hergert}}, \bibinfo {author} {\bibfnamefont {C.}~\bibnamefont {Hornung}}, \bibinfo {author} {\bibfnamefont {C.}~\bibnamefont {Jesch}}, \bibinfo {author} {\bibfnamefont {R.}~\bibnamefont {Klawitter}}, \bibinfo {author} {\bibfnamefont {Y.}~\bibnamefont {Lan}}, \bibinfo {author} {\bibfnamefont {D.}~\bibnamefont {Lascar}}, \bibinfo {author} {\bibfnamefont {K.~G.}\ \bibnamefont {Leach}}, \bibinfo {author} {\bibfnamefont {W.}~\bibnamefont {Lippert}}, \bibinfo {author} {\bibfnamefont {J.~E.}\ \bibnamefont {McKay}}, \bibinfo {author} {\bibfnamefont {S.~F.}\ \bibnamefont {Paul}}, \bibinfo {author} {\bibfnamefont {A.}~\bibnamefont {Schwenk}}, \bibinfo {author} {\bibfnamefont {D.}~\bibnamefont {Short}}, \bibinfo {author} {\bibfnamefont {J.}~\bibnamefont {Simonis}}, \bibinfo {author} {\bibfnamefont {V.}~\bibnamefont {Som\`a}}, \bibinfo {author} {\bibfnamefont
  {R.}~\bibnamefont {Steinbr\"ugge}}, \bibinfo {author} {\bibfnamefont {S.~R.}\ \bibnamefont {Stroberg}}, \bibinfo {author} {\bibfnamefont {R.}~\bibnamefont {Thompson}}, \bibinfo {author} {\bibfnamefont {M.~E.}\ \bibnamefont {Wieser}}, \bibinfo {author} {\bibfnamefont {C.}~\bibnamefont {Will}}, \bibinfo {author} {\bibfnamefont {M.}~\bibnamefont {Yavor}}, \bibinfo {author} {\bibfnamefont {C.}~\bibnamefont {Andreoiu}}, \bibinfo {author} {\bibfnamefont {T.}~\bibnamefont {Dickel}}, \bibinfo {author} {\bibfnamefont {I.}~\bibnamefont {Dillmann}}, \bibinfo {author} {\bibfnamefont {G.}~\bibnamefont {Gwinner}}, \bibinfo {author} {\bibfnamefont {W.~R.}\ \bibnamefont {Pla\ss{}}}, \bibinfo {author} {\bibfnamefont {C.}~\bibnamefont {Scheidenberger}}, \bibinfo {author} {\bibfnamefont {A.~A.}\ \bibnamefont {Kwiatkowski}},\ and\ \bibinfo {author} {\bibfnamefont {J.}~\bibnamefont {Dilling}},\ }\href {https://doi.org/10.1103/PhysRevLett.120.062503} {\bibfield  {journal} {\bibinfo  {journal} {Phys. Rev. Lett.}\ }\textbf
  {\bibinfo {volume} {120}},\ \bibinfo {pages} {062503} (\bibinfo {year} {2018})}\BibitemShut {NoStop}%
\bibitem [{\citenamefont {Porter}\ \emph {et~al.}(2022)\citenamefont {Porter}, \citenamefont {Dunling}, \citenamefont {Leistenschneider}, \citenamefont {Bergmann}, \citenamefont {Bollen}, \citenamefont {Dickel}, \citenamefont {Dietrich}, \citenamefont {Hamaker}, \citenamefont {Hockenbery}, \citenamefont {Izzo}, \citenamefont {Jacobs}, \citenamefont {Javaji}, \citenamefont {Kootte}, \citenamefont {Lan}, \citenamefont {Miskun}, \citenamefont {Mukul}, \citenamefont {Murb\"ock}, \citenamefont {Paul}, \citenamefont {Pla\ss{}}, \citenamefont {Puentes}, \citenamefont {Redshaw}, \citenamefont {Reiter}, \citenamefont {Ringle}, \citenamefont {Ringuette}, \citenamefont {Sandler}, \citenamefont {Scheidenberger}, \citenamefont {Silwal}, \citenamefont {Simpson}, \citenamefont {Sumithrarachchi}, \citenamefont {Teigelh\"ofer}, \citenamefont {Valverde}, \citenamefont {Weil}, \citenamefont {Yandow}, \citenamefont {Dilling},\ and\ \citenamefont {Kwiatkowski}}]{Porter2022}%
  \BibitemOpen
  \bibfield  {author} {\bibinfo {author} {\bibfnamefont {W.~S.}\ \bibnamefont {Porter}}, \bibinfo {author} {\bibfnamefont {E.}~\bibnamefont {Dunling}}, \bibinfo {author} {\bibfnamefont {E.}~\bibnamefont {Leistenschneider}}, \bibinfo {author} {\bibfnamefont {J.}~\bibnamefont {Bergmann}}, \bibinfo {author} {\bibfnamefont {G.}~\bibnamefont {Bollen}}, \bibinfo {author} {\bibfnamefont {T.}~\bibnamefont {Dickel}}, \bibinfo {author} {\bibfnamefont {K.~A.}\ \bibnamefont {Dietrich}}, \bibinfo {author} {\bibfnamefont {A.}~\bibnamefont {Hamaker}}, \bibinfo {author} {\bibfnamefont {Z.}~\bibnamefont {Hockenbery}}, \bibinfo {author} {\bibfnamefont {C.}~\bibnamefont {Izzo}}, \bibinfo {author} {\bibfnamefont {A.}~\bibnamefont {Jacobs}}, \bibinfo {author} {\bibfnamefont {A.}~\bibnamefont {Javaji}}, \bibinfo {author} {\bibfnamefont {B.}~\bibnamefont {Kootte}}, \bibinfo {author} {\bibfnamefont {Y.}~\bibnamefont {Lan}}, \bibinfo {author} {\bibfnamefont {I.}~\bibnamefont {Miskun}}, \bibinfo {author} {\bibfnamefont
  {I.}~\bibnamefont {Mukul}}, \bibinfo {author} {\bibfnamefont {T.}~\bibnamefont {Murb\"ock}}, \bibinfo {author} {\bibfnamefont {S.~F.}\ \bibnamefont {Paul}}, \bibinfo {author} {\bibfnamefont {W.~R.}\ \bibnamefont {Pla\ss{}}}, \bibinfo {author} {\bibfnamefont {D.}~\bibnamefont {Puentes}}, \bibinfo {author} {\bibfnamefont {M.}~\bibnamefont {Redshaw}}, \bibinfo {author} {\bibfnamefont {M.~P.}\ \bibnamefont {Reiter}}, \bibinfo {author} {\bibfnamefont {R.}~\bibnamefont {Ringle}}, \bibinfo {author} {\bibfnamefont {J.}~\bibnamefont {Ringuette}}, \bibinfo {author} {\bibfnamefont {R.}~\bibnamefont {Sandler}}, \bibinfo {author} {\bibfnamefont {C.}~\bibnamefont {Scheidenberger}}, \bibinfo {author} {\bibfnamefont {R.}~\bibnamefont {Silwal}}, \bibinfo {author} {\bibfnamefont {R.}~\bibnamefont {Simpson}}, \bibinfo {author} {\bibfnamefont {C.~S.}\ \bibnamefont {Sumithrarachchi}}, \bibinfo {author} {\bibfnamefont {A.}~\bibnamefont {Teigelh\"ofer}}, \bibinfo {author} {\bibfnamefont {A.~A.}\ \bibnamefont {Valverde}}, \bibinfo
  {author} {\bibfnamefont {R.}~\bibnamefont {Weil}}, \bibinfo {author} {\bibfnamefont {I.~T.}\ \bibnamefont {Yandow}}, \bibinfo {author} {\bibfnamefont {J.}~\bibnamefont {Dilling}},\ and\ \bibinfo {author} {\bibfnamefont {A.~A.}\ \bibnamefont {Kwiatkowski}},\ }\href {https://doi.org/10.1103/PhysRevC.106.024312} {\bibfield  {journal} {\bibinfo  {journal} {Phys. Rev. C}\ }\textbf {\bibinfo {volume} {106}},\ \bibinfo {pages} {024312} (\bibinfo {year} {2022})}\BibitemShut {NoStop}%
\bibitem [{\citenamefont {Steppenbeck}\ \emph {et~al.}(2013)\citenamefont {Steppenbeck}, \citenamefont {Takeuchi}, \citenamefont {Aoi}, \citenamefont {Doornenbal}, \citenamefont {Matsushita}, \citenamefont {Wang}, \citenamefont {Baba}, \citenamefont {Fukuda}, \citenamefont {Go}, \citenamefont {Honma}, \citenamefont {Lee}, \citenamefont {Matsui}, \citenamefont {Michimasa}, \citenamefont {Motobayashi}, \citenamefont {Nishimura}, \citenamefont {Otsuka}, \citenamefont {Sakurai}, \citenamefont {Shiga}, \citenamefont {Söderström}, \citenamefont {Sumikama}, \citenamefont {Suzuki}, \citenamefont {Taniuchi}, \citenamefont {Utsuno}, \citenamefont {Valiente-Dobón},\ and\ \citenamefont {Yoneda}}]{Steppenback2013}%
  \BibitemOpen
  \bibfield  {author} {\bibinfo {author} {\bibfnamefont {D.}~\bibnamefont {Steppenbeck}}, \bibinfo {author} {\bibfnamefont {S.}~\bibnamefont {Takeuchi}}, \bibinfo {author} {\bibfnamefont {N.}~\bibnamefont {Aoi}}, \bibinfo {author} {\bibfnamefont {P.}~\bibnamefont {Doornenbal}}, \bibinfo {author} {\bibfnamefont {M.}~\bibnamefont {Matsushita}}, \bibinfo {author} {\bibfnamefont {H.}~\bibnamefont {Wang}}, \bibinfo {author} {\bibfnamefont {H.}~\bibnamefont {Baba}}, \bibinfo {author} {\bibfnamefont {N.}~\bibnamefont {Fukuda}}, \bibinfo {author} {\bibfnamefont {S.}~\bibnamefont {Go}}, \bibinfo {author} {\bibfnamefont {M.}~\bibnamefont {Honma}}, \bibinfo {author} {\bibfnamefont {J.}~\bibnamefont {Lee}}, \bibinfo {author} {\bibfnamefont {K.}~\bibnamefont {Matsui}}, \bibinfo {author} {\bibfnamefont {S.}~\bibnamefont {Michimasa}}, \bibinfo {author} {\bibfnamefont {T.}~\bibnamefont {Motobayashi}}, \bibinfo {author} {\bibfnamefont {D.}~\bibnamefont {Nishimura}}, \bibinfo {author} {\bibfnamefont {T.}~\bibnamefont
  {Otsuka}}, \bibinfo {author} {\bibfnamefont {H.}~\bibnamefont {Sakurai}}, \bibinfo {author} {\bibfnamefont {Y.}~\bibnamefont {Shiga}}, \bibinfo {author} {\bibfnamefont {P.-A.}\ \bibnamefont {Söderström}}, \bibinfo {author} {\bibfnamefont {T.}~\bibnamefont {Sumikama}}, \bibinfo {author} {\bibfnamefont {H.}~\bibnamefont {Suzuki}}, \bibinfo {author} {\bibfnamefont {R.}~\bibnamefont {Taniuchi}}, \bibinfo {author} {\bibfnamefont {Y.}~\bibnamefont {Utsuno}}, \bibinfo {author} {\bibfnamefont {J.~J.}\ \bibnamefont {Valiente-Dobón}},\ and\ \bibinfo {author} {\bibfnamefont {K.}~\bibnamefont {Yoneda}},\ }\href {https://doi.org/https://doi.org/10.1038/nature12522} {\bibfield  {journal} {\bibinfo  {journal} {Nature}\ }\textbf {\bibinfo {volume} {502}},\ \bibinfo {pages} {207} (\bibinfo {year} {2013})}\BibitemShut {NoStop}%
\bibitem [{\citenamefont {Michimasa}\ \emph {et~al.}(2018)\citenamefont {Michimasa}, \citenamefont {Kobayashi}, \citenamefont {Kiyokawa}, \citenamefont {Ota}, \citenamefont {Ahn}, \citenamefont {Baba}, \citenamefont {Berg}, \citenamefont {Dozono}, \citenamefont {Fukuda}, \citenamefont {Furuno}, \citenamefont {Ideguchi}, \citenamefont {Inabe}, \citenamefont {Kawabata}, \citenamefont {Kawase}, \citenamefont {Kisamori}, \citenamefont {Kobayashi}, \citenamefont {Kubo}, \citenamefont {Kubota}, \citenamefont {Lee}, \citenamefont {Matsushita}, \citenamefont {Miya}, \citenamefont {Mizukami}, \citenamefont {Nagakura}, \citenamefont {Nishimura}, \citenamefont {Oikawa}, \citenamefont {Sakai}, \citenamefont {Shimizu}, \citenamefont {Stolz}, \citenamefont {Suzuki}, \citenamefont {Takaki}, \citenamefont {Takeda}, \citenamefont {Takeuchi}, \citenamefont {Tokieda}, \citenamefont {Uesaka}, \citenamefont {Yako}, \citenamefont {Yamaguchi}, \citenamefont {Yanagisawa}, \citenamefont {Yokoyama}, \citenamefont {Yoshida},\ and\
  \citenamefont {Shimoura}}]{Michimasa2018}%
  \BibitemOpen
  \bibfield  {author} {\bibinfo {author} {\bibfnamefont {S.}~\bibnamefont {Michimasa}}, \bibinfo {author} {\bibfnamefont {M.}~\bibnamefont {Kobayashi}}, \bibinfo {author} {\bibfnamefont {Y.}~\bibnamefont {Kiyokawa}}, \bibinfo {author} {\bibfnamefont {S.}~\bibnamefont {Ota}}, \bibinfo {author} {\bibfnamefont {D.~S.}\ \bibnamefont {Ahn}}, \bibinfo {author} {\bibfnamefont {H.}~\bibnamefont {Baba}}, \bibinfo {author} {\bibfnamefont {G.~P.~A.}\ \bibnamefont {Berg}}, \bibinfo {author} {\bibfnamefont {M.}~\bibnamefont {Dozono}}, \bibinfo {author} {\bibfnamefont {N.}~\bibnamefont {Fukuda}}, \bibinfo {author} {\bibfnamefont {T.}~\bibnamefont {Furuno}}, \bibinfo {author} {\bibfnamefont {E.}~\bibnamefont {Ideguchi}}, \bibinfo {author} {\bibfnamefont {N.}~\bibnamefont {Inabe}}, \bibinfo {author} {\bibfnamefont {T.}~\bibnamefont {Kawabata}}, \bibinfo {author} {\bibfnamefont {S.}~\bibnamefont {Kawase}}, \bibinfo {author} {\bibfnamefont {K.}~\bibnamefont {Kisamori}}, \bibinfo {author} {\bibfnamefont {K.}~\bibnamefont
  {Kobayashi}}, \bibinfo {author} {\bibfnamefont {T.}~\bibnamefont {Kubo}}, \bibinfo {author} {\bibfnamefont {Y.}~\bibnamefont {Kubota}}, \bibinfo {author} {\bibfnamefont {C.~S.}\ \bibnamefont {Lee}}, \bibinfo {author} {\bibfnamefont {M.}~\bibnamefont {Matsushita}}, \bibinfo {author} {\bibfnamefont {H.}~\bibnamefont {Miya}}, \bibinfo {author} {\bibfnamefont {A.}~\bibnamefont {Mizukami}}, \bibinfo {author} {\bibfnamefont {H.}~\bibnamefont {Nagakura}}, \bibinfo {author} {\bibfnamefont {D.}~\bibnamefont {Nishimura}}, \bibinfo {author} {\bibfnamefont {H.}~\bibnamefont {Oikawa}}, \bibinfo {author} {\bibfnamefont {H.}~\bibnamefont {Sakai}}, \bibinfo {author} {\bibfnamefont {Y.}~\bibnamefont {Shimizu}}, \bibinfo {author} {\bibfnamefont {A.}~\bibnamefont {Stolz}}, \bibinfo {author} {\bibfnamefont {H.}~\bibnamefont {Suzuki}}, \bibinfo {author} {\bibfnamefont {M.}~\bibnamefont {Takaki}}, \bibinfo {author} {\bibfnamefont {H.}~\bibnamefont {Takeda}}, \bibinfo {author} {\bibfnamefont {S.}~\bibnamefont {Takeuchi}},
  \bibinfo {author} {\bibfnamefont {H.}~\bibnamefont {Tokieda}}, \bibinfo {author} {\bibfnamefont {T.}~\bibnamefont {Uesaka}}, \bibinfo {author} {\bibfnamefont {K.}~\bibnamefont {Yako}}, \bibinfo {author} {\bibfnamefont {Y.}~\bibnamefont {Yamaguchi}}, \bibinfo {author} {\bibfnamefont {Y.}~\bibnamefont {Yanagisawa}}, \bibinfo {author} {\bibfnamefont {R.}~\bibnamefont {Yokoyama}}, \bibinfo {author} {\bibfnamefont {K.}~\bibnamefont {Yoshida}},\ and\ \bibinfo {author} {\bibfnamefont {S.}~\bibnamefont {Shimoura}},\ }\href {https://doi.org/10.1103/PhysRevLett.121.022506} {\bibfield  {journal} {\bibinfo  {journal} {Phys. Rev. Lett.}\ }\textbf {\bibinfo {volume} {121}},\ \bibinfo {pages} {022506} (\bibinfo {year} {2018})}\BibitemShut {NoStop}%
\bibitem [{\citenamefont {Liu}\ \emph {et~al.}(2019)\citenamefont {Liu}, \citenamefont {Obertelli}, \citenamefont {Doornenbal}, \citenamefont {Bertulani}, \citenamefont {Hagen}, \citenamefont {Holt}, \citenamefont {Jansen}, \citenamefont {Morris}, \citenamefont {Schwenk}, \citenamefont {Stroberg}, \citenamefont {Achouri}, \citenamefont {Baba}, \citenamefont {Browne}, \citenamefont {Calvet}, \citenamefont {Ch\^ateau}, \citenamefont {Chen}, \citenamefont {Chiga}, \citenamefont {Corsi}, \citenamefont {Cort\'es}, \citenamefont {Delbart}, \citenamefont {Gheller}, \citenamefont {Giganon}, \citenamefont {Gillibert}, \citenamefont {Hilaire}, \citenamefont {Isobe}, \citenamefont {Kobayashi}, \citenamefont {Kubota}, \citenamefont {Lapoux}, \citenamefont {Motobayashi}, \citenamefont {Murray}, \citenamefont {Otsu}, \citenamefont {Panin}, \citenamefont {Paul}, \citenamefont {Rodriguez}, \citenamefont {Sakurai}, \citenamefont {Sasano}, \citenamefont {Steppenbeck}, \citenamefont {Stuhl}, \citenamefont {Sun}, \citenamefont
  {Togano}, \citenamefont {Uesaka}, \citenamefont {Wimmer}, \citenamefont {Yoneda}, \citenamefont {Aktas}, \citenamefont {Aumann}, \citenamefont {Chung}, \citenamefont {Flavigny}, \citenamefont {Franchoo}, \citenamefont {Ga\ifmmode \check{s}\else \v{s}\fi{}pari\ifmmode~\acute{c}\else \'{c}\fi{}}, \citenamefont {Gerst}, \citenamefont {Gibelin}, \citenamefont {Hahn}, \citenamefont {Kim}, \citenamefont {Koiwai}, \citenamefont {Kondo}, \citenamefont {Koseoglou}, \citenamefont {Lee}, \citenamefont {Lehr}, \citenamefont {Linh}, \citenamefont {Lokotko}, \citenamefont {MacCormick}, \citenamefont {Moschner}, \citenamefont {Nakamura}, \citenamefont {Park}, \citenamefont {Rossi}, \citenamefont {Sahin}, \citenamefont {Sohler}, \citenamefont {S\"oderstr\"om}, \citenamefont {Takeuchi}, \citenamefont {T\"ornqvist}, \citenamefont {Vaquero}, \citenamefont {Wagner}, \citenamefont {Wang}, \citenamefont {Werner}, \citenamefont {Xu}, \citenamefont {Yamada}, \citenamefont {Yan}, \citenamefont {Yang}, \citenamefont {Yasuda},\ and\
  \citenamefont {Zanetti}}]{Liu2019}%
  \BibitemOpen
  \bibfield  {author} {\bibinfo {author} {\bibfnamefont {H.~N.}\ \bibnamefont {Liu}}, \bibinfo {author} {\bibfnamefont {A.}~\bibnamefont {Obertelli}}, \bibinfo {author} {\bibfnamefont {P.}~\bibnamefont {Doornenbal}}, \bibinfo {author} {\bibfnamefont {C.~A.}\ \bibnamefont {Bertulani}}, \bibinfo {author} {\bibfnamefont {G.}~\bibnamefont {Hagen}}, \bibinfo {author} {\bibfnamefont {J.~D.}\ \bibnamefont {Holt}}, \bibinfo {author} {\bibfnamefont {G.~R.}\ \bibnamefont {Jansen}}, \bibinfo {author} {\bibfnamefont {T.~D.}\ \bibnamefont {Morris}}, \bibinfo {author} {\bibfnamefont {A.}~\bibnamefont {Schwenk}}, \bibinfo {author} {\bibfnamefont {R.}~\bibnamefont {Stroberg}}, \bibinfo {author} {\bibfnamefont {N.}~\bibnamefont {Achouri}}, \bibinfo {author} {\bibfnamefont {H.}~\bibnamefont {Baba}}, \bibinfo {author} {\bibfnamefont {F.}~\bibnamefont {Browne}}, \bibinfo {author} {\bibfnamefont {D.}~\bibnamefont {Calvet}}, \bibinfo {author} {\bibfnamefont {F.}~\bibnamefont {Ch\^ateau}}, \bibinfo {author} {\bibfnamefont
  {S.}~\bibnamefont {Chen}}, \bibinfo {author} {\bibfnamefont {N.}~\bibnamefont {Chiga}}, \bibinfo {author} {\bibfnamefont {A.}~\bibnamefont {Corsi}}, \bibinfo {author} {\bibfnamefont {M.~L.}\ \bibnamefont {Cort\'es}}, \bibinfo {author} {\bibfnamefont {A.}~\bibnamefont {Delbart}}, \bibinfo {author} {\bibfnamefont {J.-M.}\ \bibnamefont {Gheller}}, \bibinfo {author} {\bibfnamefont {A.}~\bibnamefont {Giganon}}, \bibinfo {author} {\bibfnamefont {A.}~\bibnamefont {Gillibert}}, \bibinfo {author} {\bibfnamefont {C.}~\bibnamefont {Hilaire}}, \bibinfo {author} {\bibfnamefont {T.}~\bibnamefont {Isobe}}, \bibinfo {author} {\bibfnamefont {T.}~\bibnamefont {Kobayashi}}, \bibinfo {author} {\bibfnamefont {Y.}~\bibnamefont {Kubota}}, \bibinfo {author} {\bibfnamefont {V.}~\bibnamefont {Lapoux}}, \bibinfo {author} {\bibfnamefont {T.}~\bibnamefont {Motobayashi}}, \bibinfo {author} {\bibfnamefont {I.}~\bibnamefont {Murray}}, \bibinfo {author} {\bibfnamefont {H.}~\bibnamefont {Otsu}}, \bibinfo {author} {\bibfnamefont
  {V.}~\bibnamefont {Panin}}, \bibinfo {author} {\bibfnamefont {N.}~\bibnamefont {Paul}}, \bibinfo {author} {\bibfnamefont {W.}~\bibnamefont {Rodriguez}}, \bibinfo {author} {\bibfnamefont {H.}~\bibnamefont {Sakurai}}, \bibinfo {author} {\bibfnamefont {M.}~\bibnamefont {Sasano}}, \bibinfo {author} {\bibfnamefont {D.}~\bibnamefont {Steppenbeck}}, \bibinfo {author} {\bibfnamefont {L.}~\bibnamefont {Stuhl}}, \bibinfo {author} {\bibfnamefont {Y.~L.}\ \bibnamefont {Sun}}, \bibinfo {author} {\bibfnamefont {Y.}~\bibnamefont {Togano}}, \bibinfo {author} {\bibfnamefont {T.}~\bibnamefont {Uesaka}}, \bibinfo {author} {\bibfnamefont {K.}~\bibnamefont {Wimmer}}, \bibinfo {author} {\bibfnamefont {K.}~\bibnamefont {Yoneda}}, \bibinfo {author} {\bibfnamefont {O.}~\bibnamefont {Aktas}}, \bibinfo {author} {\bibfnamefont {T.}~\bibnamefont {Aumann}}, \bibinfo {author} {\bibfnamefont {L.~X.}\ \bibnamefont {Chung}}, \bibinfo {author} {\bibfnamefont {F.}~\bibnamefont {Flavigny}}, \bibinfo {author} {\bibfnamefont {S.}~\bibnamefont
  {Franchoo}}, \bibinfo {author} {\bibfnamefont {I.}~\bibnamefont {Ga\ifmmode \check{s}\else \v{s}\fi{}pari\ifmmode~\acute{c}\else \'{c}\fi{}}}, \bibinfo {author} {\bibfnamefont {R.-B.}\ \bibnamefont {Gerst}}, \bibinfo {author} {\bibfnamefont {J.}~\bibnamefont {Gibelin}}, \bibinfo {author} {\bibfnamefont {K.~I.}\ \bibnamefont {Hahn}}, \bibinfo {author} {\bibfnamefont {D.}~\bibnamefont {Kim}}, \bibinfo {author} {\bibfnamefont {T.}~\bibnamefont {Koiwai}}, \bibinfo {author} {\bibfnamefont {Y.}~\bibnamefont {Kondo}}, \bibinfo {author} {\bibfnamefont {P.}~\bibnamefont {Koseoglou}}, \bibinfo {author} {\bibfnamefont {J.}~\bibnamefont {Lee}}, \bibinfo {author} {\bibfnamefont {C.}~\bibnamefont {Lehr}}, \bibinfo {author} {\bibfnamefont {B.~D.}\ \bibnamefont {Linh}}, \bibinfo {author} {\bibfnamefont {T.}~\bibnamefont {Lokotko}}, \bibinfo {author} {\bibfnamefont {M.}~\bibnamefont {MacCormick}}, \bibinfo {author} {\bibfnamefont {K.}~\bibnamefont {Moschner}}, \bibinfo {author} {\bibfnamefont {T.}~\bibnamefont {Nakamura}},
  \bibinfo {author} {\bibfnamefont {S.~Y.}\ \bibnamefont {Park}}, \bibinfo {author} {\bibfnamefont {D.}~\bibnamefont {Rossi}}, \bibinfo {author} {\bibfnamefont {E.}~\bibnamefont {Sahin}}, \bibinfo {author} {\bibfnamefont {D.}~\bibnamefont {Sohler}}, \bibinfo {author} {\bibfnamefont {P.-A.}\ \bibnamefont {S\"oderstr\"om}}, \bibinfo {author} {\bibfnamefont {S.}~\bibnamefont {Takeuchi}}, \bibinfo {author} {\bibfnamefont {H.}~\bibnamefont {T\"ornqvist}}, \bibinfo {author} {\bibfnamefont {V.}~\bibnamefont {Vaquero}}, \bibinfo {author} {\bibfnamefont {V.}~\bibnamefont {Wagner}}, \bibinfo {author} {\bibfnamefont {S.}~\bibnamefont {Wang}}, \bibinfo {author} {\bibfnamefont {V.}~\bibnamefont {Werner}}, \bibinfo {author} {\bibfnamefont {X.}~\bibnamefont {Xu}}, \bibinfo {author} {\bibfnamefont {H.}~\bibnamefont {Yamada}}, \bibinfo {author} {\bibfnamefont {D.}~\bibnamefont {Yan}}, \bibinfo {author} {\bibfnamefont {Z.}~\bibnamefont {Yang}}, \bibinfo {author} {\bibfnamefont {M.}~\bibnamefont {Yasuda}},\ and\ \bibinfo
  {author} {\bibfnamefont {L.}~\bibnamefont {Zanetti}},\ }\href {https://doi.org/10.1103/PhysRevLett.122.072502} {\bibfield  {journal} {\bibinfo  {journal} {Phys. Rev. Lett.}\ }\textbf {\bibinfo {volume} {122}},\ \bibinfo {pages} {072502} (\bibinfo {year} {2019})}\BibitemShut {NoStop}%
\bibitem [{\citenamefont {Enciu}\ \emph {et~al.}(2022)\citenamefont {Enciu}, \citenamefont {Liu}, \citenamefont {Obertelli}, \citenamefont {Doornenbal}, \citenamefont {Nowacki}, \citenamefont {Ogata}, \citenamefont {Poves}, \citenamefont {Yoshida}, \citenamefont {Achouri}, \citenamefont {Baba}, \citenamefont {Browne}, \citenamefont {Calvet}, \citenamefont {Ch\^ateau}, \citenamefont {Chen}, \citenamefont {Chiga}, \citenamefont {Corsi}, \citenamefont {Cort\'es}, \citenamefont {Delbart}, \citenamefont {Gheller}, \citenamefont {Giganon}, \citenamefont {Gillibert}, \citenamefont {Hilaire}, \citenamefont {Isobe}, \citenamefont {Kobayashi}, \citenamefont {Kubota}, \citenamefont {Lapoux}, \citenamefont {Motobayashi}, \citenamefont {Murray}, \citenamefont {Otsu}, \citenamefont {Panin}, \citenamefont {Paul}, \citenamefont {Rodriguez}, \citenamefont {Sakurai}, \citenamefont {Sasano}, \citenamefont {Steppenbeck}, \citenamefont {Stuhl}, \citenamefont {Sun}, \citenamefont {Togano}, \citenamefont {Uesaka}, \citenamefont
  {Wimmer}, \citenamefont {Yoneda}, \citenamefont {Aktas}, \citenamefont {Aumann}, \citenamefont {Chung}, \citenamefont {Flavigny}, \citenamefont {Franchoo}, \citenamefont {Gasparic}, \citenamefont {Gerst}, \citenamefont {Gibelin}, \citenamefont {Hahn}, \citenamefont {Kim}, \citenamefont {Kondo}, \citenamefont {Koseoglou}, \citenamefont {Lee}, \citenamefont {Lehr}, \citenamefont {Li}, \citenamefont {Linh}, \citenamefont {Lokotko}, \citenamefont {MacCormick}, \citenamefont {Moschner}, \citenamefont {Nakamura}, \citenamefont {Park}, \citenamefont {Rossi}, \citenamefont {Sahin}, \citenamefont {S\"oderstr\"om}, \citenamefont {Sohler}, \citenamefont {Takeuchi}, \citenamefont {Toernqvist}, \citenamefont {Vaquero}, \citenamefont {Wagner}, \citenamefont {Wang}, \citenamefont {Werner}, \citenamefont {Xu}, \citenamefont {Yamada}, \citenamefont {Yan}, \citenamefont {Yang}, \citenamefont {Yasuda},\ and\ \citenamefont {Zanetti}}]{Enciu2022}%
  \BibitemOpen
  \bibfield  {author} {\bibinfo {author} {\bibfnamefont {M.}~\bibnamefont {Enciu}}, \bibinfo {author} {\bibfnamefont {H.~N.}\ \bibnamefont {Liu}}, \bibinfo {author} {\bibfnamefont {A.}~\bibnamefont {Obertelli}}, \bibinfo {author} {\bibfnamefont {P.}~\bibnamefont {Doornenbal}}, \bibinfo {author} {\bibfnamefont {F.}~\bibnamefont {Nowacki}}, \bibinfo {author} {\bibfnamefont {K.}~\bibnamefont {Ogata}}, \bibinfo {author} {\bibfnamefont {A.}~\bibnamefont {Poves}}, \bibinfo {author} {\bibfnamefont {K.}~\bibnamefont {Yoshida}}, \bibinfo {author} {\bibfnamefont {N.~L.}\ \bibnamefont {Achouri}}, \bibinfo {author} {\bibfnamefont {H.}~\bibnamefont {Baba}}, \bibinfo {author} {\bibfnamefont {F.}~\bibnamefont {Browne}}, \bibinfo {author} {\bibfnamefont {D.}~\bibnamefont {Calvet}}, \bibinfo {author} {\bibfnamefont {F.}~\bibnamefont {Ch\^ateau}}, \bibinfo {author} {\bibfnamefont {S.}~\bibnamefont {Chen}}, \bibinfo {author} {\bibfnamefont {N.}~\bibnamefont {Chiga}}, \bibinfo {author} {\bibfnamefont {A.}~\bibnamefont {Corsi}},
  \bibinfo {author} {\bibfnamefont {M.~L.}\ \bibnamefont {Cort\'es}}, \bibinfo {author} {\bibfnamefont {A.}~\bibnamefont {Delbart}}, \bibinfo {author} {\bibfnamefont {J.-M.}\ \bibnamefont {Gheller}}, \bibinfo {author} {\bibfnamefont {A.}~\bibnamefont {Giganon}}, \bibinfo {author} {\bibfnamefont {A.}~\bibnamefont {Gillibert}}, \bibinfo {author} {\bibfnamefont {C.}~\bibnamefont {Hilaire}}, \bibinfo {author} {\bibfnamefont {T.}~\bibnamefont {Isobe}}, \bibinfo {author} {\bibfnamefont {T.}~\bibnamefont {Kobayashi}}, \bibinfo {author} {\bibfnamefont {Y.}~\bibnamefont {Kubota}}, \bibinfo {author} {\bibfnamefont {V.}~\bibnamefont {Lapoux}}, \bibinfo {author} {\bibfnamefont {T.}~\bibnamefont {Motobayashi}}, \bibinfo {author} {\bibfnamefont {I.}~\bibnamefont {Murray}}, \bibinfo {author} {\bibfnamefont {H.}~\bibnamefont {Otsu}}, \bibinfo {author} {\bibfnamefont {V.}~\bibnamefont {Panin}}, \bibinfo {author} {\bibfnamefont {N.}~\bibnamefont {Paul}}, \bibinfo {author} {\bibfnamefont {W.}~\bibnamefont {Rodriguez}}, \bibinfo
  {author} {\bibfnamefont {H.}~\bibnamefont {Sakurai}}, \bibinfo {author} {\bibfnamefont {M.}~\bibnamefont {Sasano}}, \bibinfo {author} {\bibfnamefont {D.}~\bibnamefont {Steppenbeck}}, \bibinfo {author} {\bibfnamefont {L.}~\bibnamefont {Stuhl}}, \bibinfo {author} {\bibfnamefont {Y.~L.}\ \bibnamefont {Sun}}, \bibinfo {author} {\bibfnamefont {Y.}~\bibnamefont {Togano}}, \bibinfo {author} {\bibfnamefont {T.}~\bibnamefont {Uesaka}}, \bibinfo {author} {\bibfnamefont {K.}~\bibnamefont {Wimmer}}, \bibinfo {author} {\bibfnamefont {K.}~\bibnamefont {Yoneda}}, \bibinfo {author} {\bibfnamefont {O.}~\bibnamefont {Aktas}}, \bibinfo {author} {\bibfnamefont {T.}~\bibnamefont {Aumann}}, \bibinfo {author} {\bibfnamefont {L.~X.}\ \bibnamefont {Chung}}, \bibinfo {author} {\bibfnamefont {F.}~\bibnamefont {Flavigny}}, \bibinfo {author} {\bibfnamefont {S.}~\bibnamefont {Franchoo}}, \bibinfo {author} {\bibfnamefont {I.}~\bibnamefont {Gasparic}}, \bibinfo {author} {\bibfnamefont {R.-B.}\ \bibnamefont {Gerst}}, \bibinfo {author}
  {\bibfnamefont {J.}~\bibnamefont {Gibelin}}, \bibinfo {author} {\bibfnamefont {K.~I.}\ \bibnamefont {Hahn}}, \bibinfo {author} {\bibfnamefont {D.}~\bibnamefont {Kim}}, \bibinfo {author} {\bibfnamefont {Y.}~\bibnamefont {Kondo}}, \bibinfo {author} {\bibfnamefont {P.}~\bibnamefont {Koseoglou}}, \bibinfo {author} {\bibfnamefont {J.}~\bibnamefont {Lee}}, \bibinfo {author} {\bibfnamefont {C.}~\bibnamefont {Lehr}}, \bibinfo {author} {\bibfnamefont {P.~J.}\ \bibnamefont {Li}}, \bibinfo {author} {\bibfnamefont {B.~D.}\ \bibnamefont {Linh}}, \bibinfo {author} {\bibfnamefont {T.}~\bibnamefont {Lokotko}}, \bibinfo {author} {\bibfnamefont {M.}~\bibnamefont {MacCormick}}, \bibinfo {author} {\bibfnamefont {K.}~\bibnamefont {Moschner}}, \bibinfo {author} {\bibfnamefont {T.}~\bibnamefont {Nakamura}}, \bibinfo {author} {\bibfnamefont {S.~Y.}\ \bibnamefont {Park}}, \bibinfo {author} {\bibfnamefont {D.}~\bibnamefont {Rossi}}, \bibinfo {author} {\bibfnamefont {E.}~\bibnamefont {Sahin}}, \bibinfo {author} {\bibfnamefont
  {P.-A.}\ \bibnamefont {S\"oderstr\"om}}, \bibinfo {author} {\bibfnamefont {D.}~\bibnamefont {Sohler}}, \bibinfo {author} {\bibfnamefont {S.}~\bibnamefont {Takeuchi}}, \bibinfo {author} {\bibfnamefont {H.}~\bibnamefont {Toernqvist}}, \bibinfo {author} {\bibfnamefont {V.}~\bibnamefont {Vaquero}}, \bibinfo {author} {\bibfnamefont {V.}~\bibnamefont {Wagner}}, \bibinfo {author} {\bibfnamefont {S.}~\bibnamefont {Wang}}, \bibinfo {author} {\bibfnamefont {V.}~\bibnamefont {Werner}}, \bibinfo {author} {\bibfnamefont {X.}~\bibnamefont {Xu}}, \bibinfo {author} {\bibfnamefont {H.}~\bibnamefont {Yamada}}, \bibinfo {author} {\bibfnamefont {D.}~\bibnamefont {Yan}}, \bibinfo {author} {\bibfnamefont {Z.}~\bibnamefont {Yang}}, \bibinfo {author} {\bibfnamefont {M.}~\bibnamefont {Yasuda}},\ and\ \bibinfo {author} {\bibfnamefont {L.}~\bibnamefont {Zanetti}},\ }\href {https://doi.org/10.1103/PhysRevLett.129.262501} {\bibfield  {journal} {\bibinfo  {journal} {Phys. Rev. Lett.}\ }\textbf {\bibinfo {volume} {129}},\ \bibinfo
  {pages} {262501} (\bibinfo {year} {2022})}\BibitemShut {NoStop}%
\bibitem [{\citenamefont {Li}\ \emph {et~al.}(2024)\citenamefont {Li}, \citenamefont {Lee}, \citenamefont {Doornenbal}, \citenamefont {Chen}, \citenamefont {Wang}, \citenamefont {Obertelli}, \citenamefont {Chazono}, \citenamefont {Holt}, \citenamefont {Hu}, \citenamefont {Ogata}, \citenamefont {Utsuno}, \citenamefont {Yoshida}, \citenamefont {Achouri}, \citenamefont {Baba}, \citenamefont {Browne}, \citenamefont {Calvet}, \citenamefont {Château}, \citenamefont {Chiga}, \citenamefont {Corsi}, \citenamefont {Cortés}, \citenamefont {Delbart}, \citenamefont {Gheller}, \citenamefont {Giganon}, \citenamefont {Gillibert}, \citenamefont {Hilaire}, \citenamefont {Isobe}, \citenamefont {Kobayashi}, \citenamefont {Kubota}, \citenamefont {Lapoux}, \citenamefont {Liu}, \citenamefont {Motobayashi}, \citenamefont {Murray}, \citenamefont {Otsu}, \citenamefont {Panin}, \citenamefont {Paul}, \citenamefont {Rodriguez}, \citenamefont {Sakurai}, \citenamefont {Sasano}, \citenamefont {Steppenbeck}, \citenamefont {Stuhl},
  \citenamefont {Sun}, \citenamefont {Togano}, \citenamefont {Uesaka}, \citenamefont {Wimmer}, \citenamefont {Yoneda}, \citenamefont {Aktas}, \citenamefont {Aumann}, \citenamefont {Boretzky}, \citenamefont {Caesar}, \citenamefont {Chung}, \citenamefont {Flavigny}, \citenamefont {Franchoo}, \citenamefont {Gasparic}, \citenamefont {Gerst}, \citenamefont {Gibelin}, \citenamefont {Hahn}, \citenamefont {Kahlbow}, \citenamefont {Kim}, \citenamefont {Koiwai}, \citenamefont {Kondo}, \citenamefont {Körper}, \citenamefont {Koseoglou}, \citenamefont {Lehr}, \citenamefont {Linh}, \citenamefont {Lokotko}, \citenamefont {MacCormick}, \citenamefont {Miki}, \citenamefont {Moschner}, \citenamefont {Nakamura}, \citenamefont {Park}, \citenamefont {Rossi}, \citenamefont {Sahin}, \citenamefont {Schindler}, \citenamefont {Simon}, \citenamefont {Söderström}, \citenamefont {Sohler}, \citenamefont {Takeuchi}, \citenamefont {Toernqvist}, \citenamefont {Tscheuschner}, \citenamefont {Vaquero}, \citenamefont {Wagner}, \citenamefont
  {Werner}, \citenamefont {Xu}, \citenamefont {Yamada}, \citenamefont {Yan}, \citenamefont {Yang}, \citenamefont {Yasuda},\ and\ \citenamefont {Zanetti}}]{Li2024}%
  \BibitemOpen
  \bibfield  {author} {\bibinfo {author} {\bibfnamefont {P.~J.}\ \bibnamefont {Li}}, \bibinfo {author} {\bibfnamefont {J.}~\bibnamefont {Lee}}, \bibinfo {author} {\bibfnamefont {P.}~\bibnamefont {Doornenbal}}, \bibinfo {author} {\bibfnamefont {S.}~\bibnamefont {Chen}}, \bibinfo {author} {\bibfnamefont {S.}~\bibnamefont {Wang}}, \bibinfo {author} {\bibfnamefont {A.}~\bibnamefont {Obertelli}}, \bibinfo {author} {\bibfnamefont {Y.}~\bibnamefont {Chazono}}, \bibinfo {author} {\bibfnamefont {J.~D.}\ \bibnamefont {Holt}}, \bibinfo {author} {\bibfnamefont {B.~S.}\ \bibnamefont {Hu}}, \bibinfo {author} {\bibfnamefont {K.}~\bibnamefont {Ogata}}, \bibinfo {author} {\bibfnamefont {Y.}~\bibnamefont {Utsuno}}, \bibinfo {author} {\bibfnamefont {K.}~\bibnamefont {Yoshida}}, \bibinfo {author} {\bibfnamefont {N.~L.}\ \bibnamefont {Achouri}}, \bibinfo {author} {\bibfnamefont {H.}~\bibnamefont {Baba}}, \bibinfo {author} {\bibfnamefont {F.}~\bibnamefont {Browne}}, \bibinfo {author} {\bibfnamefont {D.}~\bibnamefont {Calvet}},
  \bibinfo {author} {\bibfnamefont {F.}~\bibnamefont {Château}}, \bibinfo {author} {\bibfnamefont {N.}~\bibnamefont {Chiga}}, \bibinfo {author} {\bibfnamefont {A.}~\bibnamefont {Corsi}}, \bibinfo {author} {\bibfnamefont {M.~L.}\ \bibnamefont {Cortés}}, \bibinfo {author} {\bibfnamefont {A.}~\bibnamefont {Delbart}}, \bibinfo {author} {\bibfnamefont {J.~M.}\ \bibnamefont {Gheller}}, \bibinfo {author} {\bibfnamefont {A.}~\bibnamefont {Giganon}}, \bibinfo {author} {\bibfnamefont {A.}~\bibnamefont {Gillibert}}, \bibinfo {author} {\bibfnamefont {C.}~\bibnamefont {Hilaire}}, \bibinfo {author} {\bibfnamefont {T.}~\bibnamefont {Isobe}}, \bibinfo {author} {\bibfnamefont {T.}~\bibnamefont {Kobayashi}}, \bibinfo {author} {\bibfnamefont {Y.}~\bibnamefont {Kubota}}, \bibinfo {author} {\bibfnamefont {V.}~\bibnamefont {Lapoux}}, \bibinfo {author} {\bibfnamefont {H.~N.}\ \bibnamefont {Liu}}, \bibinfo {author} {\bibfnamefont {T.}~\bibnamefont {Motobayashi}}, \bibinfo {author} {\bibfnamefont {I.}~\bibnamefont {Murray}},
  \bibinfo {author} {\bibfnamefont {H.}~\bibnamefont {Otsu}}, \bibinfo {author} {\bibfnamefont {V.}~\bibnamefont {Panin}}, \bibinfo {author} {\bibfnamefont {N.}~\bibnamefont {Paul}}, \bibinfo {author} {\bibfnamefont {W.}~\bibnamefont {Rodriguez}}, \bibinfo {author} {\bibfnamefont {H.}~\bibnamefont {Sakurai}}, \bibinfo {author} {\bibfnamefont {M.}~\bibnamefont {Sasano}}, \bibinfo {author} {\bibfnamefont {D.}~\bibnamefont {Steppenbeck}}, \bibinfo {author} {\bibfnamefont {L.}~\bibnamefont {Stuhl}}, \bibinfo {author} {\bibfnamefont {Y.~L.}\ \bibnamefont {Sun}}, \bibinfo {author} {\bibfnamefont {Y.}~\bibnamefont {Togano}}, \bibinfo {author} {\bibfnamefont {T.}~\bibnamefont {Uesaka}}, \bibinfo {author} {\bibfnamefont {K.}~\bibnamefont {Wimmer}}, \bibinfo {author} {\bibfnamefont {K.}~\bibnamefont {Yoneda}}, \bibinfo {author} {\bibfnamefont {O.}~\bibnamefont {Aktas}}, \bibinfo {author} {\bibfnamefont {T.}~\bibnamefont {Aumann}}, \bibinfo {author} {\bibfnamefont {K.}~\bibnamefont {Boretzky}}, \bibinfo {author}
  {\bibfnamefont {C.}~\bibnamefont {Caesar}}, \bibinfo {author} {\bibfnamefont {L.~X.}\ \bibnamefont {Chung}}, \bibinfo {author} {\bibfnamefont {F.}~\bibnamefont {Flavigny}}, \bibinfo {author} {\bibfnamefont {S.}~\bibnamefont {Franchoo}}, \bibinfo {author} {\bibfnamefont {I.}~\bibnamefont {Gasparic}}, \bibinfo {author} {\bibfnamefont {R.~B.}\ \bibnamefont {Gerst}}, \bibinfo {author} {\bibfnamefont {J.}~\bibnamefont {Gibelin}}, \bibinfo {author} {\bibfnamefont {K.~I.}\ \bibnamefont {Hahn}}, \bibinfo {author} {\bibfnamefont {J.}~\bibnamefont {Kahlbow}}, \bibinfo {author} {\bibfnamefont {D.}~\bibnamefont {Kim}}, \bibinfo {author} {\bibfnamefont {T.}~\bibnamefont {Koiwai}}, \bibinfo {author} {\bibfnamefont {Y.}~\bibnamefont {Kondo}}, \bibinfo {author} {\bibfnamefont {D.}~\bibnamefont {Körper}}, \bibinfo {author} {\bibfnamefont {P.}~\bibnamefont {Koseoglou}}, \bibinfo {author} {\bibfnamefont {C.}~\bibnamefont {Lehr}}, \bibinfo {author} {\bibfnamefont {B.~D.}\ \bibnamefont {Linh}}, \bibinfo {author} {\bibfnamefont
  {T.}~\bibnamefont {Lokotko}}, \bibinfo {author} {\bibfnamefont {M.}~\bibnamefont {MacCormick}}, \bibinfo {author} {\bibfnamefont {K.}~\bibnamefont {Miki}}, \bibinfo {author} {\bibfnamefont {K.}~\bibnamefont {Moschner}}, \bibinfo {author} {\bibfnamefont {T.}~\bibnamefont {Nakamura}}, \bibinfo {author} {\bibfnamefont {S.~Y.}\ \bibnamefont {Park}}, \bibinfo {author} {\bibfnamefont {D.}~\bibnamefont {Rossi}}, \bibinfo {author} {\bibfnamefont {E.}~\bibnamefont {Sahin}}, \bibinfo {author} {\bibfnamefont {F.}~\bibnamefont {Schindler}}, \bibinfo {author} {\bibfnamefont {H.}~\bibnamefont {Simon}}, \bibinfo {author} {\bibfnamefont {P.~A.}\ \bibnamefont {Söderström}}, \bibinfo {author} {\bibfnamefont {D.}~\bibnamefont {Sohler}}, \bibinfo {author} {\bibfnamefont {S.}~\bibnamefont {Takeuchi}}, \bibinfo {author} {\bibfnamefont {H.}~\bibnamefont {Toernqvist}}, \bibinfo {author} {\bibfnamefont {J.}~\bibnamefont {Tscheuschner}}, \bibinfo {author} {\bibfnamefont {V.}~\bibnamefont {Vaquero}}, \bibinfo {author}
  {\bibfnamefont {V.}~\bibnamefont {Wagner}}, \bibinfo {author} {\bibfnamefont {V.}~\bibnamefont {Werner}}, \bibinfo {author} {\bibfnamefont {X.}~\bibnamefont {Xu}}, \bibinfo {author} {\bibfnamefont {H.}~\bibnamefont {Yamada}}, \bibinfo {author} {\bibfnamefont {D.}~\bibnamefont {Yan}}, \bibinfo {author} {\bibfnamefont {Z.}~\bibnamefont {Yang}}, \bibinfo {author} {\bibfnamefont {M.}~\bibnamefont {Yasuda}},\ and\ \bibinfo {author} {\bibfnamefont {L.}~\bibnamefont {Zanetti}},\ }\href {https://doi.org/https://doi.org/10.1016/j.physletb.2024.138828} {\bibfield  {journal} {\bibinfo  {journal} {Physics Letters B}\ }\textbf {\bibinfo {volume} {855}},\ \bibinfo {pages} {138828} (\bibinfo {year} {2024})}\BibitemShut {NoStop}%
\bibitem [{\citenamefont {Leistenschneider}\ \emph {et~al.}(2021)\citenamefont {Leistenschneider}, \citenamefont {Dunling}, \citenamefont {Bollen}, \citenamefont {Brown}, \citenamefont {Dilling}, \citenamefont {Hamaker}, \citenamefont {Holt}, \citenamefont {Jacobs}, \citenamefont {Kwiatkowski}, \citenamefont {Miyagi}, \citenamefont {Porter}, \citenamefont {Puentes}, \citenamefont {Redshaw}, \citenamefont {Reiter}, \citenamefont {Ringle}, \citenamefont {Sandler}, \citenamefont {Sumithrarachchi}, \citenamefont {Valverde},\ and\ \citenamefont {Yandow}}]{Leistenschneider2021}%
  \BibitemOpen
  \bibfield  {author} {\bibinfo {author} {\bibfnamefont {E.}~\bibnamefont {Leistenschneider}}, \bibinfo {author} {\bibfnamefont {E.}~\bibnamefont {Dunling}}, \bibinfo {author} {\bibfnamefont {G.}~\bibnamefont {Bollen}}, \bibinfo {author} {\bibfnamefont {B.~A.}\ \bibnamefont {Brown}}, \bibinfo {author} {\bibfnamefont {J.}~\bibnamefont {Dilling}}, \bibinfo {author} {\bibfnamefont {A.}~\bibnamefont {Hamaker}}, \bibinfo {author} {\bibfnamefont {J.~D.}\ \bibnamefont {Holt}}, \bibinfo {author} {\bibfnamefont {A.}~\bibnamefont {Jacobs}}, \bibinfo {author} {\bibfnamefont {A.~A.}\ \bibnamefont {Kwiatkowski}}, \bibinfo {author} {\bibfnamefont {T.}~\bibnamefont {Miyagi}}, \bibinfo {author} {\bibfnamefont {W.~S.}\ \bibnamefont {Porter}}, \bibinfo {author} {\bibfnamefont {D.}~\bibnamefont {Puentes}}, \bibinfo {author} {\bibfnamefont {M.}~\bibnamefont {Redshaw}}, \bibinfo {author} {\bibfnamefont {M.~P.}\ \bibnamefont {Reiter}}, \bibinfo {author} {\bibfnamefont {R.}~\bibnamefont {Ringle}}, \bibinfo {author} {\bibfnamefont
  {R.}~\bibnamefont {Sandler}}, \bibinfo {author} {\bibfnamefont {C.~S.}\ \bibnamefont {Sumithrarachchi}}, \bibinfo {author} {\bibfnamefont {A.~A.}\ \bibnamefont {Valverde}},\ and\ \bibinfo {author} {\bibfnamefont {I.~T.}\ \bibnamefont {Yandow}} (\bibinfo {collaboration} {The LEBIT Collaboration and the TITAN Collaboration}),\ }\href {https://doi.org/10.1103/PhysRevLett.126.042501} {\bibfield  {journal} {\bibinfo  {journal} {Phys. Rev. Lett.}\ }\textbf {\bibinfo {volume} {126}},\ \bibinfo {pages} {042501} (\bibinfo {year} {2021})}\BibitemShut {NoStop}%
\bibitem [{\citenamefont {Steppenbeck}\ \emph {et~al.}(2017)\citenamefont {Steppenbeck}, \citenamefont {Takeuchi}, \citenamefont {Aoi}, \citenamefont {Doornenbal}, \citenamefont {Matsushita}, \citenamefont {Wang}, \citenamefont {Baba}, \citenamefont {Go}, \citenamefont {Holt}, \citenamefont {Lee}, \citenamefont {Matsui}, \citenamefont {Michimasa}, \citenamefont {Motobayashi}, \citenamefont {Nishimura}, \citenamefont {Otsuka}, \citenamefont {Sakurai}, \citenamefont {Shiga}, \citenamefont {S\"oderstr\"om}, \citenamefont {Stroberg}, \citenamefont {Sumikama}, \citenamefont {Taniuchi}, \citenamefont {Tostevin}, \citenamefont {Utsuno}, \citenamefont {Valiente-Dob\'on},\ and\ \citenamefont {Yoneda}}]{Steppenbeck2017}%
  \BibitemOpen
  \bibfield  {author} {\bibinfo {author} {\bibfnamefont {D.}~\bibnamefont {Steppenbeck}}, \bibinfo {author} {\bibfnamefont {S.}~\bibnamefont {Takeuchi}}, \bibinfo {author} {\bibfnamefont {N.}~\bibnamefont {Aoi}}, \bibinfo {author} {\bibfnamefont {P.}~\bibnamefont {Doornenbal}}, \bibinfo {author} {\bibfnamefont {M.}~\bibnamefont {Matsushita}}, \bibinfo {author} {\bibfnamefont {H.}~\bibnamefont {Wang}}, \bibinfo {author} {\bibfnamefont {H.}~\bibnamefont {Baba}}, \bibinfo {author} {\bibfnamefont {S.}~\bibnamefont {Go}}, \bibinfo {author} {\bibfnamefont {J.~D.}\ \bibnamefont {Holt}}, \bibinfo {author} {\bibfnamefont {J.}~\bibnamefont {Lee}}, \bibinfo {author} {\bibfnamefont {K.}~\bibnamefont {Matsui}}, \bibinfo {author} {\bibfnamefont {S.}~\bibnamefont {Michimasa}}, \bibinfo {author} {\bibfnamefont {T.}~\bibnamefont {Motobayashi}}, \bibinfo {author} {\bibfnamefont {D.}~\bibnamefont {Nishimura}}, \bibinfo {author} {\bibfnamefont {T.}~\bibnamefont {Otsuka}}, \bibinfo {author} {\bibfnamefont {H.}~\bibnamefont
  {Sakurai}}, \bibinfo {author} {\bibfnamefont {Y.}~\bibnamefont {Shiga}}, \bibinfo {author} {\bibfnamefont {P.-A.}\ \bibnamefont {S\"oderstr\"om}}, \bibinfo {author} {\bibfnamefont {S.~R.}\ \bibnamefont {Stroberg}}, \bibinfo {author} {\bibfnamefont {T.}~\bibnamefont {Sumikama}}, \bibinfo {author} {\bibfnamefont {R.}~\bibnamefont {Taniuchi}}, \bibinfo {author} {\bibfnamefont {J.~A.}\ \bibnamefont {Tostevin}}, \bibinfo {author} {\bibfnamefont {Y.}~\bibnamefont {Utsuno}}, \bibinfo {author} {\bibfnamefont {J.~J.}\ \bibnamefont {Valiente-Dob\'on}},\ and\ \bibinfo {author} {\bibfnamefont {K.}~\bibnamefont {Yoneda}},\ }\href {https://doi.org/10.1103/PhysRevC.96.064310} {\bibfield  {journal} {\bibinfo  {journal} {Phys. Rev. C}\ }\textbf {\bibinfo {volume} {96}},\ \bibinfo {pages} {064310} (\bibinfo {year} {2017})}\BibitemShut {NoStop}%
\bibitem [{\citenamefont {Dinca}\ \emph {et~al.}(2005)\citenamefont {Dinca}, \citenamefont {Janssens}, \citenamefont {Gade}, \citenamefont {Bazin}, \citenamefont {Broda}, \citenamefont {Brown}, \citenamefont {Campbell}, \citenamefont {Carpenter}, \citenamefont {Chowdhury}, \citenamefont {Cook}, \citenamefont {Deacon}, \citenamefont {Fornal}, \citenamefont {Freeman}, \citenamefont {Glasmacher}, \citenamefont {Honma}, \citenamefont {Kondev}, \citenamefont {Lecouey}, \citenamefont {Liddick}, \citenamefont {Mantica}, \citenamefont {Mueller}, \citenamefont {Olliver}, \citenamefont {Otsuka}, \citenamefont {Terry}, \citenamefont {Tomlin},\ and\ \citenamefont {Yoneda}}]{Dinca2005}%
  \BibitemOpen
  \bibfield  {author} {\bibinfo {author} {\bibfnamefont {D.-C.}\ \bibnamefont {Dinca}}, \bibinfo {author} {\bibfnamefont {R.~V.~F.}\ \bibnamefont {Janssens}}, \bibinfo {author} {\bibfnamefont {A.}~\bibnamefont {Gade}}, \bibinfo {author} {\bibfnamefont {D.}~\bibnamefont {Bazin}}, \bibinfo {author} {\bibfnamefont {R.}~\bibnamefont {Broda}}, \bibinfo {author} {\bibfnamefont {B.~A.}\ \bibnamefont {Brown}}, \bibinfo {author} {\bibfnamefont {C.~M.}\ \bibnamefont {Campbell}}, \bibinfo {author} {\bibfnamefont {M.~P.}\ \bibnamefont {Carpenter}}, \bibinfo {author} {\bibfnamefont {P.}~\bibnamefont {Chowdhury}}, \bibinfo {author} {\bibfnamefont {J.~M.}\ \bibnamefont {Cook}}, \bibinfo {author} {\bibfnamefont {A.~N.}\ \bibnamefont {Deacon}}, \bibinfo {author} {\bibfnamefont {B.}~\bibnamefont {Fornal}}, \bibinfo {author} {\bibfnamefont {S.~J.}\ \bibnamefont {Freeman}}, \bibinfo {author} {\bibfnamefont {T.}~\bibnamefont {Glasmacher}}, \bibinfo {author} {\bibfnamefont {M.}~\bibnamefont {Honma}}, \bibinfo {author}
  {\bibfnamefont {F.~G.}\ \bibnamefont {Kondev}}, \bibinfo {author} {\bibfnamefont {J.-L.}\ \bibnamefont {Lecouey}}, \bibinfo {author} {\bibfnamefont {S.~N.}\ \bibnamefont {Liddick}}, \bibinfo {author} {\bibfnamefont {P.~F.}\ \bibnamefont {Mantica}}, \bibinfo {author} {\bibfnamefont {W.~F.}\ \bibnamefont {Mueller}}, \bibinfo {author} {\bibfnamefont {H.}~\bibnamefont {Olliver}}, \bibinfo {author} {\bibfnamefont {T.}~\bibnamefont {Otsuka}}, \bibinfo {author} {\bibfnamefont {J.~R.}\ \bibnamefont {Terry}}, \bibinfo {author} {\bibfnamefont {B.~A.}\ \bibnamefont {Tomlin}},\ and\ \bibinfo {author} {\bibfnamefont {K.}~\bibnamefont {Yoneda}},\ }\href {https://doi.org/10.1103/PhysRevC.71.041302} {\bibfield  {journal} {\bibinfo  {journal} {Phys. Rev. C}\ }\textbf {\bibinfo {volume} {71}},\ \bibinfo {pages} {041302} (\bibinfo {year} {2005})}\BibitemShut {NoStop}%
\bibitem [{\citenamefont {Iimura}\ \emph {et~al.}(2023)\citenamefont {Iimura}, \citenamefont {Rosenbusch}, \citenamefont {Takamine}, \citenamefont {Tsunoda}, \citenamefont {Wada}, \citenamefont {Chen}, \citenamefont {Hou}, \citenamefont {Xian}, \citenamefont {Ishiyama}, \citenamefont {Yan}, \citenamefont {Schury}, \citenamefont {Crawford}, \citenamefont {Doornenbal}, \citenamefont {Hirayama}, \citenamefont {Ito}, \citenamefont {Kimura}, \citenamefont {Koiwai}, \citenamefont {Kojima}, \citenamefont {Koura}, \citenamefont {Lee}, \citenamefont {Liu}, \citenamefont {Michimasa}, \citenamefont {Miyatake}, \citenamefont {Moon}, \citenamefont {Naimi}, \citenamefont {Nishimura}, \citenamefont {Niwase}, \citenamefont {Odahara}, \citenamefont {Otsuka}, \citenamefont {Paschalis}, \citenamefont {Petri}, \citenamefont {Shimizu}, \citenamefont {Sonoda}, \citenamefont {Suzuki}, \citenamefont {Watanabe}, \citenamefont {Wimmer},\ and\ \citenamefont {Wollnik}}]{Iimura2023}%
  \BibitemOpen
  \bibfield  {author} {\bibinfo {author} {\bibfnamefont {S.}~\bibnamefont {Iimura}}, \bibinfo {author} {\bibfnamefont {M.}~\bibnamefont {Rosenbusch}}, \bibinfo {author} {\bibfnamefont {A.}~\bibnamefont {Takamine}}, \bibinfo {author} {\bibfnamefont {Y.}~\bibnamefont {Tsunoda}}, \bibinfo {author} {\bibfnamefont {M.}~\bibnamefont {Wada}}, \bibinfo {author} {\bibfnamefont {S.}~\bibnamefont {Chen}}, \bibinfo {author} {\bibfnamefont {D.~S.}\ \bibnamefont {Hou}}, \bibinfo {author} {\bibfnamefont {W.}~\bibnamefont {Xian}}, \bibinfo {author} {\bibfnamefont {H.}~\bibnamefont {Ishiyama}}, \bibinfo {author} {\bibfnamefont {S.}~\bibnamefont {Yan}}, \bibinfo {author} {\bibfnamefont {P.}~\bibnamefont {Schury}}, \bibinfo {author} {\bibfnamefont {H.}~\bibnamefont {Crawford}}, \bibinfo {author} {\bibfnamefont {P.}~\bibnamefont {Doornenbal}}, \bibinfo {author} {\bibfnamefont {Y.}~\bibnamefont {Hirayama}}, \bibinfo {author} {\bibfnamefont {Y.}~\bibnamefont {Ito}}, \bibinfo {author} {\bibfnamefont {S.}~\bibnamefont {Kimura}},
  \bibinfo {author} {\bibfnamefont {T.}~\bibnamefont {Koiwai}}, \bibinfo {author} {\bibfnamefont {T.~M.}\ \bibnamefont {Kojima}}, \bibinfo {author} {\bibfnamefont {H.}~\bibnamefont {Koura}}, \bibinfo {author} {\bibfnamefont {J.}~\bibnamefont {Lee}}, \bibinfo {author} {\bibfnamefont {J.}~\bibnamefont {Liu}}, \bibinfo {author} {\bibfnamefont {S.}~\bibnamefont {Michimasa}}, \bibinfo {author} {\bibfnamefont {H.}~\bibnamefont {Miyatake}}, \bibinfo {author} {\bibfnamefont {J.~Y.}\ \bibnamefont {Moon}}, \bibinfo {author} {\bibfnamefont {S.}~\bibnamefont {Naimi}}, \bibinfo {author} {\bibfnamefont {S.}~\bibnamefont {Nishimura}}, \bibinfo {author} {\bibfnamefont {T.}~\bibnamefont {Niwase}}, \bibinfo {author} {\bibfnamefont {A.}~\bibnamefont {Odahara}}, \bibinfo {author} {\bibfnamefont {T.}~\bibnamefont {Otsuka}}, \bibinfo {author} {\bibfnamefont {S.}~\bibnamefont {Paschalis}}, \bibinfo {author} {\bibfnamefont {M.}~\bibnamefont {Petri}}, \bibinfo {author} {\bibfnamefont {N.}~\bibnamefont {Shimizu}}, \bibinfo {author}
  {\bibfnamefont {T.}~\bibnamefont {Sonoda}}, \bibinfo {author} {\bibfnamefont {D.}~\bibnamefont {Suzuki}}, \bibinfo {author} {\bibfnamefont {Y.~X.}\ \bibnamefont {Watanabe}}, \bibinfo {author} {\bibfnamefont {K.}~\bibnamefont {Wimmer}},\ and\ \bibinfo {author} {\bibfnamefont {H.}~\bibnamefont {Wollnik}},\ }\href {https://doi.org/10.1103/PhysRevLett.130.012501} {\bibfield  {journal} {\bibinfo  {journal} {Phys. Rev. Lett.}\ }\textbf {\bibinfo {volume} {130}},\ \bibinfo {pages} {012501} (\bibinfo {year} {2023})}\BibitemShut {NoStop}%
\bibitem [{\citenamefont {Zhu}\ \emph {et~al.}(2006)\citenamefont {Zhu}, \citenamefont {Deacon}, \citenamefont {Freeman}, \citenamefont {Janssens}, \citenamefont {Fornal}, \citenamefont {Honma}, \citenamefont {Xu}, \citenamefont {Broda}, \citenamefont {Calderin}, \citenamefont {Carpenter}, \citenamefont {Chowdhury}, \citenamefont {Kondev}, \citenamefont {Kr\'olas}, \citenamefont {Lauritsen}, \citenamefont {Liddick}, \citenamefont {Lister}, \citenamefont {Mantica}, \citenamefont {Paw\l{}at}, \citenamefont {Seweryniak}, \citenamefont {Smith}, \citenamefont {Tabor}, \citenamefont {Tomlin}, \citenamefont {Varley},\ and\ \citenamefont {Wrzesi\ifmmode~\acute{n}\else \'{n}\fi{}ski}}]{Zhu2006}%
  \BibitemOpen
  \bibfield  {author} {\bibinfo {author} {\bibfnamefont {S.}~\bibnamefont {Zhu}}, \bibinfo {author} {\bibfnamefont {A.~N.}\ \bibnamefont {Deacon}}, \bibinfo {author} {\bibfnamefont {S.~J.}\ \bibnamefont {Freeman}}, \bibinfo {author} {\bibfnamefont {R.~V.~F.}\ \bibnamefont {Janssens}}, \bibinfo {author} {\bibfnamefont {B.}~\bibnamefont {Fornal}}, \bibinfo {author} {\bibfnamefont {M.}~\bibnamefont {Honma}}, \bibinfo {author} {\bibfnamefont {F.~R.}\ \bibnamefont {Xu}}, \bibinfo {author} {\bibfnamefont {R.}~\bibnamefont {Broda}}, \bibinfo {author} {\bibfnamefont {I.~R.}\ \bibnamefont {Calderin}}, \bibinfo {author} {\bibfnamefont {M.~P.}\ \bibnamefont {Carpenter}}, \bibinfo {author} {\bibfnamefont {P.}~\bibnamefont {Chowdhury}}, \bibinfo {author} {\bibfnamefont {F.~G.}\ \bibnamefont {Kondev}}, \bibinfo {author} {\bibfnamefont {W.}~\bibnamefont {Kr\'olas}}, \bibinfo {author} {\bibfnamefont {T.}~\bibnamefont {Lauritsen}}, \bibinfo {author} {\bibfnamefont {S.~N.}\ \bibnamefont {Liddick}}, \bibinfo {author}
  {\bibfnamefont {C.~J.}\ \bibnamefont {Lister}}, \bibinfo {author} {\bibfnamefont {P.~F.}\ \bibnamefont {Mantica}}, \bibinfo {author} {\bibfnamefont {T.}~\bibnamefont {Paw\l{}at}}, \bibinfo {author} {\bibfnamefont {D.}~\bibnamefont {Seweryniak}}, \bibinfo {author} {\bibfnamefont {J.~F.}\ \bibnamefont {Smith}}, \bibinfo {author} {\bibfnamefont {S.~L.}\ \bibnamefont {Tabor}}, \bibinfo {author} {\bibfnamefont {B.~E.}\ \bibnamefont {Tomlin}}, \bibinfo {author} {\bibfnamefont {B.~J.}\ \bibnamefont {Varley}},\ and\ \bibinfo {author} {\bibfnamefont {J.}~\bibnamefont {Wrzesi\ifmmode~\acute{n}\else \'{n}\fi{}ski}},\ }\href {https://doi.org/10.1103/PhysRevC.74.064315} {\bibfield  {journal} {\bibinfo  {journal} {Phys. Rev. C}\ }\textbf {\bibinfo {volume} {74}},\ \bibinfo {pages} {064315} (\bibinfo {year} {2006})}\BibitemShut {NoStop}%
\bibitem [{\citenamefont {Garcia~Ruiz}\ \emph {et~al.}(2016)\citenamefont {Garcia~Ruiz}, \citenamefont {Bissell}, \citenamefont {Blaum}, \citenamefont {Ekström}, \citenamefont {Frömmgen}, \citenamefont {Hagen}, \citenamefont {Hammen}, \citenamefont {Hebeler}, \citenamefont {Holt}, \citenamefont {Jansen}, \citenamefont {Kowalska}, \citenamefont {Kreim}, \citenamefont {Nazarewicz}, \citenamefont {Neugart}, \citenamefont {Neyens}, \citenamefont {Nörtershäuser}, \citenamefont {Papenbrock}, \citenamefont {Papuga}, \citenamefont {Schwenk}, \citenamefont {Simonis}, \citenamefont {Wendt},\ and\ \citenamefont {Yordanov}}]{GarciaRuiz2016}%
  \BibitemOpen
  \bibfield  {author} {\bibinfo {author} {\bibfnamefont {R.~F.}\ \bibnamefont {Garcia~Ruiz}}, \bibinfo {author} {\bibfnamefont {M.~L.}\ \bibnamefont {Bissell}}, \bibinfo {author} {\bibfnamefont {K.}~\bibnamefont {Blaum}}, \bibinfo {author} {\bibfnamefont {A.}~\bibnamefont {Ekström}}, \bibinfo {author} {\bibfnamefont {N.}~\bibnamefont {Frömmgen}}, \bibinfo {author} {\bibfnamefont {G.}~\bibnamefont {Hagen}}, \bibinfo {author} {\bibfnamefont {M.}~\bibnamefont {Hammen}}, \bibinfo {author} {\bibfnamefont {K.}~\bibnamefont {Hebeler}}, \bibinfo {author} {\bibfnamefont {J.~D.}\ \bibnamefont {Holt}}, \bibinfo {author} {\bibfnamefont {G.~R.}\ \bibnamefont {Jansen}}, \bibinfo {author} {\bibfnamefont {M.}~\bibnamefont {Kowalska}}, \bibinfo {author} {\bibfnamefont {K.}~\bibnamefont {Kreim}}, \bibinfo {author} {\bibfnamefont {W.}~\bibnamefont {Nazarewicz}}, \bibinfo {author} {\bibfnamefont {R.}~\bibnamefont {Neugart}}, \bibinfo {author} {\bibfnamefont {G.}~\bibnamefont {Neyens}}, \bibinfo {author} {\bibfnamefont
  {W.}~\bibnamefont {Nörtershäuser}}, \bibinfo {author} {\bibfnamefont {T.}~\bibnamefont {Papenbrock}}, \bibinfo {author} {\bibfnamefont {J.}~\bibnamefont {Papuga}}, \bibinfo {author} {\bibfnamefont {A.}~\bibnamefont {Schwenk}}, \bibinfo {author} {\bibfnamefont {J.}~\bibnamefont {Simonis}}, \bibinfo {author} {\bibfnamefont {K.~A.}\ \bibnamefont {Wendt}},\ and\ \bibinfo {author} {\bibfnamefont {D.~T.}\ \bibnamefont {Yordanov}},\ }\href {https://doi.org/https://doi.org/10.1038/nphys3645} {\bibfield  {journal} {\bibinfo  {journal} {Nature Physics}\ }\textbf {\bibinfo {volume} {12}},\ \bibinfo {pages} {594} (\bibinfo {year} {2016})}\BibitemShut {NoStop}%
\bibitem [{\citenamefont {Brown}(2022)}]{brown2022}%
  \BibitemOpen
  \bibfield  {author} {\bibinfo {author} {\bibfnamefont {B.~A.}\ \bibnamefont {Brown}},\ }\href {https://doi.org/10.3390/physics4020035} {\bibfield  {journal} {\bibinfo  {journal} {Physics}\ }\textbf {\bibinfo {volume} {4}},\ \bibinfo {pages} {525} (\bibinfo {year} {2022})}\BibitemShut {NoStop}%
\bibitem [{\citenamefont {Bonnard}\ \emph {et~al.}(2016)\citenamefont {Bonnard}, \citenamefont {Lenzi},\ and\ \citenamefont {Zuker}}]{Bonnard2016}%
  \BibitemOpen
  \bibfield  {author} {\bibinfo {author} {\bibfnamefont {J.}~\bibnamefont {Bonnard}}, \bibinfo {author} {\bibfnamefont {S.~M.}\ \bibnamefont {Lenzi}},\ and\ \bibinfo {author} {\bibfnamefont {A.~P.}\ \bibnamefont {Zuker}},\ }\href {https://doi.org/10.1103/PhysRevLett.116.212501} {\bibfield  {journal} {\bibinfo  {journal} {Phys. Rev. Lett.}\ }\textbf {\bibinfo {volume} {116}},\ \bibinfo {pages} {212501} (\bibinfo {year} {2016})}\BibitemShut {NoStop}%
\bibitem [{\citenamefont {Riley}\ \emph {et~al.}(2014)\citenamefont {Riley}, \citenamefont {Agiorgousis}, \citenamefont {Baugher}, \citenamefont {Bazin}, \citenamefont {Bowry}, \citenamefont {Cottle}, \citenamefont {DeVone}, \citenamefont {Gade}, \citenamefont {Glowacki}, \citenamefont {Kemper}, \citenamefont {Lunderberg}, \citenamefont {McPherson}, \citenamefont {Noji}, \citenamefont {Recchia}, \citenamefont {Sadler}, \citenamefont {Scott}, \citenamefont {Weisshaar},\ and\ \citenamefont {Zegers}}]{riley2014}%
  \BibitemOpen
  \bibfield  {author} {\bibinfo {author} {\bibfnamefont {L.~A.}\ \bibnamefont {Riley}}, \bibinfo {author} {\bibfnamefont {M.~L.}\ \bibnamefont {Agiorgousis}}, \bibinfo {author} {\bibfnamefont {T.~R.}\ \bibnamefont {Baugher}}, \bibinfo {author} {\bibfnamefont {D.}~\bibnamefont {Bazin}}, \bibinfo {author} {\bibfnamefont {M.}~\bibnamefont {Bowry}}, \bibinfo {author} {\bibfnamefont {P.~D.}\ \bibnamefont {Cottle}}, \bibinfo {author} {\bibfnamefont {F.~G.}\ \bibnamefont {DeVone}}, \bibinfo {author} {\bibfnamefont {A.}~\bibnamefont {Gade}}, \bibinfo {author} {\bibfnamefont {M.~T.}\ \bibnamefont {Glowacki}}, \bibinfo {author} {\bibfnamefont {K.~W.}\ \bibnamefont {Kemper}}, \bibinfo {author} {\bibfnamefont {E.}~\bibnamefont {Lunderberg}}, \bibinfo {author} {\bibfnamefont {D.~M.}\ \bibnamefont {McPherson}}, \bibinfo {author} {\bibfnamefont {S.}~\bibnamefont {Noji}}, \bibinfo {author} {\bibfnamefont {F.}~\bibnamefont {Recchia}}, \bibinfo {author} {\bibfnamefont {B.~V.}\ \bibnamefont {Sadler}}, \bibinfo {author}
  {\bibfnamefont {M.}~\bibnamefont {Scott}}, \bibinfo {author} {\bibfnamefont {D.}~\bibnamefont {Weisshaar}},\ and\ \bibinfo {author} {\bibfnamefont {R.~G.~T.}\ \bibnamefont {Zegers}},\ }\href {https://doi.org/10.1103/PhysRevC.90.011305} {\bibfield  {journal} {\bibinfo  {journal} {Phys. Rev. C}\ }\textbf {\bibinfo {volume} {90}},\ \bibinfo {pages} {011305} (\bibinfo {year} {2014})}\BibitemShut {NoStop}%
\bibitem [{FDS(2020)}]{FDSi1}%
  \BibitemOpen
  \href@noop {} {\bibinfo {title} {{FRIB Decay Station initiator proposal}}},\ \bibinfo {howpublished} {\url{https://fds.ornl.gov/wp-content/uploads/2020/09/FDSi-Proposal-May2020.pdf}} (\bibinfo {year} {2020})\BibitemShut {NoStop}%
\bibitem [{FDS(2023)}]{FDSi2}%
  \BibitemOpen
  \href@noop {} {\bibinfo {title} {{FRIB Decay Station initiator}}},\ \bibinfo {howpublished} {\url{https://fds.ornl.gov/initiator/}} (\bibinfo {year} {2023})\BibitemShut {NoStop}%
\bibitem [{\citenamefont {Paulauskas}\ \emph {et~al.}(2014)\citenamefont {Paulauskas}, \citenamefont {Madurga}, \citenamefont {Grzywacz}, \citenamefont {Miller}, \citenamefont {Padgett},\ and\ \citenamefont {Tan}}]{VANDLE1}%
  \BibitemOpen
  \bibfield  {author} {\bibinfo {author} {\bibfnamefont {S.~V.}\ \bibnamefont {Paulauskas}}, \bibinfo {author} {\bibfnamefont {M.}~\bibnamefont {Madurga}}, \bibinfo {author} {\bibfnamefont {R.}~\bibnamefont {Grzywacz}}, \bibinfo {author} {\bibfnamefont {D.}~\bibnamefont {Miller}}, \bibinfo {author} {\bibfnamefont {S.}~\bibnamefont {Padgett}},\ and\ \bibinfo {author} {\bibfnamefont {H.}~\bibnamefont {Tan}},\ }\href {https://doi.org/https://doi.org/10.1016/j.nima.2013.11.028} {\bibfield  {journal} {\bibinfo  {journal} {Nuclear Instruments and Methods in Physics Research Section A: Accelerators, Spectrometers, Detectors and Associated Equipment}\ }\textbf {\bibinfo {volume} {737}},\ \bibinfo {pages} {22} (\bibinfo {year} {2014})}\BibitemShut {NoStop}%
\bibitem [{\citenamefont {Peters}\ \emph {et~al.}(2016)\citenamefont {Peters}, \citenamefont {Ilyushkin}, \citenamefont {Madurga}, \citenamefont {Matei}, \citenamefont {Paulauskas}, \citenamefont {Grzywacz}, \citenamefont {Bardayan}, \citenamefont {Brune}, \citenamefont {Allen}, \citenamefont {Allen}, \citenamefont {Bergstrom}, \citenamefont {Blackmon}, \citenamefont {Brewer}, \citenamefont {Cizewski}, \citenamefont {Copp}, \citenamefont {Howard}, \citenamefont {Ikeyama}, \citenamefont {Kozub}, \citenamefont {Manning}, \citenamefont {Massey}, \citenamefont {Matos}, \citenamefont {Merino}, \citenamefont {O'Malley}, \citenamefont {Raiola}, \citenamefont {Reingold}, \citenamefont {Sarazin}, \citenamefont {Spassova}, \citenamefont {Taylor},\ and\ \citenamefont {Walter}}]{VANDLE2}%
  \BibitemOpen
  \bibfield  {author} {\bibinfo {author} {\bibfnamefont {W.~A.}\ \bibnamefont {Peters}}, \bibinfo {author} {\bibfnamefont {S.}~\bibnamefont {Ilyushkin}}, \bibinfo {author} {\bibfnamefont {M.}~\bibnamefont {Madurga}}, \bibinfo {author} {\bibfnamefont {C.}~\bibnamefont {Matei}}, \bibinfo {author} {\bibfnamefont {S.~V.}\ \bibnamefont {Paulauskas}}, \bibinfo {author} {\bibfnamefont {R.~K.}\ \bibnamefont {Grzywacz}}, \bibinfo {author} {\bibfnamefont {D.~W.}\ \bibnamefont {Bardayan}}, \bibinfo {author} {\bibfnamefont {C.~R.}\ \bibnamefont {Brune}}, \bibinfo {author} {\bibfnamefont {J.}~\bibnamefont {Allen}}, \bibinfo {author} {\bibfnamefont {J.~M.}\ \bibnamefont {Allen}}, \bibinfo {author} {\bibfnamefont {Z.}~\bibnamefont {Bergstrom}}, \bibinfo {author} {\bibfnamefont {J.}~\bibnamefont {Blackmon}}, \bibinfo {author} {\bibfnamefont {N.~T.}\ \bibnamefont {Brewer}}, \bibinfo {author} {\bibfnamefont {J.~A.}\ \bibnamefont {Cizewski}}, \bibinfo {author} {\bibfnamefont {P.}~\bibnamefont {Copp}}, \bibinfo {author}
  {\bibfnamefont {M.~E.}\ \bibnamefont {Howard}}, \bibinfo {author} {\bibfnamefont {R.}~\bibnamefont {Ikeyama}}, \bibinfo {author} {\bibfnamefont {R.~L.}\ \bibnamefont {Kozub}}, \bibinfo {author} {\bibfnamefont {B.}~\bibnamefont {Manning}}, \bibinfo {author} {\bibfnamefont {T.~N.}\ \bibnamefont {Massey}}, \bibinfo {author} {\bibfnamefont {M.}~\bibnamefont {Matos}}, \bibinfo {author} {\bibfnamefont {E.}~\bibnamefont {Merino}}, \bibinfo {author} {\bibfnamefont {P.~D.}\ \bibnamefont {O'Malley}}, \bibinfo {author} {\bibfnamefont {F.}~\bibnamefont {Raiola}}, \bibinfo {author} {\bibfnamefont {C.~S.}\ \bibnamefont {Reingold}}, \bibinfo {author} {\bibfnamefont {F.}~\bibnamefont {Sarazin}}, \bibinfo {author} {\bibfnamefont {I.}~\bibnamefont {Spassova}}, \bibinfo {author} {\bibfnamefont {S.}~\bibnamefont {Taylor}},\ and\ \bibinfo {author} {\bibfnamefont {D.}~\bibnamefont {Walter}},\ }\href {https://doi.org/https://doi.org/10.1016/j.nima.2016.08.054} {\bibfield  {journal} {\bibinfo  {journal} {Nuclear Instruments and
  Methods in Physics Research Section A: Accelerators, Spectrometers, Detectors and Associated Equipment}\ }\textbf {\bibinfo {volume} {836}},\ \bibinfo {pages} {122} (\bibinfo {year} {2016})}\BibitemShut {NoStop}%
\bibitem [{\citenamefont {Karny}\ \emph {et~al.}(2016)\citenamefont {Karny}, \citenamefont {Rykaczewski}, \citenamefont {Fijałkowska}, \citenamefont {Rasco}, \citenamefont {Wolińska-Cichocka}, \citenamefont {Grzywacz}, \citenamefont {Goetz}, \citenamefont {Miller},\ and\ \citenamefont {Zganjar}}]{Karny2016}%
  \BibitemOpen
  \bibfield  {author} {\bibinfo {author} {\bibfnamefont {M.}~\bibnamefont {Karny}}, \bibinfo {author} {\bibfnamefont {K.~P.}\ \bibnamefont {Rykaczewski}}, \bibinfo {author} {\bibfnamefont {A.}~\bibnamefont {Fijałkowska}}, \bibinfo {author} {\bibfnamefont {B.~C.}\ \bibnamefont {Rasco}}, \bibinfo {author} {\bibfnamefont {M.}~\bibnamefont {Wolińska-Cichocka}}, \bibinfo {author} {\bibfnamefont {R.~K.}\ \bibnamefont {Grzywacz}}, \bibinfo {author} {\bibfnamefont {K.~C.}\ \bibnamefont {Goetz}}, \bibinfo {author} {\bibfnamefont {D.}~\bibnamefont {Miller}},\ and\ \bibinfo {author} {\bibfnamefont {E.~F.}\ \bibnamefont {Zganjar}},\ }\href {https://doi.org/https://doi.org/10.1016/j.nima.2016.08.046} {\bibfield  {journal} {\bibinfo  {journal} {Nuclear Instruments and Methods in Physics Research Section A: Accelerators, Spectrometers, Detectors and Associated Equipment}\ }\textbf {\bibinfo {volume} {836}},\ \bibinfo {pages} {83} (\bibinfo {year} {2016})}\BibitemShut {NoStop}%
\bibitem [{\citenamefont {Karny}\ \emph {et~al.}(2020)\citenamefont {Karny}, \citenamefont {Fijałkowska}, \citenamefont {Grzywacz}, \citenamefont {Rasco}, \citenamefont {Rykaczewski},\ and\ \citenamefont {Stepaniuk}}]{Karny2020}%
  \BibitemOpen
  \bibfield  {author} {\bibinfo {author} {\bibfnamefont {M.}~\bibnamefont {Karny}}, \bibinfo {author} {\bibfnamefont {A.}~\bibnamefont {Fijałkowska}}, \bibinfo {author} {\bibfnamefont {R.~K.}\ \bibnamefont {Grzywacz}}, \bibinfo {author} {\bibfnamefont {B.~C.}\ \bibnamefont {Rasco}}, \bibinfo {author} {\bibfnamefont {K.~P.}\ \bibnamefont {Rykaczewski}},\ and\ \bibinfo {author} {\bibfnamefont {M.}~\bibnamefont {Stepaniuk}},\ }\href {https://doi.org/https://doi.org/10.1016/j.nimb.2019.04.045} {\bibfield  {journal} {\bibinfo  {journal} {Nuclear Instruments and Methods in Physics Research Section B: Beam Interactions with Materials and Atoms}\ }\textbf {\bibinfo {volume} {463}},\ \bibinfo {pages} {390} (\bibinfo {year} {2020})}\BibitemShut {NoStop}%
\bibitem [{\citenamefont {Yokoyama}\ \emph {et~al.}(2019)\citenamefont {Yokoyama}, \citenamefont {Singh}, \citenamefont {Grzywacz}, \citenamefont {Keeler}, \citenamefont {King}, \citenamefont {Agramunt}, \citenamefont {Brewer}, \citenamefont {Go}, \citenamefont {Heideman}, \citenamefont {Liu}, \citenamefont {Nishimura}, \citenamefont {Parkhurst}, \citenamefont {Phong}, \citenamefont {Rajabali}, \citenamefont {Rasco}, \citenamefont {Rykaczewski}, \citenamefont {Stracener}, \citenamefont {Tain}, \citenamefont {Tolosa-Delgado}, \citenamefont {Vaigneur},\ and\ \citenamefont {Woli\'{n}ska-Cichocka}}]{IMPLANT}%
  \BibitemOpen
  \bibfield  {author} {\bibinfo {author} {\bibfnamefont {R.}~\bibnamefont {Yokoyama}}, \bibinfo {author} {\bibfnamefont {M.}~\bibnamefont {Singh}}, \bibinfo {author} {\bibfnamefont {R.}~\bibnamefont {Grzywacz}}, \bibinfo {author} {\bibfnamefont {A.}~\bibnamefont {Keeler}}, \bibinfo {author} {\bibfnamefont {T.~T.}\ \bibnamefont {King}}, \bibinfo {author} {\bibfnamefont {J.}~\bibnamefont {Agramunt}}, \bibinfo {author} {\bibfnamefont {N.~T.}\ \bibnamefont {Brewer}}, \bibinfo {author} {\bibfnamefont {S.}~\bibnamefont {Go}}, \bibinfo {author} {\bibfnamefont {J.}~\bibnamefont {Heideman}}, \bibinfo {author} {\bibfnamefont {J.}~\bibnamefont {Liu}}, \bibinfo {author} {\bibfnamefont {S.}~\bibnamefont {Nishimura}}, \bibinfo {author} {\bibfnamefont {P.}~\bibnamefont {Parkhurst}}, \bibinfo {author} {\bibfnamefont {V.~H.}\ \bibnamefont {Phong}}, \bibinfo {author} {\bibfnamefont {M.~M.}\ \bibnamefont {Rajabali}}, \bibinfo {author} {\bibfnamefont {B.~C.}\ \bibnamefont {Rasco}}, \bibinfo {author} {\bibfnamefont {K.~P.}\
  \bibnamefont {Rykaczewski}}, \bibinfo {author} {\bibfnamefont {D.~W.}\ \bibnamefont {Stracener}}, \bibinfo {author} {\bibfnamefont {J.~L.}\ \bibnamefont {Tain}}, \bibinfo {author} {\bibfnamefont {A.}~\bibnamefont {Tolosa-Delgado}}, \bibinfo {author} {\bibfnamefont {K.}~\bibnamefont {Vaigneur}},\ and\ \bibinfo {author} {\bibfnamefont {M.}~\bibnamefont {Woli\'{n}ska-Cichocka}},\ }\href {https://doi.org/https://doi.org/10.1016/j.nima.2019.05.026} {\bibfield  {journal} {\bibinfo  {journal} {Nuclear Instruments and Methods in Physics Research Section A: Accelerators, Spectrometers, Detectors and Associated Equipment}\ }\textbf {\bibinfo {volume} {937}},\ \bibinfo {pages} {93} (\bibinfo {year} {2019})}\BibitemShut {NoStop}%
\bibitem [{\citenamefont {Singh}\ \emph {et~al.}(2025)\citenamefont {Singh}, \citenamefont {Yokoyama}, \citenamefont {Grzywacz}, \citenamefont {Keeler}, \citenamefont {King}, \citenamefont {Agramunt}, \citenamefont {Brewer}, \citenamefont {Go}, \citenamefont {Liu}, \citenamefont {Nishimura}, \citenamefont {Parkhurst}, \citenamefont {Phong}, \citenamefont {Rajabali}, \citenamefont {Rasco}, \citenamefont {Rykaczewski}, \citenamefont {Stracener}, \citenamefont {Tolosa-Delgado}, \citenamefont {Vaigneur},\ and\ \citenamefont {Wolińska-Cichocka}}]{singh2025}%
  \BibitemOpen
  \bibfield  {author} {\bibinfo {author} {\bibfnamefont {M.}~\bibnamefont {Singh}}, \bibinfo {author} {\bibfnamefont {R.}~\bibnamefont {Yokoyama}}, \bibinfo {author} {\bibfnamefont {R.}~\bibnamefont {Grzywacz}}, \bibinfo {author} {\bibfnamefont {A.}~\bibnamefont {Keeler}}, \bibinfo {author} {\bibfnamefont {T.}~\bibnamefont {King}}, \bibinfo {author} {\bibfnamefont {J.}~\bibnamefont {Agramunt}}, \bibinfo {author} {\bibfnamefont {N.}~\bibnamefont {Brewer}}, \bibinfo {author} {\bibfnamefont {S.}~\bibnamefont {Go}}, \bibinfo {author} {\bibfnamefont {J.}~\bibnamefont {Liu}}, \bibinfo {author} {\bibfnamefont {S.}~\bibnamefont {Nishimura}}, \bibinfo {author} {\bibfnamefont {P.}~\bibnamefont {Parkhurst}}, \bibinfo {author} {\bibfnamefont {V.}~\bibnamefont {Phong}}, \bibinfo {author} {\bibfnamefont {M.}~\bibnamefont {Rajabali}}, \bibinfo {author} {\bibfnamefont {B.}~\bibnamefont {Rasco}}, \bibinfo {author} {\bibfnamefont {K.}~\bibnamefont {Rykaczewski}}, \bibinfo {author} {\bibfnamefont {D.}~\bibnamefont {Stracener}},
  \bibinfo {author} {\bibfnamefont {A.}~\bibnamefont {Tolosa-Delgado}}, \bibinfo {author} {\bibfnamefont {K.}~\bibnamefont {Vaigneur}},\ and\ \bibinfo {author} {\bibfnamefont {M.}~\bibnamefont {Wolińska-Cichocka}},\ }\href {https://doi.org/https://doi.org/10.1016/j.nima.2025.170239} {\bibfield  {journal} {\bibinfo  {journal} {Nuclear Instruments and Methods in Physics Research Section A: Accelerators, Spectrometers, Detectors and Associated Equipment}\ ,\ \bibinfo {pages} {170239}} (\bibinfo {year} {2025})}\BibitemShut {NoStop}%
\bibitem [{\citenamefont {Crawford}\ \emph {et~al.}(2022)\citenamefont {Crawford}, \citenamefont {Tripathi}, \citenamefont {Allmond}, \citenamefont {Crider}, \citenamefont {Grzywacz}, \citenamefont {Liddick}, \citenamefont {Andalib}, \citenamefont {Argo}, \citenamefont {Benetti}, \citenamefont {Bhattacharya}, \citenamefont {Campbell}, \citenamefont {Carpenter}, \citenamefont {Chan}, \citenamefont {Chester}, \citenamefont {Christie}, \citenamefont {Clark}, \citenamefont {Cox}, \citenamefont {Doetsch}, \citenamefont {Dopfer}, \citenamefont {Duarte}, \citenamefont {Fallon}, \citenamefont {Frotscher}, \citenamefont {Gaballah}, \citenamefont {Gray}, \citenamefont {Harke}, \citenamefont {Heideman}, \citenamefont {Heugen}, \citenamefont {Jain}, \citenamefont {King}, \citenamefont {Kitamura}, \citenamefont {Kolos}, \citenamefont {Kondev}, \citenamefont {Laminack}, \citenamefont {Longfellow}, \citenamefont {Lubna}, \citenamefont {Luitel}, \citenamefont {Madurga}, \citenamefont {Mahajan}, \citenamefont {Mogannam},
  \citenamefont {Morse}, \citenamefont {Neupane}, \citenamefont {Nowicki}, \citenamefont {Ogunbeku}, \citenamefont {Ong}, \citenamefont {Porzio}, \citenamefont {Prokop}, \citenamefont {Rasco}, \citenamefont {Ronning}, \citenamefont {Rubino}, \citenamefont {Ruland}, \citenamefont {Rykaczewski}, \citenamefont {Schaedig}, \citenamefont {Seweryniak}, \citenamefont {Siegl}, \citenamefont {Singh}, \citenamefont {Tabor}, \citenamefont {Tang}, \citenamefont {Wheeler}, \citenamefont {Winger},\ and\ \citenamefont {Xu}}]{Crawford2022}%
  \BibitemOpen
  \bibfield  {author} {\bibinfo {author} {\bibfnamefont {H.~L.}\ \bibnamefont {Crawford}}, \bibinfo {author} {\bibfnamefont {V.}~\bibnamefont {Tripathi}}, \bibinfo {author} {\bibfnamefont {J.~M.}\ \bibnamefont {Allmond}}, \bibinfo {author} {\bibfnamefont {B.~P.}\ \bibnamefont {Crider}}, \bibinfo {author} {\bibfnamefont {R.}~\bibnamefont {Grzywacz}}, \bibinfo {author} {\bibfnamefont {S.~N.}\ \bibnamefont {Liddick}}, \bibinfo {author} {\bibfnamefont {A.}~\bibnamefont {Andalib}}, \bibinfo {author} {\bibfnamefont {E.}~\bibnamefont {Argo}}, \bibinfo {author} {\bibfnamefont {C.}~\bibnamefont {Benetti}}, \bibinfo {author} {\bibfnamefont {S.}~\bibnamefont {Bhattacharya}}, \bibinfo {author} {\bibfnamefont {C.~M.}\ \bibnamefont {Campbell}}, \bibinfo {author} {\bibfnamefont {M.~P.}\ \bibnamefont {Carpenter}}, \bibinfo {author} {\bibfnamefont {J.}~\bibnamefont {Chan}}, \bibinfo {author} {\bibfnamefont {A.}~\bibnamefont {Chester}}, \bibinfo {author} {\bibfnamefont {J.}~\bibnamefont {Christie}}, \bibinfo {author}
  {\bibfnamefont {B.~R.}\ \bibnamefont {Clark}}, \bibinfo {author} {\bibfnamefont {I.}~\bibnamefont {Cox}}, \bibinfo {author} {\bibfnamefont {A.~A.}\ \bibnamefont {Doetsch}}, \bibinfo {author} {\bibfnamefont {J.}~\bibnamefont {Dopfer}}, \bibinfo {author} {\bibfnamefont {J.~G.}\ \bibnamefont {Duarte}}, \bibinfo {author} {\bibfnamefont {P.}~\bibnamefont {Fallon}}, \bibinfo {author} {\bibfnamefont {A.}~\bibnamefont {Frotscher}}, \bibinfo {author} {\bibfnamefont {T.}~\bibnamefont {Gaballah}}, \bibinfo {author} {\bibfnamefont {T.~J.}\ \bibnamefont {Gray}}, \bibinfo {author} {\bibfnamefont {J.~T.}\ \bibnamefont {Harke}}, \bibinfo {author} {\bibfnamefont {J.}~\bibnamefont {Heideman}}, \bibinfo {author} {\bibfnamefont {H.}~\bibnamefont {Heugen}}, \bibinfo {author} {\bibfnamefont {R.}~\bibnamefont {Jain}}, \bibinfo {author} {\bibfnamefont {T.~T.}\ \bibnamefont {King}}, \bibinfo {author} {\bibfnamefont {N.}~\bibnamefont {Kitamura}}, \bibinfo {author} {\bibfnamefont {K.}~\bibnamefont {Kolos}}, \bibinfo {author}
  {\bibfnamefont {F.~G.}\ \bibnamefont {Kondev}}, \bibinfo {author} {\bibfnamefont {A.}~\bibnamefont {Laminack}}, \bibinfo {author} {\bibfnamefont {B.}~\bibnamefont {Longfellow}}, \bibinfo {author} {\bibfnamefont {R.~S.}\ \bibnamefont {Lubna}}, \bibinfo {author} {\bibfnamefont {S.}~\bibnamefont {Luitel}}, \bibinfo {author} {\bibfnamefont {M.}~\bibnamefont {Madurga}}, \bibinfo {author} {\bibfnamefont {R.}~\bibnamefont {Mahajan}}, \bibinfo {author} {\bibfnamefont {M.~J.}\ \bibnamefont {Mogannam}}, \bibinfo {author} {\bibfnamefont {C.}~\bibnamefont {Morse}}, \bibinfo {author} {\bibfnamefont {S.}~\bibnamefont {Neupane}}, \bibinfo {author} {\bibfnamefont {A.}~\bibnamefont {Nowicki}}, \bibinfo {author} {\bibfnamefont {T.~H.}\ \bibnamefont {Ogunbeku}}, \bibinfo {author} {\bibfnamefont {W.-J.}\ \bibnamefont {Ong}}, \bibinfo {author} {\bibfnamefont {C.}~\bibnamefont {Porzio}}, \bibinfo {author} {\bibfnamefont {C.~J.}\ \bibnamefont {Prokop}}, \bibinfo {author} {\bibfnamefont {B.~C.}\ \bibnamefont {Rasco}}, \bibinfo
  {author} {\bibfnamefont {E.~K.}\ \bibnamefont {Ronning}}, \bibinfo {author} {\bibfnamefont {E.}~\bibnamefont {Rubino}}, \bibinfo {author} {\bibfnamefont {T.~J.}\ \bibnamefont {Ruland}}, \bibinfo {author} {\bibfnamefont {K.~P.}\ \bibnamefont {Rykaczewski}}, \bibinfo {author} {\bibfnamefont {L.}~\bibnamefont {Schaedig}}, \bibinfo {author} {\bibfnamefont {D.}~\bibnamefont {Seweryniak}}, \bibinfo {author} {\bibfnamefont {K.}~\bibnamefont {Siegl}}, \bibinfo {author} {\bibfnamefont {M.}~\bibnamefont {Singh}}, \bibinfo {author} {\bibfnamefont {S.~L.}\ \bibnamefont {Tabor}}, \bibinfo {author} {\bibfnamefont {T.~L.}\ \bibnamefont {Tang}}, \bibinfo {author} {\bibfnamefont {T.}~\bibnamefont {Wheeler}}, \bibinfo {author} {\bibfnamefont {J.~A.}\ \bibnamefont {Winger}},\ and\ \bibinfo {author} {\bibfnamefont {Z.}~\bibnamefont {Xu}},\ }\href {https://doi.org/10.1103/PhysRevLett.129.212501} {\bibfield  {journal} {\bibinfo  {journal} {Phys. Rev. Lett.}\ }\textbf {\bibinfo {volume} {129}},\ \bibinfo {pages} {212501}
  (\bibinfo {year} {2022})}\BibitemShut {NoStop}%
\bibitem [{\citenamefont {Gray}\ \emph {et~al.}(2023)\citenamefont {Gray}, \citenamefont {Allmond}, \citenamefont {Xu}, \citenamefont {King}, \citenamefont {Lubna}, \citenamefont {Crawford}, \citenamefont {Tripathi}, \citenamefont {Crider}, \citenamefont {Grzywacz}, \citenamefont {Liddick}, \citenamefont {Macchiavelli}, \citenamefont {Miyagi}, \citenamefont {Poves}, \citenamefont {Andalib}, \citenamefont {Argo}, \citenamefont {Benetti}, \citenamefont {Bhattacharya}, \citenamefont {Campbell}, \citenamefont {Carpenter}, \citenamefont {Chan}, \citenamefont {Chester}, \citenamefont {Christie}, \citenamefont {Clark}, \citenamefont {Cox}, \citenamefont {Doetsch}, \citenamefont {Dopfer}, \citenamefont {Duarte}, \citenamefont {Fallon}, \citenamefont {Frotscher}, \citenamefont {Gaballah}, \citenamefont {Harke}, \citenamefont {Heideman}, \citenamefont {Huegen}, \citenamefont {Holt}, \citenamefont {Jain}, \citenamefont {Kitamura}, \citenamefont {Kolos}, \citenamefont {Kondev}, \citenamefont {Laminack}, \citenamefont
  {Longfellow}, \citenamefont {Luitel}, \citenamefont {Madurga}, \citenamefont {Mahajan}, \citenamefont {Mogannam}, \citenamefont {Morse}, \citenamefont {Neupane}, \citenamefont {Nowicki}, \citenamefont {Ogunbeku}, \citenamefont {Ong}, \citenamefont {Porzio}, \citenamefont {Prokop}, \citenamefont {Rasco}, \citenamefont {Ronning}, \citenamefont {Rubino}, \citenamefont {Ruland}, \citenamefont {Rykaczewski}, \citenamefont {Schaedig}, \citenamefont {Seweryniak}, \citenamefont {Siegl}, \citenamefont {Singh}, \citenamefont {Stuchbery}, \citenamefont {Tabor}, \citenamefont {Tang}, \citenamefont {Wheeler}, \citenamefont {Winger},\ and\ \citenamefont {Wood}}]{Gray2023}%
  \BibitemOpen
  \bibfield  {author} {\bibinfo {author} {\bibfnamefont {T.~J.}\ \bibnamefont {Gray}}, \bibinfo {author} {\bibfnamefont {J.~M.}\ \bibnamefont {Allmond}}, \bibinfo {author} {\bibfnamefont {Z.}~\bibnamefont {Xu}}, \bibinfo {author} {\bibfnamefont {T.~T.}\ \bibnamefont {King}}, \bibinfo {author} {\bibfnamefont {R.~S.}\ \bibnamefont {Lubna}}, \bibinfo {author} {\bibfnamefont {H.~L.}\ \bibnamefont {Crawford}}, \bibinfo {author} {\bibfnamefont {V.}~\bibnamefont {Tripathi}}, \bibinfo {author} {\bibfnamefont {B.~P.}\ \bibnamefont {Crider}}, \bibinfo {author} {\bibfnamefont {R.}~\bibnamefont {Grzywacz}}, \bibinfo {author} {\bibfnamefont {S.~N.}\ \bibnamefont {Liddick}}, \bibinfo {author} {\bibfnamefont {A.~O.}\ \bibnamefont {Macchiavelli}}, \bibinfo {author} {\bibfnamefont {T.}~\bibnamefont {Miyagi}}, \bibinfo {author} {\bibfnamefont {A.}~\bibnamefont {Poves}}, \bibinfo {author} {\bibfnamefont {A.}~\bibnamefont {Andalib}}, \bibinfo {author} {\bibfnamefont {E.}~\bibnamefont {Argo}}, \bibinfo {author} {\bibfnamefont
  {C.}~\bibnamefont {Benetti}}, \bibinfo {author} {\bibfnamefont {S.}~\bibnamefont {Bhattacharya}}, \bibinfo {author} {\bibfnamefont {C.~M.}\ \bibnamefont {Campbell}}, \bibinfo {author} {\bibfnamefont {M.~P.}\ \bibnamefont {Carpenter}}, \bibinfo {author} {\bibfnamefont {J.}~\bibnamefont {Chan}}, \bibinfo {author} {\bibfnamefont {A.}~\bibnamefont {Chester}}, \bibinfo {author} {\bibfnamefont {J.}~\bibnamefont {Christie}}, \bibinfo {author} {\bibfnamefont {B.~R.}\ \bibnamefont {Clark}}, \bibinfo {author} {\bibfnamefont {I.}~\bibnamefont {Cox}}, \bibinfo {author} {\bibfnamefont {A.~A.}\ \bibnamefont {Doetsch}}, \bibinfo {author} {\bibfnamefont {J.}~\bibnamefont {Dopfer}}, \bibinfo {author} {\bibfnamefont {J.~G.}\ \bibnamefont {Duarte}}, \bibinfo {author} {\bibfnamefont {P.}~\bibnamefont {Fallon}}, \bibinfo {author} {\bibfnamefont {A.}~\bibnamefont {Frotscher}}, \bibinfo {author} {\bibfnamefont {T.}~\bibnamefont {Gaballah}}, \bibinfo {author} {\bibfnamefont {J.~T.}\ \bibnamefont {Harke}}, \bibinfo {author}
  {\bibfnamefont {J.}~\bibnamefont {Heideman}}, \bibinfo {author} {\bibfnamefont {H.}~\bibnamefont {Huegen}}, \bibinfo {author} {\bibfnamefont {J.~D.}\ \bibnamefont {Holt}}, \bibinfo {author} {\bibfnamefont {R.}~\bibnamefont {Jain}}, \bibinfo {author} {\bibfnamefont {N.}~\bibnamefont {Kitamura}}, \bibinfo {author} {\bibfnamefont {K.}~\bibnamefont {Kolos}}, \bibinfo {author} {\bibfnamefont {F.~G.}\ \bibnamefont {Kondev}}, \bibinfo {author} {\bibfnamefont {A.}~\bibnamefont {Laminack}}, \bibinfo {author} {\bibfnamefont {B.}~\bibnamefont {Longfellow}}, \bibinfo {author} {\bibfnamefont {S.}~\bibnamefont {Luitel}}, \bibinfo {author} {\bibfnamefont {M.}~\bibnamefont {Madurga}}, \bibinfo {author} {\bibfnamefont {R.}~\bibnamefont {Mahajan}}, \bibinfo {author} {\bibfnamefont {M.~J.}\ \bibnamefont {Mogannam}}, \bibinfo {author} {\bibfnamefont {C.}~\bibnamefont {Morse}}, \bibinfo {author} {\bibfnamefont {S.}~\bibnamefont {Neupane}}, \bibinfo {author} {\bibfnamefont {A.}~\bibnamefont {Nowicki}}, \bibinfo {author}
  {\bibfnamefont {T.~H.}\ \bibnamefont {Ogunbeku}}, \bibinfo {author} {\bibfnamefont {W.-J.}\ \bibnamefont {Ong}}, \bibinfo {author} {\bibfnamefont {C.}~\bibnamefont {Porzio}}, \bibinfo {author} {\bibfnamefont {C.~J.}\ \bibnamefont {Prokop}}, \bibinfo {author} {\bibfnamefont {B.~C.}\ \bibnamefont {Rasco}}, \bibinfo {author} {\bibfnamefont {E.~K.}\ \bibnamefont {Ronning}}, \bibinfo {author} {\bibfnamefont {E.}~\bibnamefont {Rubino}}, \bibinfo {author} {\bibfnamefont {T.~J.}\ \bibnamefont {Ruland}}, \bibinfo {author} {\bibfnamefont {K.~P.}\ \bibnamefont {Rykaczewski}}, \bibinfo {author} {\bibfnamefont {L.}~\bibnamefont {Schaedig}}, \bibinfo {author} {\bibfnamefont {D.}~\bibnamefont {Seweryniak}}, \bibinfo {author} {\bibfnamefont {K.}~\bibnamefont {Siegl}}, \bibinfo {author} {\bibfnamefont {M.}~\bibnamefont {Singh}}, \bibinfo {author} {\bibfnamefont {A.~E.}\ \bibnamefont {Stuchbery}}, \bibinfo {author} {\bibfnamefont {S.~L.}\ \bibnamefont {Tabor}}, \bibinfo {author} {\bibfnamefont {T.~L.}\ \bibnamefont {Tang}},
  \bibinfo {author} {\bibfnamefont {T.}~\bibnamefont {Wheeler}}, \bibinfo {author} {\bibfnamefont {J.~A.}\ \bibnamefont {Winger}},\ and\ \bibinfo {author} {\bibfnamefont {J.~L.}\ \bibnamefont {Wood}},\ }\href {https://doi.org/10.1103/PhysRevLett.130.242501} {\bibfield  {journal} {\bibinfo  {journal} {Phys. Rev. Lett.}\ }\textbf {\bibinfo {volume} {130}},\ \bibinfo {pages} {242501} (\bibinfo {year} {2023})}\BibitemShut {NoStop}%
\bibitem [{\citenamefont {Cox}\ \emph {et~al.}(2024)\citenamefont {Cox}, \citenamefont {Xu}, \citenamefont {Grzywacz}, \citenamefont {Ong}, \citenamefont {Rasco}, \citenamefont {Kitamura}, \citenamefont {Hoskins}, \citenamefont {Neupane}, \citenamefont {Ruland}, \citenamefont {Allmond}, \citenamefont {King}, \citenamefont {Lubna}, \citenamefont {Rykaczewski}, \citenamefont {Schatz}, \citenamefont {Sherrill}, \citenamefont {Tarasov}, \citenamefont {Ayangeakaa}, \citenamefont {Berg}, \citenamefont {Bleuel}, \citenamefont {Cerizza}, \citenamefont {Christie}, \citenamefont {Chester}, \citenamefont {Davis}, \citenamefont {Dembski}, \citenamefont {Doetsch}, \citenamefont {Duarte}, \citenamefont {Estrade}, \citenamefont {Fija\l{}kowska}, \citenamefont {Gray}, \citenamefont {Good}, \citenamefont {Haak}, \citenamefont {Hanai}, \citenamefont {Harke}, \citenamefont {Harris}, \citenamefont {Hermansen}, \citenamefont {Hoff}, \citenamefont {Jain}, \citenamefont {Karny}, \citenamefont {Kolos}, \citenamefont {Laminack},
  \citenamefont {Liddick}, \citenamefont {Longfellow}, \citenamefont {Lyons}, \citenamefont {Madurga}, \citenamefont {Mogannam}, \citenamefont {Nowicki}, \citenamefont {Ogunbeku}, \citenamefont {Owens-Fryar}, \citenamefont {Rajabali}, \citenamefont {Richard}, \citenamefont {Ronning}, \citenamefont {Rose}, \citenamefont {Siegl}, \citenamefont {Singh}, \citenamefont {Spyrou}, \citenamefont {Sweet}, \citenamefont {Tsantiri}, \citenamefont {Walters},\ and\ \citenamefont {Yokoyama}}]{Cox2024}%
  \BibitemOpen
  \bibfield  {author} {\bibinfo {author} {\bibfnamefont {I.}~\bibnamefont {Cox}}, \bibinfo {author} {\bibfnamefont {Z.~Y.}\ \bibnamefont {Xu}}, \bibinfo {author} {\bibfnamefont {R.}~\bibnamefont {Grzywacz}}, \bibinfo {author} {\bibfnamefont {W.-J.}\ \bibnamefont {Ong}}, \bibinfo {author} {\bibfnamefont {B.~C.}\ \bibnamefont {Rasco}}, \bibinfo {author} {\bibfnamefont {N.}~\bibnamefont {Kitamura}}, \bibinfo {author} {\bibfnamefont {D.}~\bibnamefont {Hoskins}}, \bibinfo {author} {\bibfnamefont {S.}~\bibnamefont {Neupane}}, \bibinfo {author} {\bibfnamefont {T.~J.}\ \bibnamefont {Ruland}}, \bibinfo {author} {\bibfnamefont {J.~M.}\ \bibnamefont {Allmond}}, \bibinfo {author} {\bibfnamefont {T.~T.}\ \bibnamefont {King}}, \bibinfo {author} {\bibfnamefont {R.~S.}\ \bibnamefont {Lubna}}, \bibinfo {author} {\bibfnamefont {K.~P.}\ \bibnamefont {Rykaczewski}}, \bibinfo {author} {\bibfnamefont {H.}~\bibnamefont {Schatz}}, \bibinfo {author} {\bibfnamefont {B.~M.}\ \bibnamefont {Sherrill}}, \bibinfo {author} {\bibfnamefont
  {O.~B.}\ \bibnamefont {Tarasov}}, \bibinfo {author} {\bibfnamefont {A.~D.}\ \bibnamefont {Ayangeakaa}}, \bibinfo {author} {\bibfnamefont {H.~C.}\ \bibnamefont {Berg}}, \bibinfo {author} {\bibfnamefont {D.~L.}\ \bibnamefont {Bleuel}}, \bibinfo {author} {\bibfnamefont {G.}~\bibnamefont {Cerizza}}, \bibinfo {author} {\bibfnamefont {J.}~\bibnamefont {Christie}}, \bibinfo {author} {\bibfnamefont {A.}~\bibnamefont {Chester}}, \bibinfo {author} {\bibfnamefont {J.}~\bibnamefont {Davis}}, \bibinfo {author} {\bibfnamefont {C.}~\bibnamefont {Dembski}}, \bibinfo {author} {\bibfnamefont {A.~A.}\ \bibnamefont {Doetsch}}, \bibinfo {author} {\bibfnamefont {J.~G.}\ \bibnamefont {Duarte}}, \bibinfo {author} {\bibfnamefont {A.}~\bibnamefont {Estrade}}, \bibinfo {author} {\bibfnamefont {A.}~\bibnamefont {Fija\l{}kowska}}, \bibinfo {author} {\bibfnamefont {T.~J.}\ \bibnamefont {Gray}}, \bibinfo {author} {\bibfnamefont {E.~C.}\ \bibnamefont {Good}}, \bibinfo {author} {\bibfnamefont {K.}~\bibnamefont {Haak}}, \bibinfo {author}
  {\bibfnamefont {S.}~\bibnamefont {Hanai}}, \bibinfo {author} {\bibfnamefont {J.~T.}\ \bibnamefont {Harke}}, \bibinfo {author} {\bibfnamefont {C.}~\bibnamefont {Harris}}, \bibinfo {author} {\bibfnamefont {K.}~\bibnamefont {Hermansen}}, \bibinfo {author} {\bibfnamefont {D.~E.~M.}\ \bibnamefont {Hoff}}, \bibinfo {author} {\bibfnamefont {R.}~\bibnamefont {Jain}}, \bibinfo {author} {\bibfnamefont {M.}~\bibnamefont {Karny}}, \bibinfo {author} {\bibfnamefont {K.}~\bibnamefont {Kolos}}, \bibinfo {author} {\bibfnamefont {A.}~\bibnamefont {Laminack}}, \bibinfo {author} {\bibfnamefont {S.~N.}\ \bibnamefont {Liddick}}, \bibinfo {author} {\bibfnamefont {B.}~\bibnamefont {Longfellow}}, \bibinfo {author} {\bibfnamefont {S.}~\bibnamefont {Lyons}}, \bibinfo {author} {\bibfnamefont {M.}~\bibnamefont {Madurga}}, \bibinfo {author} {\bibfnamefont {M.~J.}\ \bibnamefont {Mogannam}}, \bibinfo {author} {\bibfnamefont {A.}~\bibnamefont {Nowicki}}, \bibinfo {author} {\bibfnamefont {T.~H.}\ \bibnamefont {Ogunbeku}}, \bibinfo {author}
  {\bibfnamefont {G.}~\bibnamefont {Owens-Fryar}}, \bibinfo {author} {\bibfnamefont {M.~M.}\ \bibnamefont {Rajabali}}, \bibinfo {author} {\bibfnamefont {A.~L.}\ \bibnamefont {Richard}}, \bibinfo {author} {\bibfnamefont {E.~K.}\ \bibnamefont {Ronning}}, \bibinfo {author} {\bibfnamefont {G.~E.}\ \bibnamefont {Rose}}, \bibinfo {author} {\bibfnamefont {K.}~\bibnamefont {Siegl}}, \bibinfo {author} {\bibfnamefont {M.}~\bibnamefont {Singh}}, \bibinfo {author} {\bibfnamefont {A.}~\bibnamefont {Spyrou}}, \bibinfo {author} {\bibfnamefont {A.}~\bibnamefont {Sweet}}, \bibinfo {author} {\bibfnamefont {A.}~\bibnamefont {Tsantiri}}, \bibinfo {author} {\bibfnamefont {W.~B.}\ \bibnamefont {Walters}},\ and\ \bibinfo {author} {\bibfnamefont {R.}~\bibnamefont {Yokoyama}},\ }\href {https://doi.org/10.1103/PhysRevLett.132.152503} {\bibfield  {journal} {\bibinfo  {journal} {Phys. Rev. Lett.}\ }\textbf {\bibinfo {volume} {132}},\ \bibinfo {pages} {152503} (\bibinfo {year} {2024})}\BibitemShut {NoStop}%
\bibitem [{\citenamefont {Mantica}\ \emph {et~al.}(2008)\citenamefont {Mantica}, \citenamefont {Broda}, \citenamefont {Crawford}, \citenamefont {Damaske}, \citenamefont {Fornal}, \citenamefont {Hecht}, \citenamefont {Hoffman}, \citenamefont {Horoi}, \citenamefont {Hoteling}, \citenamefont {Janssens}, \citenamefont {Pereira}, \citenamefont {Pinter}, \citenamefont {Stoker}, \citenamefont {Tabor}, \citenamefont {Sumikama}, \citenamefont {Walters}, \citenamefont {Wang},\ and\ \citenamefont {Zhu}}]{Mantica2008}%
  \BibitemOpen
  \bibfield  {author} {\bibinfo {author} {\bibfnamefont {P.~F.}\ \bibnamefont {Mantica}}, \bibinfo {author} {\bibfnamefont {R.}~\bibnamefont {Broda}}, \bibinfo {author} {\bibfnamefont {H.~L.}\ \bibnamefont {Crawford}}, \bibinfo {author} {\bibfnamefont {A.}~\bibnamefont {Damaske}}, \bibinfo {author} {\bibfnamefont {B.}~\bibnamefont {Fornal}}, \bibinfo {author} {\bibfnamefont {A.~A.}\ \bibnamefont {Hecht}}, \bibinfo {author} {\bibfnamefont {C.}~\bibnamefont {Hoffman}}, \bibinfo {author} {\bibfnamefont {M.}~\bibnamefont {Horoi}}, \bibinfo {author} {\bibfnamefont {N.}~\bibnamefont {Hoteling}}, \bibinfo {author} {\bibfnamefont {R.~V.~F.}\ \bibnamefont {Janssens}}, \bibinfo {author} {\bibfnamefont {J.}~\bibnamefont {Pereira}}, \bibinfo {author} {\bibfnamefont {J.~S.}\ \bibnamefont {Pinter}}, \bibinfo {author} {\bibfnamefont {J.~B.}\ \bibnamefont {Stoker}}, \bibinfo {author} {\bibfnamefont {S.~L.}\ \bibnamefont {Tabor}}, \bibinfo {author} {\bibfnamefont {T.}~\bibnamefont {Sumikama}}, \bibinfo {author} {\bibfnamefont
  {W.~B.}\ \bibnamefont {Walters}}, \bibinfo {author} {\bibfnamefont {X.}~\bibnamefont {Wang}},\ and\ \bibinfo {author} {\bibfnamefont {S.}~\bibnamefont {Zhu}},\ }\href {https://doi.org/10.1103/PhysRevC.77.014313} {\bibfield  {journal} {\bibinfo  {journal} {Phys. Rev. C}\ }\textbf {\bibinfo {volume} {77}},\ \bibinfo {pages} {014313} (\bibinfo {year} {2008})}\BibitemShut {NoStop}%
\bibitem [{\citenamefont {Crawford}\ \emph {et~al.}(2010)\citenamefont {Crawford}, \citenamefont {Janssens}, \citenamefont {Mantica}, \citenamefont {Berryman}, \citenamefont {Broda}, \citenamefont {Carpenter}, \citenamefont {Cieplicka}, \citenamefont {Fornal}, \citenamefont {Grinyer}, \citenamefont {Hoteling}, \citenamefont {Kay}, \citenamefont {Lauritsen}, \citenamefont {Minamisono}, \citenamefont {Stefanescu}, \citenamefont {Stoker}, \citenamefont {Walters},\ and\ \citenamefont {Zhu}}]{Crawford2010}%
  \BibitemOpen
  \bibfield  {author} {\bibinfo {author} {\bibfnamefont {H.~L.}\ \bibnamefont {Crawford}}, \bibinfo {author} {\bibfnamefont {R.~V.~F.}\ \bibnamefont {Janssens}}, \bibinfo {author} {\bibfnamefont {P.~F.}\ \bibnamefont {Mantica}}, \bibinfo {author} {\bibfnamefont {J.~S.}\ \bibnamefont {Berryman}}, \bibinfo {author} {\bibfnamefont {R.}~\bibnamefont {Broda}}, \bibinfo {author} {\bibfnamefont {M.~P.}\ \bibnamefont {Carpenter}}, \bibinfo {author} {\bibfnamefont {N.}~\bibnamefont {Cieplicka}}, \bibinfo {author} {\bibfnamefont {B.}~\bibnamefont {Fornal}}, \bibinfo {author} {\bibfnamefont {G.~F.}\ \bibnamefont {Grinyer}}, \bibinfo {author} {\bibfnamefont {N.}~\bibnamefont {Hoteling}}, \bibinfo {author} {\bibfnamefont {B.~P.}\ \bibnamefont {Kay}}, \bibinfo {author} {\bibfnamefont {T.}~\bibnamefont {Lauritsen}}, \bibinfo {author} {\bibfnamefont {K.}~\bibnamefont {Minamisono}}, \bibinfo {author} {\bibfnamefont {I.}~\bibnamefont {Stefanescu}}, \bibinfo {author} {\bibfnamefont {J.~B.}\ \bibnamefont {Stoker}}, \bibinfo
  {author} {\bibfnamefont {W.~B.}\ \bibnamefont {Walters}},\ and\ \bibinfo {author} {\bibfnamefont {S.}~\bibnamefont {Zhu}},\ }\href {https://doi.org/10.1103/PhysRevC.82.014311} {\bibfield  {journal} {\bibinfo  {journal} {Phys. Rev. C}\ }\textbf {\bibinfo {volume} {82}},\ \bibinfo {pages} {014311} (\bibinfo {year} {2010})}\BibitemShut {NoStop}%
\bibitem [{\citenamefont {Grzywacz}\ \emph {et~al.}(1998)\citenamefont {Grzywacz}, \citenamefont {B\'eraud}, \citenamefont {Borcea}, \citenamefont {Emsallem}, \citenamefont {Glogowski}, \citenamefont {Grawe}, \citenamefont {Guillemaud-Mueller}, \citenamefont {Hjorth-Jensen}, \citenamefont {Houry}, \citenamefont {Lewitowicz}, \citenamefont {Mueller}, \citenamefont {Nowak}, \citenamefont {P\l{}ochocki}, \citenamefont {Pf\"utzner}, \citenamefont {Rykaczewski}, \citenamefont {Saint-Laurent}, \citenamefont {Sauvestre}, \citenamefont {Schaefer}, \citenamefont {Sorlin}, \citenamefont {Szerypo}, \citenamefont {Trinder}, \citenamefont {Viteritti},\ and\ \citenamefont {Winfield}}]{Grzywacz1998}%
  \BibitemOpen
  \bibfield  {author} {\bibinfo {author} {\bibfnamefont {R.}~\bibnamefont {Grzywacz}}, \bibinfo {author} {\bibfnamefont {R.}~\bibnamefont {B\'eraud}}, \bibinfo {author} {\bibfnamefont {C.}~\bibnamefont {Borcea}}, \bibinfo {author} {\bibfnamefont {A.}~\bibnamefont {Emsallem}}, \bibinfo {author} {\bibfnamefont {M.}~\bibnamefont {Glogowski}}, \bibinfo {author} {\bibfnamefont {H.}~\bibnamefont {Grawe}}, \bibinfo {author} {\bibfnamefont {D.}~\bibnamefont {Guillemaud-Mueller}}, \bibinfo {author} {\bibfnamefont {M.}~\bibnamefont {Hjorth-Jensen}}, \bibinfo {author} {\bibfnamefont {M.}~\bibnamefont {Houry}}, \bibinfo {author} {\bibfnamefont {M.}~\bibnamefont {Lewitowicz}}, \bibinfo {author} {\bibfnamefont {A.~C.}\ \bibnamefont {Mueller}}, \bibinfo {author} {\bibfnamefont {A.}~\bibnamefont {Nowak}}, \bibinfo {author} {\bibfnamefont {A.}~\bibnamefont {P\l{}ochocki}}, \bibinfo {author} {\bibfnamefont {M.}~\bibnamefont {Pf\"utzner}}, \bibinfo {author} {\bibfnamefont {K.}~\bibnamefont {Rykaczewski}}, \bibinfo {author}
  {\bibfnamefont {M.~G.}\ \bibnamefont {Saint-Laurent}}, \bibinfo {author} {\bibfnamefont {J.~E.}\ \bibnamefont {Sauvestre}}, \bibinfo {author} {\bibfnamefont {M.}~\bibnamefont {Schaefer}}, \bibinfo {author} {\bibfnamefont {O.}~\bibnamefont {Sorlin}}, \bibinfo {author} {\bibfnamefont {J.}~\bibnamefont {Szerypo}}, \bibinfo {author} {\bibfnamefont {W.}~\bibnamefont {Trinder}}, \bibinfo {author} {\bibfnamefont {S.}~\bibnamefont {Viteritti}},\ and\ \bibinfo {author} {\bibfnamefont {J.}~\bibnamefont {Winfield}},\ }\href {https://doi.org/10.1103/PhysRevLett.81.766} {\bibfield  {journal} {\bibinfo  {journal} {Phys. Rev. Lett.}\ }\textbf {\bibinfo {volume} {81}},\ \bibinfo {pages} {766} (\bibinfo {year} {1998})}\BibitemShut {NoStop}%
\bibitem [{\citenamefont {Poves}\ \emph {et~al.}(2001)\citenamefont {Poves}, \citenamefont {Sánchez-Solano}, \citenamefont {Caurier},\ and\ \citenamefont {Nowacki}}]{Poves2001}%
  \BibitemOpen
  \bibfield  {author} {\bibinfo {author} {\bibfnamefont {A.}~\bibnamefont {Poves}}, \bibinfo {author} {\bibfnamefont {J.}~\bibnamefont {Sánchez-Solano}}, \bibinfo {author} {\bibfnamefont {E.}~\bibnamefont {Caurier}},\ and\ \bibinfo {author} {\bibfnamefont {F.}~\bibnamefont {Nowacki}},\ }\href {https://doi.org/https://doi.org/10.1016/S0375-9474(01)00967-8} {\bibfield  {journal} {\bibinfo  {journal} {Nuclear Physics A}\ }\textbf {\bibinfo {volume} {694}},\ \bibinfo {pages} {157} (\bibinfo {year} {2001})}\BibitemShut {NoStop}%
\bibitem [{\citenamefont {Honma}\ \emph {et~al.}(2005)\citenamefont {Honma}, \citenamefont {Otsuka}, \citenamefont {Brown},\ and\ \citenamefont {Mizusaki}}]{Honma2005}%
  \BibitemOpen
  \bibfield  {author} {\bibinfo {author} {\bibfnamefont {M.}~\bibnamefont {Honma}}, \bibinfo {author} {\bibfnamefont {T.}~\bibnamefont {Otsuka}}, \bibinfo {author} {\bibfnamefont {B.~A.}\ \bibnamefont {Brown}},\ and\ \bibinfo {author} {\bibfnamefont {T.}~\bibnamefont {Mizusaki}},\ }\href {https://doi.org/https://doi.org/10.1140/epjad/i2005-06-032-2} {\bibfield  {journal} {\bibinfo  {journal} {Eur. Phys. J. A}\ }\textbf {\bibinfo {volume} {25}},\ \bibinfo {pages} {499} (\bibinfo {year} {2005})}\BibitemShut {NoStop}%
\bibitem [{\citenamefont {Magilligan}\ \emph {et~al.}(2021)\citenamefont {Magilligan}, \citenamefont {Brown},\ and\ \citenamefont {Stroberg}}]{Magilligan2021}%
  \BibitemOpen
  \bibfield  {author} {\bibinfo {author} {\bibfnamefont {A.}~\bibnamefont {Magilligan}}, \bibinfo {author} {\bibfnamefont {B.~A.}\ \bibnamefont {Brown}},\ and\ \bibinfo {author} {\bibfnamefont {S.~R.}\ \bibnamefont {Stroberg}},\ }\href {https://doi.org/10.1103/PhysRevC.104.L051302} {\bibfield  {journal} {\bibinfo  {journal} {Phys. Rev. C}\ }\textbf {\bibinfo {volume} {104}},\ \bibinfo {pages} {L051302} (\bibinfo {year} {2021})}\BibitemShut {NoStop}%
\bibitem [{\citenamefont {Coraggio}\ \emph {et~al.}(2020)\citenamefont {Coraggio}, \citenamefont {De~Gregorio}, \citenamefont {Gargano}, \citenamefont {Itaco}, \citenamefont {Fukui}, \citenamefont {Ma},\ and\ \citenamefont {Xu}}]{Coraggio2020}%
  \BibitemOpen
  \bibfield  {author} {\bibinfo {author} {\bibfnamefont {L.}~\bibnamefont {Coraggio}}, \bibinfo {author} {\bibfnamefont {G.}~\bibnamefont {De~Gregorio}}, \bibinfo {author} {\bibfnamefont {A.}~\bibnamefont {Gargano}}, \bibinfo {author} {\bibfnamefont {N.}~\bibnamefont {Itaco}}, \bibinfo {author} {\bibfnamefont {T.}~\bibnamefont {Fukui}}, \bibinfo {author} {\bibfnamefont {Y.~Z.}\ \bibnamefont {Ma}},\ and\ \bibinfo {author} {\bibfnamefont {F.~R.}\ \bibnamefont {Xu}},\ }\href {https://doi.org/10.1103/PhysRevC.102.054326} {\bibfield  {journal} {\bibinfo  {journal} {Phys. Rev. C}\ }\textbf {\bibinfo {volume} {102}},\ \bibinfo {pages} {054326} (\bibinfo {year} {2020})}\BibitemShut {NoStop}%
\bibitem [{\citenamefont {Coraggio}\ \emph {et~al.}(2021)\citenamefont {Coraggio}, \citenamefont {De~Gregorio}, \citenamefont {Gargano}, \citenamefont {Itaco}, \citenamefont {Fukui}, \citenamefont {Ma},\ and\ \citenamefont {Xu}}]{Coraggio2021}%
  \BibitemOpen
  \bibfield  {author} {\bibinfo {author} {\bibfnamefont {L.}~\bibnamefont {Coraggio}}, \bibinfo {author} {\bibfnamefont {G.}~\bibnamefont {De~Gregorio}}, \bibinfo {author} {\bibfnamefont {A.}~\bibnamefont {Gargano}}, \bibinfo {author} {\bibfnamefont {N.}~\bibnamefont {Itaco}}, \bibinfo {author} {\bibfnamefont {T.}~\bibnamefont {Fukui}}, \bibinfo {author} {\bibfnamefont {Y.~Z.}\ \bibnamefont {Ma}},\ and\ \bibinfo {author} {\bibfnamefont {F.~R.}\ \bibnamefont {Xu}},\ }\href {https://doi.org/10.1103/PhysRevC.104.054304} {\bibfield  {journal} {\bibinfo  {journal} {Phys. Rev. C}\ }\textbf {\bibinfo {volume} {104}},\ \bibinfo {pages} {054304} (\bibinfo {year} {2021})}\BibitemShut {NoStop}%
\bibitem [{\citenamefont {Hergert}\ \emph {et~al.}(2016)\citenamefont {Hergert}, \citenamefont {Bogner}, \citenamefont {Morris}, \citenamefont {Schwenk},\ and\ \citenamefont {Tsukiyama}}]{hergert2016}%
  \BibitemOpen
  \bibfield  {author} {\bibinfo {author} {\bibfnamefont {H.}~\bibnamefont {Hergert}}, \bibinfo {author} {\bibfnamefont {S.~K.}\ \bibnamefont {Bogner}}, \bibinfo {author} {\bibfnamefont {T.~D.}\ \bibnamefont {Morris}}, \bibinfo {author} {\bibfnamefont {A.}~\bibnamefont {Schwenk}},\ and\ \bibinfo {author} {\bibfnamefont {K.}~\bibnamefont {Tsukiyama}},\ }\href {https://doi.org/https://doi.org/10.1016/j.physrep.2015.12.007} {\bibfield  {journal} {\bibinfo  {journal} {Physics Reports}\ }\textbf {\bibinfo {volume} {621}},\ \bibinfo {pages} {165} (\bibinfo {year} {2016})},\ \bibinfo {note} {{M}emorial Volume in Honor of Gerald E. Brown}\BibitemShut {NoStop}%
\bibitem [{\citenamefont {Stroberg}\ \emph {et~al.}(2019)\citenamefont {Stroberg}, \citenamefont {Hergert}, \citenamefont {Bogner},\ and\ \citenamefont {Holt}}]{Stroberg2019}%
  \BibitemOpen
  \bibfield  {author} {\bibinfo {author} {\bibfnamefont {S.~R.}\ \bibnamefont {Stroberg}}, \bibinfo {author} {\bibfnamefont {H.}~\bibnamefont {Hergert}}, \bibinfo {author} {\bibfnamefont {S.~K.}\ \bibnamefont {Bogner}},\ and\ \bibinfo {author} {\bibfnamefont {J.~D.}\ \bibnamefont {Holt}},\ }\href {https://doi.org/10.1146/annurev-nucl-101917-021120} {\bibfield  {journal} {\bibinfo  {journal} {Annu. Rev. Nucl. Part. Sci.}\ }\textbf {\bibinfo {volume} {69}},\ \bibinfo {pages} {307} (\bibinfo {year} {2019})}\BibitemShut {NoStop}%
\bibitem [{\citenamefont {Stroberg}(2023)}]{imsrg}%
  \BibitemOpen
  \bibfield  {author} {\bibinfo {author} {\bibfnamefont {S.~R.}\ \bibnamefont {Stroberg}},\ }\href {https://github.com/ragnarstroberg/imsrg} {\bibinfo {title} {https://github.com/ragnarstroberg/imsrg}} (\bibinfo {year} {2023})\BibitemShut {NoStop}%
\bibitem [{\citenamefont {Shimizu}\ \emph {et~al.}(2019)\citenamefont {Shimizu}, \citenamefont {Mizusaki}, \citenamefont {Utsuno},\ and\ \citenamefont {Tsunoda}}]{Shimizu2019}%
  \BibitemOpen
  \bibfield  {author} {\bibinfo {author} {\bibfnamefont {N.}~\bibnamefont {Shimizu}}, \bibinfo {author} {\bibfnamefont {T.}~\bibnamefont {Mizusaki}}, \bibinfo {author} {\bibfnamefont {Y.}~\bibnamefont {Utsuno}},\ and\ \bibinfo {author} {\bibfnamefont {Y.}~\bibnamefont {Tsunoda}},\ }\href {https://doi.org/10.1016/j.cpc.2019.06.011} {\bibfield  {journal} {\bibinfo  {journal} {Comput. Phys. Commun.}\ }\textbf {\bibinfo {volume} {244}},\ \bibinfo {pages} {372} (\bibinfo {year} {2019})}\BibitemShut {NoStop}%
\bibitem [{\citenamefont {Koiwai}\ \emph {et~al.}(2022)\citenamefont {Koiwai}, \citenamefont {Wimmer}, \citenamefont {Doornenbal}, \citenamefont {Obertelli}, \citenamefont {Barbieri}, \citenamefont {Duguet}, \citenamefont {Holt}, \citenamefont {Miyagi}, \citenamefont {Navrátil}, \citenamefont {Ogata}, \citenamefont {Shimizu}, \citenamefont {Somà}, \citenamefont {Utsuno}, \citenamefont {Yoshida}, \citenamefont {Achouri}, \citenamefont {Baba}, \citenamefont {Browne}, \citenamefont {Calvet}, \citenamefont {Château}, \citenamefont {Chen}, \citenamefont {Chiga}, \citenamefont {Corsi}, \citenamefont {Cortés}, \citenamefont {Delbart}, \citenamefont {Gheller}, \citenamefont {Giganon}, \citenamefont {Gillibert}, \citenamefont {Hilaire}, \citenamefont {Isobe}, \citenamefont {Kobayashi}, \citenamefont {Kubota}, \citenamefont {Lapoux}, \citenamefont {Liu}, \citenamefont {Motobayashi}, \citenamefont {Murray}, \citenamefont {Otsu}, \citenamefont {Panin}, \citenamefont {Paul}, \citenamefont {Rodriguez}, \citenamefont
  {Sakurai}, \citenamefont {Sasano}, \citenamefont {Steppenbeck}, \citenamefont {Stuhl}, \citenamefont {Sun}, \citenamefont {Togano}, \citenamefont {Uesaka}, \citenamefont {Yoneda}, \citenamefont {Aktas}, \citenamefont {Aumann}, \citenamefont {Chung}, \citenamefont {Flavigny}, \citenamefont {Franchoo}, \citenamefont {Gasparic}, \citenamefont {Gerst}, \citenamefont {Gibelin}, \citenamefont {Hahn}, \citenamefont {Kim}, \citenamefont {Kondo}, \citenamefont {Koseoglou}, \citenamefont {Lee}, \citenamefont {Lehr}, \citenamefont {Linh}, \citenamefont {Lokotko}, \citenamefont {MacCormick}, \citenamefont {Moschner}, \citenamefont {Nakamura}, \citenamefont {Park}, \citenamefont {Rossi}, \citenamefont {Sahin}, \citenamefont {Söderström}, \citenamefont {Sohler}, \citenamefont {Takeuchi}, \citenamefont {Toernqvist}, \citenamefont {Vaquero}, \citenamefont {Wagner}, \citenamefont {Wang}, \citenamefont {Werner}, \citenamefont {Xu}, \citenamefont {Yamada}, \citenamefont {Yan}, \citenamefont {Yang}, \citenamefont {Yasuda},\
  and\ \citenamefont {Zanetti}}]{Koiwai2022}%
  \BibitemOpen
  \bibfield  {author} {\bibinfo {author} {\bibfnamefont {T.}~\bibnamefont {Koiwai}}, \bibinfo {author} {\bibfnamefont {K.}~\bibnamefont {Wimmer}}, \bibinfo {author} {\bibfnamefont {P.}~\bibnamefont {Doornenbal}}, \bibinfo {author} {\bibfnamefont {A.}~\bibnamefont {Obertelli}}, \bibinfo {author} {\bibfnamefont {C.}~\bibnamefont {Barbieri}}, \bibinfo {author} {\bibfnamefont {T.}~\bibnamefont {Duguet}}, \bibinfo {author} {\bibfnamefont {J.~D.}\ \bibnamefont {Holt}}, \bibinfo {author} {\bibfnamefont {T.}~\bibnamefont {Miyagi}}, \bibinfo {author} {\bibfnamefont {P.}~\bibnamefont {Navrátil}}, \bibinfo {author} {\bibfnamefont {K.}~\bibnamefont {Ogata}}, \bibinfo {author} {\bibfnamefont {N.}~\bibnamefont {Shimizu}}, \bibinfo {author} {\bibfnamefont {V.}~\bibnamefont {Somà}}, \bibinfo {author} {\bibfnamefont {Y.}~\bibnamefont {Utsuno}}, \bibinfo {author} {\bibfnamefont {K.}~\bibnamefont {Yoshida}}, \bibinfo {author} {\bibfnamefont {N.~L.}\ \bibnamefont {Achouri}}, \bibinfo {author} {\bibfnamefont {H.}~\bibnamefont
  {Baba}}, \bibinfo {author} {\bibfnamefont {F.}~\bibnamefont {Browne}}, \bibinfo {author} {\bibfnamefont {D.}~\bibnamefont {Calvet}}, \bibinfo {author} {\bibfnamefont {F.}~\bibnamefont {Château}}, \bibinfo {author} {\bibfnamefont {S.}~\bibnamefont {Chen}}, \bibinfo {author} {\bibfnamefont {N.}~\bibnamefont {Chiga}}, \bibinfo {author} {\bibfnamefont {A.}~\bibnamefont {Corsi}}, \bibinfo {author} {\bibfnamefont {M.~L.}\ \bibnamefont {Cortés}}, \bibinfo {author} {\bibfnamefont {A.}~\bibnamefont {Delbart}}, \bibinfo {author} {\bibfnamefont {J.~M.}\ \bibnamefont {Gheller}}, \bibinfo {author} {\bibfnamefont {A.}~\bibnamefont {Giganon}}, \bibinfo {author} {\bibfnamefont {A.}~\bibnamefont {Gillibert}}, \bibinfo {author} {\bibfnamefont {C.}~\bibnamefont {Hilaire}}, \bibinfo {author} {\bibfnamefont {T.}~\bibnamefont {Isobe}}, \bibinfo {author} {\bibfnamefont {T.}~\bibnamefont {Kobayashi}}, \bibinfo {author} {\bibfnamefont {Y.}~\bibnamefont {Kubota}}, \bibinfo {author} {\bibfnamefont {V.}~\bibnamefont {Lapoux}},
  \bibinfo {author} {\bibfnamefont {H.}~\bibnamefont {Liu}}, \bibinfo {author} {\bibfnamefont {T.}~\bibnamefont {Motobayashi}}, \bibinfo {author} {\bibfnamefont {I.}~\bibnamefont {Murray}}, \bibinfo {author} {\bibfnamefont {H.}~\bibnamefont {Otsu}}, \bibinfo {author} {\bibfnamefont {V.}~\bibnamefont {Panin}}, \bibinfo {author} {\bibfnamefont {N.}~\bibnamefont {Paul}}, \bibinfo {author} {\bibfnamefont {W.}~\bibnamefont {Rodriguez}}, \bibinfo {author} {\bibfnamefont {H.}~\bibnamefont {Sakurai}}, \bibinfo {author} {\bibfnamefont {M.}~\bibnamefont {Sasano}}, \bibinfo {author} {\bibfnamefont {D.}~\bibnamefont {Steppenbeck}}, \bibinfo {author} {\bibfnamefont {L.}~\bibnamefont {Stuhl}}, \bibinfo {author} {\bibfnamefont {Y.~L.}\ \bibnamefont {Sun}}, \bibinfo {author} {\bibfnamefont {Y.}~\bibnamefont {Togano}}, \bibinfo {author} {\bibfnamefont {T.}~\bibnamefont {Uesaka}}, \bibinfo {author} {\bibfnamefont {K.}~\bibnamefont {Yoneda}}, \bibinfo {author} {\bibfnamefont {O.}~\bibnamefont {Aktas}}, \bibinfo {author}
  {\bibfnamefont {T.}~\bibnamefont {Aumann}}, \bibinfo {author} {\bibfnamefont {L.~X.}\ \bibnamefont {Chung}}, \bibinfo {author} {\bibfnamefont {F.}~\bibnamefont {Flavigny}}, \bibinfo {author} {\bibfnamefont {S.}~\bibnamefont {Franchoo}}, \bibinfo {author} {\bibfnamefont {I.}~\bibnamefont {Gasparic}}, \bibinfo {author} {\bibfnamefont {R.~B.}\ \bibnamefont {Gerst}}, \bibinfo {author} {\bibfnamefont {J.}~\bibnamefont {Gibelin}}, \bibinfo {author} {\bibfnamefont {K.~I.}\ \bibnamefont {Hahn}}, \bibinfo {author} {\bibfnamefont {D.}~\bibnamefont {Kim}}, \bibinfo {author} {\bibfnamefont {Y.}~\bibnamefont {Kondo}}, \bibinfo {author} {\bibfnamefont {P.}~\bibnamefont {Koseoglou}}, \bibinfo {author} {\bibfnamefont {J.}~\bibnamefont {Lee}}, \bibinfo {author} {\bibfnamefont {C.}~\bibnamefont {Lehr}}, \bibinfo {author} {\bibfnamefont {B.~D.}\ \bibnamefont {Linh}}, \bibinfo {author} {\bibfnamefont {T.}~\bibnamefont {Lokotko}}, \bibinfo {author} {\bibfnamefont {M.}~\bibnamefont {MacCormick}}, \bibinfo {author} {\bibfnamefont
  {K.}~\bibnamefont {Moschner}}, \bibinfo {author} {\bibfnamefont {T.}~\bibnamefont {Nakamura}}, \bibinfo {author} {\bibfnamefont {S.~Y.}\ \bibnamefont {Park}}, \bibinfo {author} {\bibfnamefont {D.}~\bibnamefont {Rossi}}, \bibinfo {author} {\bibfnamefont {E.}~\bibnamefont {Sahin}}, \bibinfo {author} {\bibfnamefont {P.~A.}\ \bibnamefont {Söderström}}, \bibinfo {author} {\bibfnamefont {D.}~\bibnamefont {Sohler}}, \bibinfo {author} {\bibfnamefont {S.}~\bibnamefont {Takeuchi}}, \bibinfo {author} {\bibfnamefont {H.}~\bibnamefont {Toernqvist}}, \bibinfo {author} {\bibfnamefont {V.}~\bibnamefont {Vaquero}}, \bibinfo {author} {\bibfnamefont {V.}~\bibnamefont {Wagner}}, \bibinfo {author} {\bibfnamefont {S.}~\bibnamefont {Wang}}, \bibinfo {author} {\bibfnamefont {V.}~\bibnamefont {Werner}}, \bibinfo {author} {\bibfnamefont {X.}~\bibnamefont {Xu}}, \bibinfo {author} {\bibfnamefont {H.}~\bibnamefont {Yamada}}, \bibinfo {author} {\bibfnamefont {D.}~\bibnamefont {Yan}}, \bibinfo {author} {\bibfnamefont {Z.}~\bibnamefont
  {Yang}}, \bibinfo {author} {\bibfnamefont {M.}~\bibnamefont {Yasuda}},\ and\ \bibinfo {author} {\bibfnamefont {L.}~\bibnamefont {Zanetti}},\ }\href {https://doi.org/https://doi.org/10.1016/j.physletb.2022.136953} {\bibfield  {journal} {\bibinfo  {journal} {Physics Letters B}\ }\textbf {\bibinfo {volume} {827}},\ \bibinfo {pages} {136953} (\bibinfo {year} {2022})}\BibitemShut {NoStop}%
\bibitem [{\citenamefont {Brown}\ \emph {et~al.}(1974)\citenamefont {Brown}, \citenamefont {Fossan}, \citenamefont {McDonald},\ and\ \citenamefont {Snover}}]{brown1974}%
  \BibitemOpen
  \bibfield  {author} {\bibinfo {author} {\bibfnamefont {B.~A.}\ \bibnamefont {Brown}}, \bibinfo {author} {\bibfnamefont {D.~B.}\ \bibnamefont {Fossan}}, \bibinfo {author} {\bibfnamefont {J.~M.}\ \bibnamefont {McDonald}},\ and\ \bibinfo {author} {\bibfnamefont {K.~A.}\ \bibnamefont {Snover}},\ }\href {https://doi.org/10.1103/PhysRevC.9.1033} {\bibfield  {journal} {\bibinfo  {journal} {Phys. Rev. C}\ }\textbf {\bibinfo {volume} {9}},\ \bibinfo {pages} {1033} (\bibinfo {year} {1974})}\BibitemShut {NoStop}%
\bibitem [{\citenamefont {Lisetskiy}\ \emph {et~al.}(2004)\citenamefont {Lisetskiy}, \citenamefont {Brown}, \citenamefont {Horoi},\ and\ \citenamefont {Grawe}}]{Lisetskiy2004}%
  \BibitemOpen
  \bibfield  {author} {\bibinfo {author} {\bibfnamefont {A.~F.}\ \bibnamefont {Lisetskiy}}, \bibinfo {author} {\bibfnamefont {B.~A.}\ \bibnamefont {Brown}}, \bibinfo {author} {\bibfnamefont {M.}~\bibnamefont {Horoi}},\ and\ \bibinfo {author} {\bibfnamefont {H.}~\bibnamefont {Grawe}},\ }\href {https://doi.org/10.1103/PhysRevC.70.044314} {\bibfield  {journal} {\bibinfo  {journal} {Phys. Rev. C}\ }\textbf {\bibinfo {volume} {70}},\ \bibinfo {pages} {044314} (\bibinfo {year} {2004})}\BibitemShut {NoStop}%
\bibitem [{\citenamefont {Blomqvist}\ and\ \citenamefont {Molinari}(1968)}]{BLOMQVIST1968}%
  \BibitemOpen
  \bibfield  {author} {\bibinfo {author} {\bibfnamefont {J.}~\bibnamefont {Blomqvist}}\ and\ \bibinfo {author} {\bibfnamefont {A.}~\bibnamefont {Molinari}},\ }\href {https://doi.org/https://doi.org/10.1016/0375-9474(68)90515-0} {\bibfield  {journal} {\bibinfo  {journal} {Nuclear Physics A}\ }\textbf {\bibinfo {volume} {106}},\ \bibinfo {pages} {545} (\bibinfo {year} {1968})}\BibitemShut {NoStop}%
\bibitem [{\citenamefont {Garcia~Ruiz}\ \emph {et~al.}(2015)\citenamefont {Garcia~Ruiz}, \citenamefont {Bissell}, \citenamefont {Blaum}, \citenamefont {Fr\"ommgen}, \citenamefont {Hammen}, \citenamefont {Holt}, \citenamefont {Kowalska}, \citenamefont {Kreim}, \citenamefont {Men\'endez}, \citenamefont {Neugart}, \citenamefont {Neyens}, \citenamefont {N\"ortersh\"auser}, \citenamefont {Nowacki}, \citenamefont {Papuga}, \citenamefont {Poves}, \citenamefont {Schwenk}, \citenamefont {Simonis},\ and\ \citenamefont {Yordanov}}]{GarcaRuiz2015}%
  \BibitemOpen
  \bibfield  {author} {\bibinfo {author} {\bibfnamefont {R.~F.}\ \bibnamefont {Garcia~Ruiz}}, \bibinfo {author} {\bibfnamefont {M.~L.}\ \bibnamefont {Bissell}}, \bibinfo {author} {\bibfnamefont {K.}~\bibnamefont {Blaum}}, \bibinfo {author} {\bibfnamefont {N.}~\bibnamefont {Fr\"ommgen}}, \bibinfo {author} {\bibfnamefont {M.}~\bibnamefont {Hammen}}, \bibinfo {author} {\bibfnamefont {J.~D.}\ \bibnamefont {Holt}}, \bibinfo {author} {\bibfnamefont {M.}~\bibnamefont {Kowalska}}, \bibinfo {author} {\bibfnamefont {K.}~\bibnamefont {Kreim}}, \bibinfo {author} {\bibfnamefont {J.}~\bibnamefont {Men\'endez}}, \bibinfo {author} {\bibfnamefont {R.}~\bibnamefont {Neugart}}, \bibinfo {author} {\bibfnamefont {G.}~\bibnamefont {Neyens}}, \bibinfo {author} {\bibfnamefont {W.}~\bibnamefont {N\"ortersh\"auser}}, \bibinfo {author} {\bibfnamefont {F.}~\bibnamefont {Nowacki}}, \bibinfo {author} {\bibfnamefont {J.}~\bibnamefont {Papuga}}, \bibinfo {author} {\bibfnamefont {A.}~\bibnamefont {Poves}}, \bibinfo {author} {\bibfnamefont
  {A.}~\bibnamefont {Schwenk}}, \bibinfo {author} {\bibfnamefont {J.}~\bibnamefont {Simonis}},\ and\ \bibinfo {author} {\bibfnamefont {D.~T.}\ \bibnamefont {Yordanov}},\ }\href {https://doi.org/10.1103/PhysRevC.91.041304} {\bibfield  {journal} {\bibinfo  {journal} {Phys. Rev. C}\ }\textbf {\bibinfo {volume} {91}},\ \bibinfo {pages} {041304} (\bibinfo {year} {2015})}\BibitemShut {NoStop}%
\bibitem [{\citenamefont {du~Rietz}\ \emph {et~al.}(2004)\citenamefont {du~Rietz}, \citenamefont {Ekman}, \citenamefont {Rudolph}, \citenamefont {Fahlander}, \citenamefont {Dewald}, \citenamefont {M\"oller}, \citenamefont {Saha}, \citenamefont {Axiotis}, \citenamefont {Bentley}, \citenamefont {Chandler}, \citenamefont {de~Angelis}, \citenamefont {Della~Vedova}, \citenamefont {Gadea}, \citenamefont {Hammond}, \citenamefont {Lenzi}, \citenamefont {M\ifmmode~\u{a}\else \u{a}\fi{}rginean}, \citenamefont {Napoli}, \citenamefont {Nespolo}, \citenamefont {Rusu},\ and\ \citenamefont {Tonev}}]{duRietz2004}%
  \BibitemOpen
  \bibfield  {author} {\bibinfo {author} {\bibfnamefont {R.}~\bibnamefont {du~Rietz}}, \bibinfo {author} {\bibfnamefont {J.}~\bibnamefont {Ekman}}, \bibinfo {author} {\bibfnamefont {D.}~\bibnamefont {Rudolph}}, \bibinfo {author} {\bibfnamefont {C.}~\bibnamefont {Fahlander}}, \bibinfo {author} {\bibfnamefont {A.}~\bibnamefont {Dewald}}, \bibinfo {author} {\bibfnamefont {O.}~\bibnamefont {M\"oller}}, \bibinfo {author} {\bibfnamefont {B.}~\bibnamefont {Saha}}, \bibinfo {author} {\bibfnamefont {M.}~\bibnamefont {Axiotis}}, \bibinfo {author} {\bibfnamefont {M.~A.}\ \bibnamefont {Bentley}}, \bibinfo {author} {\bibfnamefont {C.}~\bibnamefont {Chandler}}, \bibinfo {author} {\bibfnamefont {G.}~\bibnamefont {de~Angelis}}, \bibinfo {author} {\bibfnamefont {F.}~\bibnamefont {Della~Vedova}}, \bibinfo {author} {\bibfnamefont {A.}~\bibnamefont {Gadea}}, \bibinfo {author} {\bibfnamefont {G.}~\bibnamefont {Hammond}}, \bibinfo {author} {\bibfnamefont {S.~M.}\ \bibnamefont {Lenzi}}, \bibinfo {author} {\bibfnamefont
  {N.}~\bibnamefont {M\ifmmode~\u{a}\else \u{a}\fi{}rginean}}, \bibinfo {author} {\bibfnamefont {D.~R.}\ \bibnamefont {Napoli}}, \bibinfo {author} {\bibfnamefont {M.}~\bibnamefont {Nespolo}}, \bibinfo {author} {\bibfnamefont {C.}~\bibnamefont {Rusu}},\ and\ \bibinfo {author} {\bibfnamefont {D.}~\bibnamefont {Tonev}},\ }\href {https://doi.org/10.1103/PhysRevLett.93.222501} {\bibfield  {journal} {\bibinfo  {journal} {Phys. Rev. Lett.}\ }\textbf {\bibinfo {volume} {93}},\ \bibinfo {pages} {222501} (\bibinfo {year} {2004})}\BibitemShut {NoStop}%
\bibitem [{\citenamefont {Allmond}\ \emph {et~al.}(2014)\citenamefont {Allmond}, \citenamefont {Brown}, \citenamefont {Stuchbery}, \citenamefont {Galindo-Uribarri}, \citenamefont {Padilla-Rodal}, \citenamefont {Radford}, \citenamefont {Batchelder}, \citenamefont {Howard}, \citenamefont {Liang}, \citenamefont {Manning}, \citenamefont {Varner},\ and\ \citenamefont {Yu}}]{allmond2014}%
  \BibitemOpen
  \bibfield  {author} {\bibinfo {author} {\bibfnamefont {J.~M.}\ \bibnamefont {Allmond}}, \bibinfo {author} {\bibfnamefont {B.~A.}\ \bibnamefont {Brown}}, \bibinfo {author} {\bibfnamefont {A.~E.}\ \bibnamefont {Stuchbery}}, \bibinfo {author} {\bibfnamefont {A.}~\bibnamefont {Galindo-Uribarri}}, \bibinfo {author} {\bibfnamefont {E.}~\bibnamefont {Padilla-Rodal}}, \bibinfo {author} {\bibfnamefont {D.~C.}\ \bibnamefont {Radford}}, \bibinfo {author} {\bibfnamefont {J.~C.}\ \bibnamefont {Batchelder}}, \bibinfo {author} {\bibfnamefont {M.~E.}\ \bibnamefont {Howard}}, \bibinfo {author} {\bibfnamefont {J.~F.}\ \bibnamefont {Liang}}, \bibinfo {author} {\bibfnamefont {B.}~\bibnamefont {Manning}}, \bibinfo {author} {\bibfnamefont {R.~L.}\ \bibnamefont {Varner}},\ and\ \bibinfo {author} {\bibfnamefont {C.-H.}\ \bibnamefont {Yu}},\ }\href {https://doi.org/10.1103/PhysRevC.90.034309} {\bibfield  {journal} {\bibinfo  {journal} {Phys. Rev. C}\ }\textbf {\bibinfo {volume} {90}},\ \bibinfo {pages} {034309} (\bibinfo {year}
  {2014})}\BibitemShut {NoStop}%
\bibitem [{\citenamefont {Gray}\ \emph {et~al.}(2024)\citenamefont {Gray}, \citenamefont {Allmond}, \citenamefont {Benetti}, \citenamefont {Wibisono}, \citenamefont {Baby}, \citenamefont {Gargano}, \citenamefont {Miyagi}, \citenamefont {Macchiavelli}, \citenamefont {Stuchbery}, \citenamefont {Wood}, \citenamefont {Ajayi}, \citenamefont {Aragon}, \citenamefont {Asher}, \citenamefont {Barber}, \citenamefont {Bhattacharya}, \citenamefont {Boisseau}, \citenamefont {Christie}, \citenamefont {Conley}, \citenamefont {{De Rosa}}, \citenamefont {Dowling}, \citenamefont {Esparza}, \citenamefont {Gibbons}, \citenamefont {Hanselman}, \citenamefont {Holt}, \citenamefont {Lopez-Caceres}, \citenamefont {{Lopez Saavedra}}, \citenamefont {McCann}, \citenamefont {Morelock}, \citenamefont {Kelly}, \citenamefont {King}, \citenamefont {Rasco}, \citenamefont {Sitaraman}, \citenamefont {Tabor}, \citenamefont {Temanson}, \citenamefont {Tripathi}, \citenamefont {Wiedenhöver},\ and\ \citenamefont {Yadav}}]{gray2024}%
  \BibitemOpen
  \bibfield  {author} {\bibinfo {author} {\bibfnamefont {T.~J.}\ \bibnamefont {Gray}}, \bibinfo {author} {\bibfnamefont {J.~M.}\ \bibnamefont {Allmond}}, \bibinfo {author} {\bibfnamefont {C.}~\bibnamefont {Benetti}}, \bibinfo {author} {\bibfnamefont {C.}~\bibnamefont {Wibisono}}, \bibinfo {author} {\bibfnamefont {L.}~\bibnamefont {Baby}}, \bibinfo {author} {\bibfnamefont {A.}~\bibnamefont {Gargano}}, \bibinfo {author} {\bibfnamefont {T.}~\bibnamefont {Miyagi}}, \bibinfo {author} {\bibfnamefont {A.~O.}\ \bibnamefont {Macchiavelli}}, \bibinfo {author} {\bibfnamefont {A.~E.}\ \bibnamefont {Stuchbery}}, \bibinfo {author} {\bibfnamefont {J.~L.}\ \bibnamefont {Wood}}, \bibinfo {author} {\bibfnamefont {S.}~\bibnamefont {Ajayi}}, \bibinfo {author} {\bibfnamefont {J.}~\bibnamefont {Aragon}}, \bibinfo {author} {\bibfnamefont {B.~W.}\ \bibnamefont {Asher}}, \bibinfo {author} {\bibfnamefont {P.}~\bibnamefont {Barber}}, \bibinfo {author} {\bibfnamefont {S.}~\bibnamefont {Bhattacharya}}, \bibinfo {author} {\bibfnamefont
  {R.}~\bibnamefont {Boisseau}}, \bibinfo {author} {\bibfnamefont {J.~M.}\ \bibnamefont {Christie}}, \bibinfo {author} {\bibfnamefont {A.~L.}\ \bibnamefont {Conley}}, \bibinfo {author} {\bibfnamefont {P.}~\bibnamefont {{De Rosa}}}, \bibinfo {author} {\bibfnamefont {D.~T.}\ \bibnamefont {Dowling}}, \bibinfo {author} {\bibfnamefont {C.}~\bibnamefont {Esparza}}, \bibinfo {author} {\bibfnamefont {J.}~\bibnamefont {Gibbons}}, \bibinfo {author} {\bibfnamefont {K.}~\bibnamefont {Hanselman}}, \bibinfo {author} {\bibfnamefont {J.~D.}\ \bibnamefont {Holt}}, \bibinfo {author} {\bibfnamefont {S.}~\bibnamefont {Lopez-Caceres}}, \bibinfo {author} {\bibfnamefont {E.}~\bibnamefont {{Lopez Saavedra}}}, \bibinfo {author} {\bibfnamefont {G.~W.}\ \bibnamefont {McCann}}, \bibinfo {author} {\bibfnamefont {A.}~\bibnamefont {Morelock}}, \bibinfo {author} {\bibfnamefont {B.}~\bibnamefont {Kelly}}, \bibinfo {author} {\bibfnamefont {T.~T.}\ \bibnamefont {King}}, \bibinfo {author} {\bibfnamefont {B.~C.}\ \bibnamefont {Rasco}}, \bibinfo
  {author} {\bibfnamefont {V.}~\bibnamefont {Sitaraman}}, \bibinfo {author} {\bibfnamefont {S.~L.}\ \bibnamefont {Tabor}}, \bibinfo {author} {\bibfnamefont {E.}~\bibnamefont {Temanson}}, \bibinfo {author} {\bibfnamefont {V.}~\bibnamefont {Tripathi}}, \bibinfo {author} {\bibfnamefont {I.}~\bibnamefont {Wiedenhöver}},\ and\ \bibinfo {author} {\bibfnamefont {R.~B.}\ \bibnamefont {Yadav}},\ }\href {https://doi.org/https://doi.org/10.1016/j.physletb.2024.138856} {\bibfield  {journal} {\bibinfo  {journal} {Physics Letters B}\ }\textbf {\bibinfo {volume} {855}},\ \bibinfo {pages} {138856} (\bibinfo {year} {2024})}\BibitemShut {NoStop}%
\bibitem [{\citenamefont {Richter}\ \emph {et~al.}(2008)\citenamefont {Richter}, \citenamefont {Mkhize},\ and\ \citenamefont {Brown}}]{Richter2008}%
  \BibitemOpen
  \bibfield  {author} {\bibinfo {author} {\bibfnamefont {W.~A.}\ \bibnamefont {Richter}}, \bibinfo {author} {\bibfnamefont {S.}~\bibnamefont {Mkhize}},\ and\ \bibinfo {author} {\bibfnamefont {B.~A.}\ \bibnamefont {Brown}},\ }\href {https://doi.org/10.1103/PhysRevC.78.064302} {\bibfield  {journal} {\bibinfo  {journal} {Phys. Rev. C}\ }\textbf {\bibinfo {volume} {78}},\ \bibinfo {pages} {064302} (\bibinfo {year} {2008})}\BibitemShut {NoStop}%
\bibitem [{\citenamefont {Dufour}\ and\ \citenamefont {Zuker}(1996)}]{dufour1996}%
  \BibitemOpen
  \bibfield  {author} {\bibinfo {author} {\bibfnamefont {M.}~\bibnamefont {Dufour}}\ and\ \bibinfo {author} {\bibfnamefont {A.~P.}\ \bibnamefont {Zuker}},\ }\href {https://doi.org/10.1103/PhysRevC.54.1641} {\bibfield  {journal} {\bibinfo  {journal} {Phys. Rev. C}\ }\textbf {\bibinfo {volume} {54}},\ \bibinfo {pages} {1641} (\bibinfo {year} {1996})}\BibitemShut {NoStop}%
\bibitem [{\citenamefont {Dao}\ and\ \citenamefont {Nowacki}(2022)}]{dao2022}%
  \BibitemOpen
  \bibfield  {author} {\bibinfo {author} {\bibfnamefont {D.~D.}\ \bibnamefont {Dao}}\ and\ \bibinfo {author} {\bibfnamefont {F.}~\bibnamefont {Nowacki}},\ }\href {https://doi.org/10.1103/PhysRevC.105.054314} {\bibfield  {journal} {\bibinfo  {journal} {Phys. Rev. C}\ }\textbf {\bibinfo {volume} {105}},\ \bibinfo {pages} {054314} (\bibinfo {year} {2022})}\BibitemShut {NoStop}%
\bibitem [{\citenamefont {Caurier}\ \emph {et~al.}(2014)\citenamefont {Caurier}, \citenamefont {Nowacki},\ and\ \citenamefont {Poves}}]{caurier2014}%
  \BibitemOpen
  \bibfield  {author} {\bibinfo {author} {\bibfnamefont {E.}~\bibnamefont {Caurier}}, \bibinfo {author} {\bibfnamefont {F.}~\bibnamefont {Nowacki}},\ and\ \bibinfo {author} {\bibfnamefont {A.}~\bibnamefont {Poves}},\ }\href {https://doi.org/10.1103/PhysRevC.90.014302} {\bibfield  {journal} {\bibinfo  {journal} {Phys. Rev. C}\ }\textbf {\bibinfo {volume} {90}},\ \bibinfo {pages} {014302} (\bibinfo {year} {2014})}\BibitemShut {NoStop}%
\bibitem [{\citenamefont {Poves}\ \emph {et~al.}(2020)\citenamefont {Poves}, \citenamefont {Nowacki},\ and\ \citenamefont {Alhassid}}]{poves2020}%
  \BibitemOpen
  \bibfield  {author} {\bibinfo {author} {\bibfnamefont {A.}~\bibnamefont {Poves}}, \bibinfo {author} {\bibfnamefont {F.}~\bibnamefont {Nowacki}},\ and\ \bibinfo {author} {\bibfnamefont {Y.}~\bibnamefont {Alhassid}},\ }\href {https://doi.org/10.1103/PhysRevC.101.054307} {\bibfield  {journal} {\bibinfo  {journal} {Phys. Rev. C}\ }\textbf {\bibinfo {volume} {101}},\ \bibinfo {pages} {054307} (\bibinfo {year} {2020})}\BibitemShut {NoStop}%
\bibitem [{\citenamefont {Crawford}\ \emph {et~al.}(2013)\citenamefont {Crawford}, \citenamefont {Clark}, \citenamefont {Fallon}, \citenamefont {Macchiavelli}, \citenamefont {Baugher}, \citenamefont {Bazin}, \citenamefont {Beausang}, \citenamefont {Berryman}, \citenamefont {Bleuel}, \citenamefont {Campbell}, \citenamefont {Cromaz}, \citenamefont {de~Angelis}, \citenamefont {Gade}, \citenamefont {Hughes}, \citenamefont {Lee}, \citenamefont {Lenzi}, \citenamefont {Nowacki}, \citenamefont {Paschalis}, \citenamefont {Petri}, \citenamefont {Poves}, \citenamefont {Ratkiewicz}, \citenamefont {Ross}, \citenamefont {Sahin}, \citenamefont {Weisshaar}, \citenamefont {Wimmer},\ and\ \citenamefont {Winkler}}]{crawford2013}%
  \BibitemOpen
  \bibfield  {author} {\bibinfo {author} {\bibfnamefont {H.~L.}\ \bibnamefont {Crawford}}, \bibinfo {author} {\bibfnamefont {R.~M.}\ \bibnamefont {Clark}}, \bibinfo {author} {\bibfnamefont {P.}~\bibnamefont {Fallon}}, \bibinfo {author} {\bibfnamefont {A.~O.}\ \bibnamefont {Macchiavelli}}, \bibinfo {author} {\bibfnamefont {T.}~\bibnamefont {Baugher}}, \bibinfo {author} {\bibfnamefont {D.}~\bibnamefont {Bazin}}, \bibinfo {author} {\bibfnamefont {C.~W.}\ \bibnamefont {Beausang}}, \bibinfo {author} {\bibfnamefont {J.~S.}\ \bibnamefont {Berryman}}, \bibinfo {author} {\bibfnamefont {D.~L.}\ \bibnamefont {Bleuel}}, \bibinfo {author} {\bibfnamefont {C.~M.}\ \bibnamefont {Campbell}}, \bibinfo {author} {\bibfnamefont {M.}~\bibnamefont {Cromaz}}, \bibinfo {author} {\bibfnamefont {G.}~\bibnamefont {de~Angelis}}, \bibinfo {author} {\bibfnamefont {A.}~\bibnamefont {Gade}}, \bibinfo {author} {\bibfnamefont {R.~O.}\ \bibnamefont {Hughes}}, \bibinfo {author} {\bibfnamefont {I.~Y.}\ \bibnamefont {Lee}}, \bibinfo {author}
  {\bibfnamefont {S.~M.}\ \bibnamefont {Lenzi}}, \bibinfo {author} {\bibfnamefont {F.}~\bibnamefont {Nowacki}}, \bibinfo {author} {\bibfnamefont {S.}~\bibnamefont {Paschalis}}, \bibinfo {author} {\bibfnamefont {M.}~\bibnamefont {Petri}}, \bibinfo {author} {\bibfnamefont {A.}~\bibnamefont {Poves}}, \bibinfo {author} {\bibfnamefont {A.}~\bibnamefont {Ratkiewicz}}, \bibinfo {author} {\bibfnamefont {T.~J.}\ \bibnamefont {Ross}}, \bibinfo {author} {\bibfnamefont {E.}~\bibnamefont {Sahin}}, \bibinfo {author} {\bibfnamefont {D.}~\bibnamefont {Weisshaar}}, \bibinfo {author} {\bibfnamefont {K.}~\bibnamefont {Wimmer}},\ and\ \bibinfo {author} {\bibfnamefont {R.}~\bibnamefont {Winkler}},\ }\href {https://doi.org/10.1103/PhysRevLett.110.242701} {\bibfield  {journal} {\bibinfo  {journal} {Phys. Rev. Lett.}\ }\textbf {\bibinfo {volume} {110}},\ \bibinfo {pages} {242701} (\bibinfo {year} {2013})}\BibitemShut {NoStop}%
\bibitem [{\citenamefont {Nowacki}\ \emph {et~al.}(2016)\citenamefont {Nowacki}, \citenamefont {Poves}, \citenamefont {Caurier},\ and\ \citenamefont {Bounthong}}]{nowacki2016}%
  \BibitemOpen
  \bibfield  {author} {\bibinfo {author} {\bibfnamefont {F.}~\bibnamefont {Nowacki}}, \bibinfo {author} {\bibfnamefont {A.}~\bibnamefont {Poves}}, \bibinfo {author} {\bibfnamefont {E.}~\bibnamefont {Caurier}},\ and\ \bibinfo {author} {\bibfnamefont {B.}~\bibnamefont {Bounthong}},\ }\href {https://doi.org/10.1103/PhysRevLett.117.272501} {\bibfield  {journal} {\bibinfo  {journal} {Phys. Rev. Lett.}\ }\textbf {\bibinfo {volume} {117}},\ \bibinfo {pages} {272501} (\bibinfo {year} {2016})}\BibitemShut {NoStop}%
\end{thebibliography}
\end{document}